\documentclass[%
article, two column,
superscriptaddress,
preprintnumbers,
amsmath,amssymb,
aps,
prd,
longbibliography]{revtex4-2}
\usepackage{graphicx}
\usepackage{dcolumn}
\usepackage{bm}
\usepackage{bbold}
\usepackage{calc}
\usepackage{algpseudocode}
\usepackage{changes}
\usepackage[braket, qm]{qcircuit}
\usepackage{xcolor}
\usepackage{braket}
\usepackage[caption=false]{subfig}
\usepackage{amsthm}
\usepackage{ulem}
\usepackage{hyperref}
\usepackage{url}

\usepackage{color}
\usepackage{xcolor,colortbl}
\usepackage{comment}

\hypersetup{
	colorlinks=true,
	linkcolor=blue,
	citecolor=blue,
	filecolor=magenta,      
	urlcolor=blue,
	pdftitle={Overleaf Example},
	pdfpagemode=FullScreen,
}

\newcommand*{\vcenteredhbox}[1]{\begingroup
	\setbox0=\hbox{#1}\parbox{\wd0}{\box0}\endgroup}

\begin{document}
	
	\preprint{N3AS-24-010}
	
	\preprint{RIKEN-iTHEMS-Report-25}
	
	\preprint{NT@UW-25-4}

	\title{Collective Neutrino Oscillations in Three Flavors on Qubit and Qutrit Processors}
	\author{Luca Spagnoli}
	\affiliation{Physics Department, University of Trento, Via Sommarive 14, I-38123 Trento, Italy}
	\affiliation{INFN-TIFPA Trento Institute of Fundamental Physics and Applications,  , Via Sommarive 14, I-38123 Trento, Italy}
	\author{Noah Goss}
	\affiliation{Quantum Nanoelectronics Laboratory, Department of Physics, University of California at Berkeley, Berkeley, CA 94720, USA}
	\affiliation{Applied Mathematics and Computational Research Division, Lawrence Berkeley National Laboratory, Berkeley, CA 94720, USA}
	\author{Alessandro Roggero}
	\affiliation{Physics Department, University of Trento, Via Sommarive 14, I-38123 Trento, Italy}
	\affiliation{INFN-TIFPA Trento Institute of Fundamental Physics and Applications,  Trento, Italy}
	\author{Ermal Rrapaj}
	\affiliation{National Energy Research Scientific Computing Center, Lawrence Berkeley National Laboratory, Berkeley, CA 94720, USA }
	\affiliation{Department of Physics, University of California, Berkeley, CA 94720, USA}
	\affiliation{RIKEN iTHEMS, Wako, Saitama 351-0198, Japan}
	\author{Michael J.~Cervia}
	\affiliation{Department of Physics, University of Washington, Seattle, WA 98195, USA}
	\author{Amol V. Patwardhan}
	\affiliation{School of Physics and Astronomy, University of Minnesota, Minneapolis, MN 55455, USA}
	\affiliation{Department of Physics, New York Institute of Technology, New York, NY 10023, USA}
	\author{Ravi K. Naik}
	\affiliation{Quantum Nanoelectronics Laboratory, Department of Physics, University of California at Berkeley, Berkeley, CA 94720, USA}
	\affiliation{Applied Mathematics and Computational Research Division, Lawrence Berkeley National Laboratory, Berkeley, CA 94720, USA}
	\author{A. Baha Balantekin}
	\affiliation{Department of Physics, University of Wisconsin, Madison WI 53706, USA}
	\author{Ed Younis}
	\affiliation{Applied Mathematics and Computational Research Division, Lawrence Berkeley National Laboratory, Berkeley, CA 94720, USA}
	\author{David I. Santiago}
	\affiliation{Quantum Nanoelectronics Laboratory, Department of Physics, University of California at Berkeley, Berkeley, CA 94720, USA}
	\affiliation{Applied Mathematics and Computational Research Division, Lawrence Berkeley National Laboratory, Berkeley, CA 94720, USA}
	\author{Irfan Siddiqi}
	\affiliation{Quantum Nanoelectronics Laboratory, Department of Physics, University of California at Berkeley, Berkeley, CA 94720, USA}
	\affiliation{Applied Mathematics and Computational Research Division, Lawrence Berkeley National Laboratory, Berkeley, CA 94720, USA}
	
	\author{Sheakha Aldaihan}
	\affiliation{Department of Physics and Astronomy, College of Science, King Saud University, Riyadh 11451, Saudi Arabia}

	\begin{abstract}
		Collective neutrino flavor oscillations are of primary importance in understanding the dynamic evolution of core-collapse supernovae and subsequent terrestrial detection, but also among the most challenging aspects of numerical simulations. This situation is complicated by the quantum many-body nature of the problem due to neutrino-neutrino interactions which demands a quantum treatment. An additional complication is the presence of three flavors, which often is approximated by the electron flavor and a heavy lepton flavor. In this work, we provide both qubit and qutrit encodings for all three flavors, and develop optimized quantum circuits for the time evolution and analyze the Trotter error. We conclude our study with a hardware experiment of a system of two neutrinos with superconducting hardware: the IBM Torino device for qubits and Advanced Quantum Testbed device at the Lawrence Berkeley National Laboratory for qutrits. We find that error mitigation greatly helps in obtaining a signal consistent with simulations. While hardware results are comparable at this stage, we expect the qutrit setup to be more convenient for large-scale simulations since it does not suffer from probability leakage into nonphsycial qubit space, unlike the qubit setup. 
	\end{abstract}
	
	\maketitle
	
	\section{Introduction}
	
	Within the last several decades our knowledge of neutrino properties has grown significantly \cite{Balantekin:2018ppj,Huber:2022lpm}.
	However, even state-of-the-art three-dimensional hydrodynamical simulations of core-collapse supernova do not yet include the physics of neutrino flavor oscillations. Therefore, the precise extent to which these oscillations can affect the dynamics of this explosion and the accompanying nucleosynthesis remains an open question~\cite{Mirizzi:2015eza, Janka:2017vlw, Burrows:2020qrp}. 
	Indeed, flavor oscillations in such dense media are complex phenomena, because neutrinos not only interact with matter, yielding a one-body interaction term, but also with each other, introducing the two-body interaction term to a Hamiltonian that accounts for the collective neutrino oscillations (for reviews, see Refs.~\cite{Duan:2010bg, Chakraborty:2016yeg, Balantekin:2018mpq, Tamborra:2020cul, Richers:2022zug, Balantekin:2023qvm, Volpe:2023met}). The latter contribution makes the neutrino flavor oscillations a manifestly dynamic many-body problem \cite{PANTALEONE1992128,Balantekin:2006tg}. 
	
	There are roughly $10^{58}$ neutrinos emitted from a core-collapse supernova within 10s of seconds.  { We can expect the neutrino momenta in supernovae to span the range from KeV up to $100$ MeV, with the majority being around $10$ MeV~\cite{Saez:2024}. The neutrino density is highest near the core with about $10^{57}$ neutrinos between emitted over a radius of approximately $10$ km on average, and decrease quartically with the distance from center~\cite{Rrapaj:2021_nfml}. As such, when looking at the supernova radial profile we can find regions dominated by the two-body interactions, regions with equal strength in one and tow-body terms, and the outer regions where we can neglect any neutrino-neutrino interactions.}
	Solving a quantum many-body problem for such a large number of particles is computationally formidable. To circumvent this hindrance, it is tempting to make approximations such as with a mean field treatment, in which a many-body problem is translated to an effective one-body problem. The mean-field approximation is a powerful tool to tackle the problem and learn about the role of collective oscillations, but it comes at a cost of losing quantum entanglement, which can be measured in the form of quantum correlations and can even impact single body observables such as the flavor composition of the neutrinos detected from these events. To assess the validity of the mean-field approach, several simplified many-body models in a two-flavor framework have been explored~\cite{Bell:2003mg, Friedland:2003dv, Friedland:2003eh, Friedland:2006ke, McKellar:2009py, Balantekin:2006tg, Pehlivan:2011hp, Birol:2018qhx, Patwardhan:2019zta, Cervia:2019res, Rrapaj:2019pxz, Roggero:2021asb, Roggero:2021fyo, Xiong:2021evk, Martin:2021bri, Patwardhan:2021rej, Roggero:2022hpy, Cervia:2022pro, Illa:2022zgu, Lacroix:2022krq, PhysRevC.110.045801, Martin:2023gbo, Martin:2023ljq,PhysRevD.110.103027} and differences have been found in the evolution of certain observables, compared to the corresponding mean-field predictions. For simple geometries, such as a two-beam setup with a constant neutrino gas density, one can efficiently perform computations with tensor network methods~\cite{Roggero:2021asb}, but in more general settings with higher entanglement only a system of few tens of neutrinos could be simulated~\cite{Cervia:2022pro}. Aside from the challenges of adding more neutrinos, recent explorations have also sought to ask different questions---for instance, whether these deviations from the mean-field behavior could persist if the effects of finite interaction time~\cite{Shalgar:2023ooi, Kost:2024esc}, non-forward scattering~\cite{Johns:2023ewj,PhysRevD.110.123028}, or Pauli blocking~\cite{Goimil-Garcia:2024wgw} are considered.
	
	A more realistic treatment of this problem also warrants the inclusion of all three neutrino flavors in this formalism~\cite{Pehlivan:2014zua}. It has been observed from the mean-field calculations that the three-flavor collective neutrino oscillations exhibit unique features like multiple spectral splits. From the many-body calculations in the three-flavor settings, it was found that the neutrinos are even more entangled than the two-flavor case, further questioning the reliability of the mean-field approximation in this realistic setting \cite{Siwach:2022xhx}. As shown in recent works, quantum magic (in essence a measure of the computational resources required for the simulation) as quantified by the stabilizer Renyi entropy can reach its maximal value in the dynamics of three-flavor collective oscillations~\cite{Chernyshev:2024}. 
	However, such simulations were limited to only a few neutrinos due to the memory requirements for the large Hilbert space. 
	
	Quantum computers are presumably a natural choice to simulate the quantum many-body problems \cite{Klco_2022,Bauer:2022hpo,PRXQuantum.5.037001}. Several efforts have been made to simulate collective neutrino oscillations on a quantum computer in the two-flavor setting \cite{Hall:2021rbv,Yeter-Aydeniz:2021olz,PhysRevA.106.052605,Amitrano:2022yyn,PhysRevLett.130.221003,Siwach:2023wzy} and, more recently, also with the full three flavors~\cite{Turro:2025,PhysRevD.111.043017}. 
	Here, we seek to develop optimized circuits for both qubit and qutrit representations based upon the specific properties of the Hamiltonian and the two-body interactions.
	In Sec.~\ref{sec:theory} we represent the Hamiltonian and the resulting SU(3) flavor algebra properties, irrespective of encoding. Then, we proceed to analyze the Trotter error in Sec.~\ref{sec:trotter} and outline the qubit and qutrit representations of our model in Sec.~\ref{sec:reps}.
	We optimize the resulting quantum circuits in Sec.~\ref{sec:circuit}. In Sec.~\ref{sec:experiments} we apply the computed circuits to study a system of two neutrinos on the IBM Torino superconducting quantum computing device for the qubit setup, and on the Advanced Quantum Testbed (AQT) device at the Lawrence Berkeley National Laboratory for the qutrit setup. Finally, we summarize and remark upon findings and future outlook in Sec.~\ref{sec:conclusion}. 
	
	\section{Algebra of collective three-flavor oscillations}
	\label{sec:theory}
	
	Here, we introduce the Hamiltonian model for collective neutrino oscillations in three flavors. 
	Accounting only for forward scattering, the Hamiltonian governing the flavor evolution for a system of $N$ neutrinos can be expressed as $H = H_{\nu}  + H_{\nu \nu}$, where $H_\nu$ encompasses the one-body neutrino interaction terms, including vacuum oscillations and interactions with the matter background. The full two-body term including momentum exchanges~\cite{PhysRevD.110.123028} is beyond the scope of this article and will be treated in future work. 
	
	Ignoring neutrino-matter interactions for simplicity (as they can be nearly rotated away with an appropriate change-of-basis transformation), the contribution from vacuum oscillations can be written as
	\begin{equation}
		H_{\nu}=\sum_{q=1}^N\omega_q\vec{B}\cdot\vec{\lambda}_q\;.
		\label{eq:vacuum_H}
	\end{equation}
	Within the Hilbert subspace of each neutrino indexed by $q$, we denote by $\vec{\lambda}$ the vector of Gell-Mann matrices, and $\vec B$ is likewise a constant, eight-component unit vector. Defining the neutrino mass-squared differences, $\Delta_{ij} = m^2_i - m^2_j$, we express the vacuum oscillation frequencies $\omega_q$ and the non-zero entries of $\vec{B}$ in the mass basis as follows:
	\begin{equation}
		\begin{split}
			\omega_q &=\frac{ \sqrt{ \Delta^2_{12}+( \Delta_{13}+\Delta_{23})^2/3}}{4p_q} \\
			B_3 &= \frac{\Delta_{12}}{\sqrt{\Delta^2_{12}+(\Delta_{13}+\Delta_{23})^2/3}} \\
			B_8 &= \frac{\Delta_{13}+\Delta_{23}}{\sqrt{3\Delta^2_{12}+(\Delta_{13}+\Delta_{23})^2}}\;. \\
		\end{split}
	\end{equation}
	{Here we denote $p_q$ the momentum of the neutrino indexed by the label $q$.} All other elements of $\vec{B}$ are zero. 
	
	Similarly, we may write the two-body term 
	\begin{equation}
		H_{\nu\nu}=\frac{\mu}{2N}\sum_{q<k}^N(1-\cos(\theta_{qk}))\vec{\lambda}_q\cdot\vec{\lambda}_k:=\sum_{q<k}^NJ_{qk}\vec{\lambda}_q\cdot\vec{\lambda}_k, 
		\label{eq:twoBody_H}
	\end{equation}
	where coupling constant $\mu=\sqrt{2} G_F n_\nu$ depends on both Fermi's constant $G_F$ and the local neutrino number density $n_\nu$ ($= N/V$ for $N$ neutrinos quantized in a box of volume $V$), { and $\theta_{qk} = \hat{p}_q \cdot \hat{p}_k$ is the angle between the momenta $\vec{p}_q$ and $\vec{p}_k$.} A full derivation of the Hamiltonian can be found in Appendix \ref{app:Hamiltonian_derivation}. 
	
	The Gell-Mann matrices satify the following algebra,
	\begin{equation}
		\begin{split}
			[\lambda_i,\lambda_j] &= 2if_{ijk}\lambda_k,\\
			f^{147} &= - f^{156} = f^{246} = f^{257} = f^{345} = - f^{367} = \frac{1}{2} \\
			f^{458} &= f^{678} = \frac{\sqrt{3}}{2}, \ f^{123} = 1
		\end{split}
	\end{equation}
	We note that there are three embedded SU(2) sub-algebras,
	\begin{equation}
		\begin{split}
			&i\rightarrow\{ \lambda_1,\lambda_2,\lambda_3\}\\
			&j\rightarrow\{ \lambda_4,\lambda_5, \lambda_+ \}\\
			&k\rightarrow\{ \lambda_6, \lambda_7, \lambda_-\},
		\end{split}
		\label{eq:S(2)}
	\end{equation}
	where $\lambda_{\pm}=\frac{1}{2}(\pm \lambda_3 + \sqrt{3} \lambda_8)$ and have denoted them by $(i,j,k)$. The Casimir operators $(\vec{\lambda} \cdot \vec{\lambda})_{i,j,k}$ of these sub-algebras mutually commute. Another important property of the SU$(n)$ quadratic Casimir operator is its relation to the SWAP operator,
	\begin{equation}
		\label{eq:swap}
		\frac{1}{n}\mathbb{1}_n + \frac{1}{2}\vec{g}_q \cdot \vec{g}_k = \text{SWAP}_{qk}\;,
	\end{equation}
	where $\text{SWAP}|x\rangle |y\rangle =|y\rangle |x\rangle $ for any pair vectors in the respective $n$-dimensional qudit spaces, and $\vec{g}$ are the generators of the algebra, i.e., Pauli matrices for SU(2) and Gell-Mann matrices for SU(3).
	
	\section{Trotter Error}
	\label{sec:trotter}
	Given the Hamiltonian model for oscillations outlined in the previous section, we are ready to describe how we simulate real-time evolution of a neutrino system. 
	In order to implement the time evolution operator $e^{-iHt}$ we will use a simple Trotterization method. 
	
	First, we can approximate the evolution operator by breaking it into the product of the vacuum interaction and the $2$-body term as follows:
	\begin{equation}
		\label{eq:trotter_firstorder}
		e^{-iHt} \approx e^{-iH_{\nu}t}e^{-iH_{\nu\nu}t}
	\end{equation}
	and the error in doing this approximation is~\cite{PhysRevX.11.011020} 
	\begin{equation}
		\left\|e^{-iHt} - e^{-iH_{\nu}t}e^{-iH_{\nu\nu}t}\right\|  \le \frac{t^2}{2}\left|[H_{\nu},H_{\nu\nu}]\right|\;. 
	\end{equation}
	More precisely, in Appendix~\ref{app:Trotter_error} we calculate the bound 
	\begin{equation}
		\|e^{-iHt} - e^{-iH_{\nu}t}e^{-iH_{\nu\nu}t}\|  \le t^2 \mu N \max_{q,k}|\omega_k-\omega_q|\;. 
	\end{equation}
	
	The one-body term $H_\nu$ is itself a sum of commuting single-body operators, so we may simply factor 
	\begin{equation}
		e^{-iH_{\nu}t} = \prod_{q=1}^N e^{-it\omega_q \vec{B} \cdot \vec{\lambda}_q },
	\end{equation}
	without involving any further approximation. 
	Moreover, for the two-body interaction, we can use again the first order Trotter formula, factoring individual interaction terms;
	\begin{equation}
		\begin{split}
			&\left\| e^{iH_{\nu\nu}t} - \prod_{q<k}e^{-itJ_{qk}\vec{\lambda}_q\cdot\vec{\lambda}_k} \right\| \le \\&\frac{t^2}{2} \sum_{K=1}^{\binom{N}{2}}J_K\left\|\left[\vec{\lambda}_{q_K}\cdot\vec{\lambda}_{k_K},\sum_{L>K}\vec{\lambda}_{q_L}\cdot\vec{\lambda}_{k_L}\right]\right\| =: \frac{t^2}{2}\| C_{22} \|
		\end{split}
	\end{equation}
	where $J_K$ is the coupling strength for the pair $K=1,\ldots,\binom{N}{2}$. 
	As shown in detail in the full derivation presented in Appendix \ref{app:Trotter_error} we have
	\begin{equation}
		\|C_{22}\|<\sqrt{3}\mu^2N\max_{q,k,l}\left|\cos(\theta_{lq})-\cos(\theta_{lk})\right|\;.
	\end{equation}
	In general, by doing $r$ steps with time $t/r$, the full Trotter error will bounded by
	\begin{equation}
		\frac{t^2}{2r}(2 \mu N \max_{q,k}|\omega_k-\omega_q| + \| C_{22} \|).
	\end{equation} 
	By increasing the number of steps $r$, and thus the number of operations, we can reduce the approximation error.
	Notably, our above analysis is independent of the representation of neutrino flavor on hardware, applying equally well to simulations on on qubit and qubit hardware. 
	
	\section{Qutrit and Qubit Representations}
	\label{sec:reps}
	Having outlined the Hamiltonian for collective oscillations and how to decompose the evolution operator into its individual terms, it remains to be shown how they are simulated on quantum hardware. 
	Before showing the operations on qubits or qutrits to simulate our system, let us introduce how to encode our problem on each kind of hardware.
	
	When considering qutrit-based hardware, we can straightforwardly encode each neutrino with a single qutrit, by associating each flavor of a neutrino with a different state of the qutrit. Our convention is as follows:
	\begin{equation}
		|\nu_e\rangle = |0\rangle, \ |\nu_\mu\rangle = |1\rangle, \ |\nu_\tau\rangle = |2\rangle.
	\end{equation}
	This encoding is particularly easy to use, since it preserves the locality of the Hamiltonian: one-body terms become single-qutrit gates, and two-body terms becomes two-qutrit gates. Consequently, the depth and gate count will be much lower here than in the qubit case.
	
	Next, we explain the mapping from neutrinos onto qubits. A three-flavor neutrino has three possible states, which means that we need two qubits for each neutrino, entailing one extra \lq unphysical\rq\ state, which ideally will never be used. Considering the computational basis of two qubits, we choose the following encoding:
	\begin{equation}
		\begin{split}
			|\nu_e\rangle =& |01\rangle, \ |\nu_\mu\rangle = |10\rangle\\
			|\nu_\tau\rangle =& |11\rangle,\ |\tilde{\nu}\rangle = |00\rangle
		\end{split}
	\end{equation}
	Since the state $|\tilde{\nu}\rangle$ is non-physical, no wave-function amplitude should propagate into this state throughout the entire time evolution. 
	
	Likewise, we map the generators of the SU$(3)$ algebra, the Gell-Mann matrices that we denoted with $\vec{\lambda}$ above, into a two-qubit operator space spanned by Pauli strings of the form $\{\sigma_i \otimes \sigma_j\}_{i,j=0,1,2,3}$. Note that this mapping must be consistent with the mapping of states above, so that the new two-qubit operators reduce correctly to the Gell-Mann matrices if we delete the $|00\rangle$ state. This mapping is given by the following definitions,
	
	\begin{equation}
		\label{eq:qubit_generators}
		\begin{split}
			2 Q_1 &= X_0 X_1 + Y_0 Y_1 \quad  2 Q_2 = Y_0 X_1 - X_0 Y_1\\
			2 Q_3 &= Z_0\mathbb{1}_1 - \mathbb{1}_0Z_1 \quad\; 2 Q_4 = X_0\mathbb{1}_1 - X_0 Z_1\\
			2 Q_5 &= Y_0\mathbb{1}_1 - Y_0 Z_1 \quad\; 2 Q_6 = \mathbb{1}_0 X_1 - Z_0 X_1 \\
			2 Q_7 &= \mathbb{1}_0 Y_1 - Z_0 Y_1 \\ 2 Q_8 &= \frac{\mathbb{1}_0 Z_1 + Z_0 \mathbb{1}_1 - 2 Z_0 Z_1}{\sqrt{3}}.
		\end{split}
	\end{equation}
	Here we denote the $2\times 2$ identity by $\mathbb{1}$ and the Pauli matrices ${X,Y,Z} = \sigma_{\{x,y,z\}}$. The subscript on the Pauli operators denotes the qubit on which operator acts, and we omitted the tensor product for brevity. In short, we have $Q_i = 0 \oplus \lambda_i$ (there the $0$ entry is meant to be on the $\ket{00}\bra{00}$ entry of the matrix). With this definition, the $Q_i$ matrices satisfy the same algebra of the Gell-Mann matrices.
	Thus we 
	encode the SU$(3)$ algebra in a larger two-qubit space containing the non-physical $\ket{\tilde{\nu}}$ state. The qubit-encoded Hamiltonian behaves in the same way as the original one in the subspace generated by $\{\ket{01}, \ket{10}, \ket{11}\}$ and does not involve the non-physical $\ket{00}$ state.

	\section{Hardware Implementation}
	\label{sec:circuit}
	
	We describe now the quantum circuit implementation of the time evolution operator, providing versions for both an all-to-all and a nearest-neighbors connectivity in the qubit case.

	\subsection{Qubit  Circuit}
	\label{subsec:qubit}
	Since we want to observe the flavor oscillations of neutrinos, we would like to encode and measure neutrinos in the flavor basis. However, the Hamiltonian is more concisely expressed in the mass basis, so we start our quantum circuit computations in the flavor basis, transform to the mass basis through the Pontecorvo-Maki-Nakagawa-Sakata (PMNS) matrix, apply time evolution, and finally transform back to the flavor basis before performing measurements. To be able to switch from one basis to the other we need to implement the PMNS matrix as a gate. This is a two-qubit gate that directly depends on the mixing angles, as explained in Appendix \ref{app:PMNSmatrix}, where we also provide its circuit. Our results are similar to those found in Ref.~\cite{PhysRevD.105.056024}, which used a different qubit mapping for the three flavors, and we also provide a version in terms of cross resonance gates.
	Having computed the error associated with the Trotter approximation, here we focus on the gate implementation of the unitary operations $e^{-it\omega_q\vec{B}\cdot \vec{Q}_q}$ and $e^{-itJ_{qk}\vec{Q}_q\cdot \vec{Q}_k}$.
	
	The exponent of the one-body term can be written as a linear combination of $Z_0, Z_1$ and $Z_0 Z_1$, which  are mutually commuting operators. As demonstrated in Appendix~\ref{app:one-body}, in the physical subspace the unitary operator is equivalent to a product of single qubit rotations $R_{Z_0}(\alpha)R_{Z_1}(\beta)$ up to a global phase, for an appropriate choice of the angles $\alpha$ and $\beta$. We can then implement the evolution under the one-body term without using entangling operations, a strategy also employed recently in Ref.~\cite{Turro:2025}.
	

	
	Using the commutation relations given by the subalgebras in Eq.~\eqref{eq:S(2)}, one finds that $\vec{Q}_q \cdot \vec{Q}_k$ can be broken into the following commuting pieces (denoting $Q_i\otimes Q_i = Q_{ii}$):
	\begin{equation}
		\begin{split}
			\sum_i Q_{ii} = (Q_{11} + Q_{22}) &+ (Q_{44} + Q_{55}) \\
			+ (Q_{66} + Q_{77}) &+ (Q_{33}) + (Q_{88})
		\end{split}
	\end{equation}
	where each parenthetical term commutes with every other. Therefore we can break the exponential \emph{exactly}, as follows:
	\begin{equation}
		\begin{split}
			e^{-itJ_{qk} \vec{Q}_q \cdot \vec{Q}_k} =& e^{-itJ_{qk} (Q_{11} + Q_{22})_{qk} } \times \\
			&e^{-itJ_{qk} ( Q_{44} + Q_{55})_{qk} } \times\\
			& e^{-itJ_{qk} ( Q_{66} + Q_{77})_{qk} }\times\\
			&e^{-itJ_{qk} (Q_{33})_{qk} } e^{-itJ_{qk} (Q_{88})_{qk} }\;,
		\end{split}
	\end{equation}
	where we used the subscript $qk$ in the parenthesis surrounding the operator to clarify on which qubits the operators actually act. For instance with the notation $(Q_{33})_{qk}$ we intend the tensor product between $Q_3$ acting on the pair of qubits representing neutrino $k$ and $Q_3$ on the pair for neutrino $q$.
	
	Now, for example let us consider in detail the second exponential $e^{-itJ_{qk} ( Q_{44} + Q_{55})_{qk} }$. We can write $Q_{44} + Q_{55}$ in terms of Pauli matrices using the mapping from Eq.~\eqref{eq:qubit_generators}
	\begin{equation}
		\begin{split}
			Q_4\otimes Q_4 &=\frac{X_1X_3}{4}\left(1-Z_4-Z_2+Z_2Z_4\right),\\
			Q_5\otimes Q_5 &=\frac{Y_1Y_3}{4}\left(1-Z_4-Z_2+Z_2Z_4\right).\\
		\end{split}
	\end{equation}
	{ We note that the first two indices (1, 2) refer to the first two qubits and the first neutrino, and the second two indices (3,4) refer the second two qubits and second neutrino, respectively.}
	Then, we define the following operator:
	\begin{equation}
		\vcenteredhbox{\Qcircuit @C=1em @R=1em {
				\lstick{q_a} &\multigate{1}{G_{ab}} &\qw\\
				\lstick{q_b} &\ghost{G_{ab}}      &\qw\\
		}}\;:=\vcenteredhbox{\Qcircuit @C=1em @R=1em { 
				&\gate{H} &\ctrl{1}&\qw\\
				&\qw      &\targ   &\qw\\
		}}\;
	\end{equation}
	It is straightforward to prove that $G_{13}$ diagonalizes the operator $Q_{44}+Q_{55}$ (see Appendix \ref{app:diagonalizingRelations} for details), and thus
	\begin{equation}
		\begin{split}
			G_{13}\left(Q_{44}+ Q_{55}\right)G^\dagger_{13}=& \frac{Z_1-Z_1Z_3}{4}\times\\
			&\left(1-Z_4-Z_2+Z_2Z_4\right)
		\end{split}
	\end{equation}
	Using this unitary transformation, the exponential term $e^{-itJ_{qk} ( Q_{44} + Q_{55})_{qk} }$ can then be implemented as,
	\begin{equation}
		G_{13}^\dag e^{-itJ_{qk}\left(\frac{Z_1-Z_1Z_3}{4}\left(1-Z_4-Z_2+Z_2Z_4\right)\right)_{qk}} G_{13}.
	\end{equation}
	
	We can implement $e^{-itJ_{qk} ( Q_{66} + Q_{77})_{qk} }$ and $e^{-itJ_{qk} ( Q_{11} + Q_{22})_{qk} }$ in a similar fashion. The diagonal unitary operations $e^{-itJ_{qk} Q_{33} }$ and $ e^{-itJ_{qk} Q_{88} }$ are four-qubit operators whose weight can be reduced by introducing gadgets similar to the diagonalizing operator $G_{ab}$. Implementation details (i.e., in terms of elementary gates) for all these terms can be found in Appendices \ref{app:2BodyInteraction} and \ref{app:diagonalizingRelations}. Our circuit implementation requires $34$ CNOT gates when qubits have all-to-all connectivity and $39$ CNOT gates for the $T$ connectivity present on IBM devices. It is likely that further optimization could be done since the authors of Ref.~\cite{Turro:2025} obtain circuits with only $18$ CNOT with all-to-all connectivity and $30$ CNOT for qubits on a chain, even though the mapping of the flavor space is slightly different from ours.

	\subsection{Qutrit circuit}
	\label{subsec:qutrit}
	By considering qutrits instead of qubits, each neutrino can be encoded with a single qutrit. The PMNS matrix is then a single-qutrit gate, which is in general easy to implement, as are the single-body Hamiltonian terms. So, in the following we will only focus on the two-body term, which requires the use of two-qutrit gates.
	
	Let us introduce the $C\tilde X$ gate, introduced in Ref.~\cite{Wang_2020}, defined as follows:
	\begin{equation}
		C\tilde X \ket{x}\ket{y} = \ket{x}\ket{-x-y}
	\end{equation}
	This gate is self-inverse and can be decomposed into three two-level controlled gates as
	\begin{equation}
		\Qcircuit @C=1em @R=.7em {
			& \ctrl{1} & \qw & & & \ustick{0} \qw & \ustick{1} \qw & \ustick{2} \qw & \qw \\
			& \gate{C\tilde X} & \qw & \push{\rule{.3em}{0em}=\rule{.3em}{0em}} & & \gate{X^{12}}\qwx{-1} & \gate{X^{02}}\qwx{-1} & \gate{X^{01}}\qwx{-1} & \qw 
		}
		\label{eq:generalization_CX}
	\end{equation}
	It can be implemented with a qutrit controlled-$Z$ gate (defined as $CZ_{3} = \sum_{k,l \in \mathbb{Z}_3} \omega^{kl}\ket{kl}\bra{kl}$, where $\omega = e^{2\pi i/3}$ is the third root of unity) and a single-qutrit quantum Fourier transform, which is nothing but the generalization of the Hadamard gate. By using this gate, the two-qutrit SWAP becomes the standard circuit if we replace the CNOTs with $C\tilde X$:
	\begin{equation}
		\text{SWAP} = \vcenteredhbox{
			\Qcircuit @C=1em @R=.7em {
				& \ctrl{1} & \gate{C\tilde X} & \ctrl{1} & \qw \\
				& \gate{C\tilde X} & \ctrl{-1} & \gate{C\tilde X} & \qw 
			}
		}
	\end{equation}
	Since we need a partial swap to simulate the interaction, we can generalize the procedure of qubits, and replace the middle $CX$ with a $CR$. Here, $CR$ is a controlled-$SU(3)$ gate where the $SU(3)$ operation is the product of a two-level phase gate $P_z$ gate and a two-level rotation $R_X$ gate,
	\begin{equation}
		\begin{split}
			R^{12}(\theta)=&\left(
			\begin{array}{ccc}
				e^{-i \theta } & 0 & 0 \\
				0 & \cos (\theta ) & -i \sin (\theta ) \\
				0 & -i \sin (\theta ) & \cos (\theta ) \\
			\end{array}
			\right)\\
			=&P_z^{01}(\theta)R_x^{12}(2\theta)\\
			R^{02}(\theta)=&P_z^{12}(\theta)R_x^{02}(2\theta) \\
			R^{01}(\theta)=&P_z^{21}(\theta)R_x^{01}(2\theta) \\
		\end{split}
	\end{equation}
	The resulting circuit, which applies $e^{-2i \theta \text{SWAP}}$, is:
	\begin{equation}
		\vcenteredhbox{\resizebox{0.4\textwidth}{!}{
				\Qcircuit @C=1em @R=.7em {
					& \ctrl{1} & \gate{R^{12}(2\theta)} & \gate{R^{02}(2\theta)} & \gate{R^{01}(2\theta)} & \ctrl{1} & \qw \\
					& \gate{C\tilde X} & \dstick{0} \qw \qwx{-1} & \dstick{1} \qw \qwx{-1} & \dstick{2} \qw \qwx{-1} & \gate{C\tilde X} & \qw \\
		}}}
		\label{eq:partial_swap}
	\end{equation}
	Given this quantum circuit, there are many ways one can implement it on hardware. We use { the BQSkit compiler~\cite{osti_1785933}} to optimize the quantum circuit in terms of $CZ_3$ gates and single-qutrit rotations. 
	By leveraging our compiler with a numerical synthesis tolerance of $10^{-6}$ (significantly lower than any gate infidelities on the device), we construct circuits that require four $CZ_3$ gates plus local $SU(3)$ rotations per Trotter step. Our $SU(3)$ rotations were implemented by embedded $SU(2)$ rotations with subspace Rabi oscillations \cite{PhysRevLett.126.210504, PhysRevA.97.022328}. We note that the gate cost of our qutrit circuit is the same as the corresponding construction from Ref.~\cite{Turro:2025}.
	
	\section{Hardware experiments}
	\label{sec:experiments}
	
	We can now proceed to study the circuits developed in the previous section on the IBM Torino device and the AQT qutrit device. The AQT qutrit device is a Transmon-based quantum processor designed and operated to perform qudit operations and implement arbitrary single qudit gates and multi-qudit gates such as $CZ_3$ \cite{goss-high, empowering, Blok2021}. Based on the past success with similar algorithms, we implement an error mitigation technique based on noise renormalization~\cite{PhysRevLett.127.270502,PhysRevD.106.074502,PRXQuantum.5.020315,PhysRevD.111.034504}. We first apply Pauli twirling to the compiled circuits in order to approximate the error channel of the device as global depolarizing noise
	\begin{equation}
		\Lambda[\rho] = (1-p) \rho + \frac{p}{\text{Tr}[\mathbb{1}]}\mathbb{1}\;,
	\end{equation}
	with $\mathbb{1}$ the global identity matrix, $\rho$ the noiseless quantum state and $p$ the noise strength ~\cite{PhysRevX.11.041039, extending}. We added 30 twirled circuits to the original circuit, and took 5120 shots per circuit. Strictly speaking, Pauli twirling does not convert a generic quantum channel into a depolarizing channel, but into a stochastic Pauli channel, which in practice may be approximated by the depolarizing channel rather well.
	
\begin{figure}
		\includegraphics[width=0.5\textwidth]{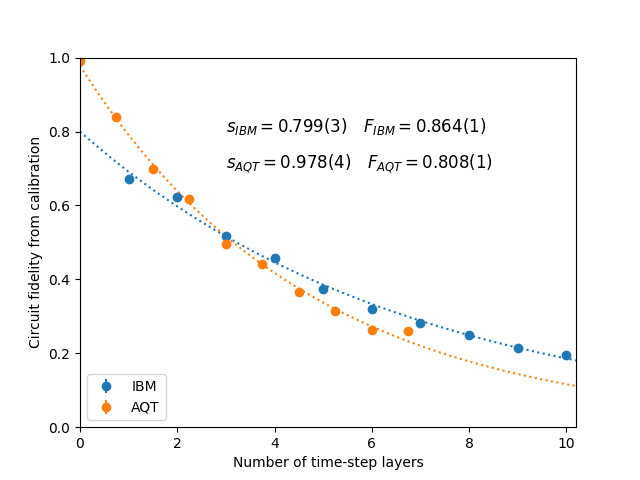}
		\caption{Fidelity normalization factor estimation by Clifford version of the circuit assuming depolarizing noise. Blue dotted points depict results from the IBM device, and orange points from the AQT device. The curves plotted are fits of the aforementioned data to an exponential function, as in Eq.~\eqref{eq:fit_func_pl}. We took over 150,000 shots per point in the plot for IBM device and over 30,000 for the AQT device. We show uncertainties by including the standard error with vertical lines, which are rather small.
			Since the AQT device used three $CZ$ gates for calibration and four $CZ$ gates for time evolution, the time steps have been rescaled by a factor of $3/4$. For $t=0$ we show the initial state preparation results for both devices. As state preparation in the qubit case requires entangling gates, SPAM error is higher than the qutrit case.
		}
		\label{fig:fidelity}
	\end{figure}

	We perform simulations of the time-evolution using the first-order product formula from Eq.~\eqref{eq:trotter_firstorder} with a fixed time-step $dt=0.5\mu^{-1}$. With this strategy to reach a total evolution time $t=Ldt$, we need to perform $L$ layers of single steps involving evolution with both the one-body and two-body Hamiltonians. Since we expect the noise strength to depend on the circuit depth, we expect to have the following noisy state
	\begin{equation}
		\Lambda\left[\rho(t=Ldt)\right] = (1-p_L) \rho(t=Ldt) + \frac{p_L}{\text{Tr}[\mathbb{1}]}\mathbb{1}\;,
	\end{equation}
	with $p_L\to1$ as $L\to\infty$. We estimate $p_L$, for a given depth, using a calibration circuit involving only Clifford gates and use the results to remove the error contribution. {The Clifford circuit has the same gate structure as the target circuit, but performs a SWAP operation by leveraging Eq.~\eqref{eq:swap}. For instance}, if we are interested in expectation values $\left\langle\Pi_k(t)\right\rangle$ of some projector operator $\Pi_k$ at time $t=Ldt$, we can use the calibration circuit to get
	\begin{equation}
		\label{eq:calib_depol}
		\left\langle\Pi_c(L)\right\rangle_{calib} = (1-p_L) + \frac{p_L}{\text{Tr}[\mathbb{1}]}\;,
	\end{equation}
	where $\Pi_c$ is a rank-1 projector on the state we expect to observe in the absence of noise. The resulting expectation value can be used to determine empirically the circuit fidelity as a function of depth. The results for two three-flavor neutrinos on both devices is shown in Fig.~\ref{fig:fidelity}. 

	{
		Further, we model the results 
		obtained with an exponential fit of the form
		\begin{equation}
			\label{eq:fit_func_pl}
			\left\langle\Pi_c(L)\right\rangle_{calib} = s F^L\;,
		\end{equation}
		where the base $F$ represents a fidelity per step and the coefficient $s$ accounts for state-preparation and measurement (SPAM) errors. The best fit parameters for both devices are reported inside the plot in Fig.~\eqref{fig:fidelity}. The much larger SPAM error observed in the qubit case---even when no time-steps are performed (ie. $L=0$)---is in large part a consequence of the transformations between the flavor and mass bases, which involve entangling operations. In contrast, on the qutrit implementation, state preparation involves only local $SU(3)$ transformations instead.
	}
	
	{The error mitigation proceeds then as follows: we first estimate the depolarizing noise strength $p_L$ using Eq.~\eqref{eq:calib_depol} obtaining
		\begin{equation}
			p_L = \frac{\text{Tr}[\mathbb{1}]}{\text{Tr}[\mathbb{1}]-1}\left(1-\left\langle\Pi_c(L)\right\rangle_{calib}\right)\;.
		\end{equation}
		For the qubit simulations we estimate $\left\langle\Pi_c(L)\right\rangle_{calib}$ directly from the measured calibration circuits. For the qutrit case, owing to the difference in gate counts per layer between the simulation circuits and the calibration ones, we use instead the fit from Eq.~\eqref{eq:fit_func_pl} which provides a good model for the measured data. Using this estimator for the noise strength we can then estimate the noise mitigated expectation value of a rank-1 projector $\Pi_k$ as follows
		\begin{equation}
			\label{eq:mitigated_proj}
			\left\langle\Pi_k(L)\right\rangle_{mit} = \frac{\left\langle\Pi_k(L)\right\rangle-\frac{p_L}{\text{Tr}[\mathbb{1}]}}{1-p_L}\;.
		\end{equation}
		In the qubit case we can express the mitigated expectation value directly in terms of the bare expectation value and the result of calibration at step $L$ as
		\begin{equation}
			\label{eq:mitigated_proj_qubits}
			\left\langle\Pi_k(L)\right\rangle_{mit} = \frac{15\left\langle\Pi_k(L)\right\rangle+\left\langle\Pi_c(L)\right\rangle_{calib}-1}{16\left\langle\Pi_c(L)\right\rangle_{calib}-1}\;,
		\end{equation}
		where we used $\text{Tr}[\mathbb{1}]=16$ for our four qubits. {We have checked that the  mitigated values obtained from the fit in the qubit case agree quite well with the extrapolation from the bare Clifford circuits.}
	}
	
	We present a selection of results from our simulations in Fig.~\ref{fig:some_results}. A complete set of results is provided in the Appendix~\ref{app:complete_results}. The system is initially prepared in the state $\ket{\nu_e\nu_\mu}$ and evolved under the full flavor Hamiltonian $H=H_\nu+H_{\nu\nu}$ using a first order Trotter formula as discussed in Sec.~\ref{sec:trotter}. {The neutrino frequencies are {$\omega=\{2, 2.5\} \ \mu$}. 
		The mass dependent parameters are $\{B_3,B_8\}= \{0.025483,0.999567\}$. The mixing angles are $\theta_{12}=33.44^{\circ},\ \theta_{13}=8.57^{\circ},\ \theta_{23}=49.2^{\circ}, \delta_{cp}=0$}. 
	Here we show the survival probabilities of $\ket{\nu_e\nu_\mu}$ and $\ket{\nu_\mu\nu_e}$ evolved for up to ten time steps, before and after performing error mitigation techniques described earlier. 
	We can observe that bare results do not match expected theoretical values beyond the first couple of Trotter steps, and as time increases resembles an equipartition of probabilities among all states. The match to the theoretical results at later times is more likely coincidence. Mitigation improves results rather drastically, with both devices agreeing with theoretical expectations for up to five Trotter steps, particularly for the flavor probability of the $|\nu_\mu \nu_e \rangle $ state, which is very small throughout the time evolution and therefore more challenging to resolve. 
	\begin{figure*}
		\centering
		\includegraphics[width=1\textwidth]{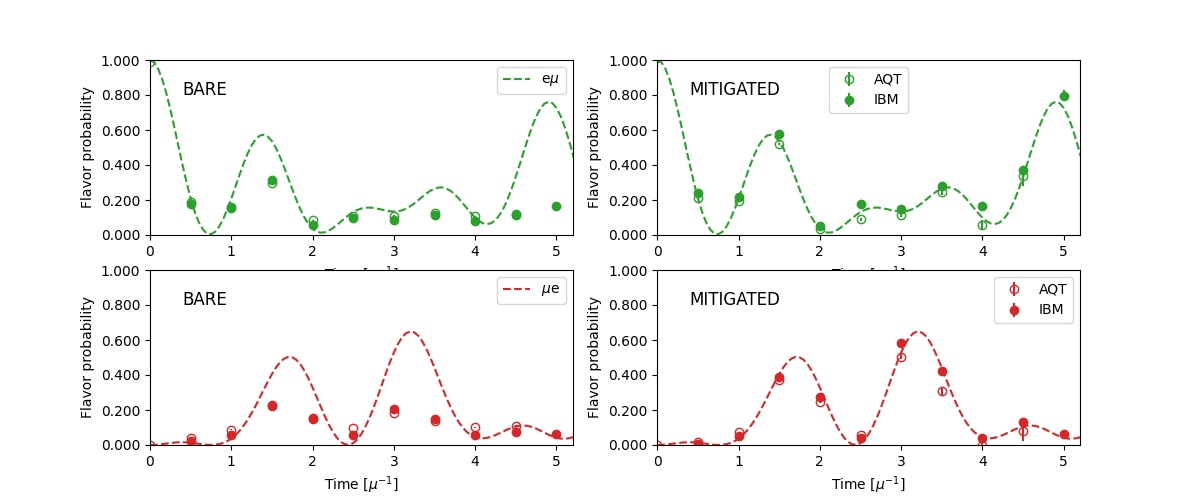}
		
		\caption{{Survival probability for the initial state $|\nu_e\nu_\mu\rangle$ (top) and its swapped counterpart $|\nu_\mu\nu_e\rangle$ (bottom) as function of time evolution. 
				We show the results from measurements both directly on the device (left) and after error mitigation (right) for up to ten time steps. Solid dots are the results from the (qubit-based) IBM Torino device and empty dots are the results from the (qutrit-based) AQT device. 
				Uncertainty bars are estimated by the standard error from each simulation on hardware. 
				Exact results are plotted as curves for reference. 
				The improvement due to error mitigation is quite drastic, and depolarizing noise, after Pauli twirling has been performed, seems to be a significant fraction of the noise.}}
	\label{fig:some_results}
\end{figure*}



\section{Conclusion and Discussion}
\label{sec:conclusion}
{In this work, we have developed and compared quantum circuits to perform collective neutrino flavor oscillations on both qubit and qutrit quantum hardware. Time evolution was computed through the first-order Trotter approximation and we find that, for a fixed total time, the error scales linearly with the number of neutrinos and inversely with the number of Trotter steps. In the qubit encoding, the one-body term becomes a two-qubit operator as does the PMNS matrix, and as such even initial product state preparation requires entangling gates. Meanwhile, in the qutrit case no such need arises. For noisy, intermediate-scale quantum (NISQ) hardware, this translates to additional noise for state preparation on qubits. Similarly, the time evolution will have more entangling gates for the qubit setup, as two-body interactions become four-qubit gates, which are decomposed into a sequence of one- and two-qubit gates. Even with the optimized circuits we construct through various gadgets (see Sec.~\ref{sec:reps} and Appendix~\ref{app:QC}), there is significant overhead for a neutrino pair time evolution step. For both hardware, in larger systems, SWAP operations will be required to represent the all-to-all two-body interactions, which fortunately we can absorb in the two-body operation due to Eq.~\eqref{eq:swap}. On the other hand, qutrit gates require higher control capabilities and can be prone to higher infidelities as a result. For both setups we need to perform error mitigation on hardware to achieve the quality of results shown in Fig.~\ref{fig:some_results}. We note that there are still large margins of improvement for fidelity of qutrit gates, and it is remarkable to see the level of control on these more complex architectures. Limited hardware connectivity issues will start to play an important role for larger systems, so the qutrit-based design is expected to have advantages in terms of overall fidelity.}

{One main difference between the two encodings is the mapping of the physical Hilbert space to the qudit space on device. In the case of qubits, there is one neutrino to two qubits, with one of the amplitudes being non-physical and decoupled from the rest during simulations. We take advantage of this fact to further improve our circuits and reduce the gate count. For a system of $N$ neutrinos, the fraction of nonphysical states on the qubit device is $1-(3/4)^N$, which becomes the vast majority of states as the number of particles increases. The presence of depolarizing noise will inevitably result in leakage to the nonphysical space. On the qutrit device there is a one-to-one mapping between physical amplitudes and device amplitudes. Thus, despite the presence of noise, such leakage cannot occur. In supernovae simulations, the main observables are single-neutrino flavor measurements, and the overall flavor composition might be robust against leakage. However, correlations and higher order operators might be more sensitive to this type of noise. One could take into consideration hardware noise in devising encodings with noise-resistant observables, provided the noise is well-categorized. For now, we leave such endeavors for future research. } 

\begin{acknowledgments}
	We thank Alessandro Baroni and Pooja Siwach for the useful discussions.
	This research was supported by the U.S. Department of Energy (DOE) under Contract No. DE-AC02-05CH11231, through the National Energy Research Scientific Computing Center (NERSC), an Office of Science User Facility located at Lawrence Berkeley National Laboratory.
	This research was supported in part by the U.S.~Department of Energy, Office of Science, Office of High Energy Physics, under Award Nos.~DE-SC0019465 and DE-SC0021143 and in part by the U.S. National Science Foundation Grants  Nos.~PHY-2020275 and PHY-2411495 at the University of Wisconsin. Experiments on AQT were supported by the Quantum Testbed Program of the Advanced Scientific Computing Research Division, Office of Science of the U.S. Department of Energy under Contract No. DE-AC02-05CH11231. This work is part of the activities of the Quantum Computing for High-Energy Physics (QC4HEP) working group. E.Y. was supported by the U.S. Department of Energy, Office of Science, Office of Advanced Scientific Computing Research under Contract No. DE-AC05-00OR22725 through the Accelerated Research in Quantum Computing Program MACH-Q project. This research used resources of the Oak Ridge Leadership Computing Facility, which is a DOE Office of Science User Facility supported under Contract DE-AC05-00OR22725. MJC is supported by the U.S.~Department of Energy Grant No. DE-FG02-97ER-41014 (UW Nuclear Theory). AVP was supported by the U.S. Department of Energy (DOE) under grant DE-FG02-87ER40328 at the University of Minnesota. 
\end{acknowledgments}

%


\appendix

\section{Hamiltonian derivation}
\label{app:Hamiltonian_derivation}

In Sec.~\ref{sec:theory}, we outlined the contents of our model Hamiltonian for three-flavor neutrino oscillations. 
Here, we expand on its derivation, first for the vacuum oscillations (one-body) term in Appendix~\ref{app:derive-vac} and then for the neutrino interactions (two-body) term in Appendix~\ref{app:derive-int}. 

\subsection{Vacuum part}
\label{app:derive-vac}
We obtain the term describing neutrino oscillations in vacuum from 
\begin{equation}
	\begin{split}
		H_\nu&=\sum_{q=1}^N \sum_{i=1}^3\sqrt{p_q^2+m_i^2}T_{ii}^q\\
		&\approx\sum_{q=1}^N \sum_{i=1}^3\left(p_q+\frac{m_i^2}{2p_q}\right)T_{ii}^q\;,
		\label{eq:free-nu}
	\end{split}
\end{equation}
where in the second step we used the ultra-relativistic limit. The $T_{ij}^q$ operators are defined as
\begin{equation}
	\label{eq:tijq_def}
	T_{ij}^q=\left(\bigotimes_{r=1}^{q-1}\mathbb{1}^{(r)}\right)\otimes t_{ij}^{(q)} \otimes \left(\bigotimes_{r=q+1}^N\mathbb{1}^{(r)}\right)\;,
\end{equation}
where $\mathbb{1}$ is the $3\times3$ identity matrix, while $t_{ij}$ are $3\times3$ matrices with the $(i,j)^\text{th}$ element set to one and all others set to $0$. The parenthetical index in each superscript denotes the subspace of the individual neutrino in which that particular operator resides. Thus, the index $q$ in the superscript indicates that this composite operator acts non-trivially only on the $q^\text{th}$ neutrino. The nine $t_{ij}$ matrices can be expressed as linear combinations of nine orthonormal matrices, like the eight Gell-Mann matrices $\{\lambda_k\}_{k=1,\ldots,8}$ plus the suitably normalized identity matrix $\lambda_0=\sqrt{2/3}\mathbb{1}$. Using the completeness relation
\begin{equation}
	\sum_{k=0}^8 [\lambda_k]_{\alpha\beta}[\lambda_k]_{\gamma\delta}=2\delta_{\alpha\delta}\delta_{\beta\gamma}\;,
\end{equation}
where $[M]_{\alpha\beta}$ denotes the $(\alpha,\beta)^\text{th}$ element of a matrix $M$, one can easily find that
\begin{align}
	2[t_{ij}]_{\alpha\beta} 
	&=\sum_{\gamma,\delta=1}^3\sum_{k=0}^8 [\lambda_k]_{\alpha\beta}[\lambda_k]_{\gamma\delta}[t_{ij}]_{\delta\gamma} \nonumber\\
	&=\sum_{k=0}^8 [\lambda_k]_{\alpha\beta}\text{Tr}\left[\lambda_k t_{ij}\right]\;.
\end{align}
Suppressing the explicit notation of the matrix indices, we have
\begin{equation}
	t_{ij}=\frac{1}{3}\delta_{ij}\mathbb{1}+\frac{1}{2}\sum_{k=1}^8[\lambda_k]_{ji}\lambda_k\;. \label{eq:single-entry-to-GM}
\end{equation}
Using the results above, together with the fact that the Gell-Mann matrices are traceless, one  finds $\sum_{i=1}^3 T_{ii}^k=\mathbb{1}$ for any $k$.

Returning to the one-body Hamiltonian, we can separate out a piece proportional to the identity as
\begin{equation}
	H_\nu = \sum_{q=1}^N \left(p_q+\sum_{i=1}^3\frac{m_i^2}{6p_q}\right)\mathbb{1}+\sum_{q=1}^N\sum_{i=1}^3\left(\sum_{j\neq i}\frac{m_i^2-m_j^2}{6p_q}\right)T_{ii}^q\;.
\end{equation}
From Eq.~\eqref{eq:single-entry-to-GM} we see the diagonal operators take the form
\begin{equation}
	\begin{split}
		T_{11}^q&=\frac{\lambda^q_3}{2}+\frac{\lambda^q_8}{2\sqrt{3}}+\frac{1}{3}\mathbb{1},\ T_{22}^q=-\frac{\lambda^q_3}{2}+\frac{\lambda^q_8}{2\sqrt{3}}+\frac{1}{3}\mathbb{1},\\
		T_{33}^q&=-\frac{\lambda^q_8}{\sqrt{3}}+\frac{1}{3}\mathbb{1}\;,
	\end{split}
\end{equation}
where the upper index $q$ on the Gell-Mann matrices indicates the relevant neutrino subspace as in Eq.~\eqref{eq:tijq_def}.
The final expression for the vaccum term is,
\begin{equation}
	\begin{split}
		H_\nu =& 
		\sum_{q=1}^N \left(p_q+\sum_{i=1}^3\frac{m_i^2}{6p_q}\right)\mathbb{1}+\sum_{q=1}^N\frac{\Delta_{12}}{4p_q}\lambda_3^q\\
		&+\sum_{q=1}^N\frac{\Delta_{13}+\Delta_{23}}{4\sqrt{3}p_q}\lambda_8^q\;,
	\end{split}
\end{equation}
where we introduced the squared mass splitting $\Delta^2_{ij}=m_i^2-m_j^2$ for convenience.
Using a more compact notation and removing the irrelevant piece proportional to the identity we find
\begin{equation}
	H_\nu=\sum_{q=1}^N \omega_q\vec{B}\cdot\vec{\lambda}_q, \ \omega_q=\frac{\sqrt{\Delta^2_{12}+(\Delta_{13}+\Delta_{23})^2/3}}{4p_q}\;.
\end{equation}
The vector $\vec{\lambda}_q$ is an eight component vector containing the Gell-Mann matrices acting on the $q^\text{th}$ neutrino, and
\begin{equation}
	\begin{split}
		\vec{B} =& B_3 \vec{e}_3+ B_8 \vec{e}_8,\\
		B_3 =& \frac{\Delta_{12}^2}{\sqrt{\Delta_{12}+(\Delta_{13}+\Delta_{23})^2/3}},\\
		B_8 =& \frac{\Delta_{13}+\Delta_{23}}{\sqrt{3\Delta_{12}+(\Delta_{13}+\Delta_{23})^2}}. 
	\end{split}
\end{equation}
We denote by $\vec{e}_i$ the elementary vectors with components $\vec{e}_i^{(j)}=\delta^j_i$.

\subsection{Neutrino-neutrino interaction}
\label{app:derive-int}
Using the same notation as the previous subsection, the forward scattering neutrino-neutrino interaction can be written as
\begin{equation}
	H_{\nu\nu}=\frac{G_F}{\sqrt{2}V}\sum_{q\neq k}\left(1-\cos(\theta_{qk})\right)\sum_{i,j=1}^3T^q_{ij}T^k_{ji}\;,
\end{equation}
where the coupling depends on the direction of the neutrino momenta through
\begin{equation}
	\cos(\theta_{qk})=\frac{\vec{p}_q\cdot\vec{p}_k}{p_qp_k}\;.
\end{equation}
Using the fact that the Gell-Mann matrices are traceless and that $\text{Tr}[\lambda_m\lambda_n]=2\delta_{mn}$ one finds
\begin{equation}
	\sum_{i,j=1}^3T^q_{ij}T^k_{ji}=\frac{1}{3}\mathbb{1}+\frac{1}{2}\vec{\lambda}_q\cdot\vec{\lambda}_k\;.
\end{equation}
Removing the inconsequential part proportional to the identity, the full interaction Hamiltonian takes the form
\begin{equation}
	\begin{split}
		H_{\nu\nu}=&\frac{G_F}{2\sqrt{2}V}\sum_{q\neq k}^N\left(1-\cos(\theta_{qk})\right)   \vec{\lambda}_q\cdot\vec{\lambda}_k\\
		=& \frac{\mu}{4N}\sum_{q\neq k}^N\left(1-\cos(\theta_{qk})\right)   \vec{\lambda}_q\cdot\vec{\lambda}_k\;,
	\end{split}
\end{equation}
where in the second equality we introduced the coupling constant $\mu=\sqrt{2}G_F n_\nu$, where $n_\nu$ is the neutrino density.

\section{Trotter Error}
\label{app:Trotter_error}


In Sec.~\ref{sec:trotter}, we cite upper bounds on the error associated with Trotterizing our time evolution operator in terms of the coupling strength, oscillation frequencies, and system size. Below we elaborate on how these bounds are obtained, considering separately commutators between two-body terms and those between two- and one-body terms of our Hamiltonian. 

The commutator between the one- and two-body terms is,
\begin{equation}
	\begin{split}
		C_{12}=[H_\nu,H_{\nu\nu}]&=\sum_{j=1}^N\sum_{q<k}^N J_{qk} \omega_j[\vec{B}\cdot\vec{\lambda}_j,\vec{\lambda}_q\cdot\vec{\lambda}_k]\\
		&=\sum_{q<k}^N J_{qk} [\omega_k\vec{B}\cdot\vec{\lambda}_k+\omega_q\vec{B}\cdot\vec{\lambda}_q,\vec{\lambda}_q\cdot\vec{\lambda}_k]\\
	\end{split}
\end{equation}
where Gell-Mann matrices acting on different neutrinos commute. Since $\vec{B}$ has only two non-zero components, 
\begin{equation}
	[\vec{B}\cdot\vec{\lambda}_k,\vec{\lambda}_q\cdot\vec{\lambda}_k]=B_3[\lambda_3^k,\vec{\lambda}_q\cdot\vec{\lambda}_k]+B_8[\lambda_8^k,\vec{\lambda}_q\cdot\vec{\lambda}_k]\;.
\end{equation}
One can easily verify that
\begin{equation}
	\|[\lambda_3^k,\vec{\lambda}_q\cdot\vec{\lambda}_k]\|=4\quad\quad\|[\lambda_8^k,\vec{\lambda}_q\cdot\vec{\lambda}_k]\|=2\sqrt{3}\;.
\end{equation}
Additionally,
\begin{equation}
	[\vec{B}\cdot\vec{\lambda}_k,\vec{\lambda}_q\cdot\vec{\lambda}_k]=-[\vec{B}\cdot\vec{\lambda}_q,\vec{\lambda}_q\cdot\vec{\lambda}_k]\;.
\end{equation}
Then, the norm of the commutator is,
\begin{equation}
	\begin{split}
		\|C_{12}\|&\leq \sum_{q<k}^N J_{qk} \|[\omega_k\vec{B}\cdot\vec{\lambda}_k+\omega_q\vec{B}\cdot\vec{\lambda}_q,\vec{\lambda}_q\cdot\vec{\lambda}_k]\|\\
		&=\sum_{q<k}^N J_{qk} \|(\omega_k-\omega_q)[\vec{B}\cdot\vec{\lambda}_k,\vec{\lambda}_q\cdot\vec{\lambda}_k]\| \\
		&=\sum_{q<k}^N J_{qk} |\omega_k-\omega_q|\|[\vec{B}\cdot\vec{\lambda}_k,\vec{\lambda}_q\cdot\vec{\lambda}_k]\| \\
		&\leq\left(4B_3+2\sqrt{3}B_8\right)\sum_{q<k}^N J_{qk} |\omega_k-\omega_q| \\
		&=\frac{4\Delta_{12}^2+2(\Delta^2_{13}+\Delta^2_{23})}{\sqrt{\Delta^4_{12}+(\Delta^2_{13}+\Delta^2_{23})^2/3}}\sum_{q<k}^N J_{qk} |\omega_k-\omega_q|.
	\end{split}
\end{equation}
We can find a simpler, though less strict, upper bound by using
\begin{equation}
	J_{qk}\leq \frac{\mu}{N}\quad\quad\Delta^2_{12}<0.1\max[\Delta^2_{23},\Delta^2_{13}]\;,
\end{equation}
to write
\begin{equation}
	\|C_{12}\|<4 \frac{2\mu}{2N}\binom{N}{2} \max_{q,k}|\omega_k-\omega_q|<2 \mu N \max_{q,k}|\omega_k-\omega_q|\;.
\end{equation}

We now consider the purely two-body commutators' size: 
\begin{equation}
	C_{22}=\sum_{K=1}^{\binom{N}{2}}J_K\left\|\left[\vec{\lambda}_{q_K}\cdot\vec{\lambda}_{k_K},\sum_{L>K}\vec{\lambda}_{q_L}\cdot\vec{\lambda}_{k_L}\right]\right\|
\end{equation}
where $J_K$ is the coupling strength for the pair $K=1,\ldots,\binom{N}{2}$. The five types of contributions that result in nontrivial the commutators are the following (see also the two flavor derivation in Ref.~\cite{Amitrano:2022yyn} for more details)
\begin{itemize}
	\item $q_L<q_K$ and $k_L=q_K>q_L$
	\item $q_L<q_K$ and $k_L=k_K>q_L$
	\item $q_L=q_K$ and $k_L\neq k_K$ and $k_L>q_L$
	\item $k_K>q_L>q_K$ and $k_L=k_K>q_L$
	\item $q_L=k_K$ and $k_L>q_L$
\end{itemize}
We can therefore rewrite the sum of commutators as follows
\begin{widetext}
	\begin{equation}
		\begin{split}
			&\sum_{q<k}^NJ_{qk}\left\|\left[\vec{\lambda}_q\cdot\vec{\lambda}_k,\sum_{l=1}^{q-1}J_{lq}\vec{\lambda}_l\cdot\vec{\lambda}_q+\sum_{l=1}^{q-1}J_{lk}\vec{\lambda}_l\cdot\vec{\lambda}_k+\sum_{l=q+1}^{k-1}J_{lq}\vec{\lambda}_q\cdot\vec{\lambda}_l+\sum_{l=k+1}^NJ_{lq}\vec{\lambda}_q\cdot\vec{\lambda}_l+\sum_{l=q+1}^{k-1}J_{lk}\vec{\lambda}_l\cdot\vec{\lambda}_k+\sum_{l=k+1}^NJ_{lk}\vec{\lambda}_k\cdot\vec{\lambda}_l\right]\right\|\\
			=&\sum_{q<k}^NJ_{qk}\left\|\left[\vec{\lambda}_q\cdot\vec{\lambda}_k,\sum_{l=1}^{q-1}\vec{\lambda}_l\cdot\left(J_{lq}\vec{\lambda}_q+J_{lk}\vec{\lambda}_k\right)+\sum_{l=q+1}^{k-1}\vec{\lambda}_l\cdot\left(J_{lq}\vec{\lambda}_q+J_{lk}\vec{\lambda}_k\right)+\sum_{l=k+1}^N\vec{\lambda}_l\cdot\left(J_{lq}\vec{\lambda}_q+J_{lk}\vec{\lambda}_k\right)\right]\right\|\\
		\end{split}
	\end{equation}
\end{widetext}
At this point we need the commutator identity
\begin{equation}
	\begin{split}
		\left[\vec{\lambda}_q\cdot\vec{\lambda}_k,\vec{\lambda}_q\cdot\vec{\lambda}_l\right]&=
		\sum_{n=1}^8\sum_{m=1}^8\left[\lambda_m^q,\lambda^q_n\right]\lambda_m^k\lambda_n^l\\
		=&2i\sum_{n=1}^8\sum_{m=1}^8\sum_{j=1}^8f^{mnj}\lambda^q_j\lambda_m^k\lambda_n^l\\
		=&\sum_{n<m}^8\left[\lambda_m^q,\lambda^q_n\right]\left(\lambda_m^k\lambda_n^l-\lambda_n^k\lambda_m^l\right)\\
	\end{split}
\end{equation}
From the last result in the second line it is clear using the anti-symmetry of the structure constants that
\begin{equation}
	\left[\vec{\lambda}_q\cdot\vec{\lambda}_k,\vec{\lambda}_k\cdot\vec{\lambda}_l\right]=-\left[\vec{\lambda}_q\cdot\vec{\lambda}_k,\vec{\lambda}_q\cdot\vec{\lambda}_l\right]\;.
\end{equation}
A direct calculations then shows that
\begin{equation}
	\begin{split}
		\left\|\left[\vec{\lambda}_q\cdot\vec{\lambda}_k,\vec{\lambda}_q\cdot\vec{\lambda}_l\right]\right\|=
		4\sqrt{3}\;.
	\end{split}
\end{equation}
Using these results we can then find the upper bound
\begin{equation}
	\begin{split}
		\|C_{22}\| \leq& 4\sqrt{3}\sum_{q<k}^NJ_{qk}\left(\sum_{l=1}^{q-1}\big|J_{lq}-J_{lk}\big|\right.\\
		&\quad\quad\quad\quad\quad+\sum_{l=q+1}^{k-1}\left|J_{lq}-J_{lk}\right|\\
		&\quad\quad\quad\quad\quad\left.+\sum_{l=k+1}^N\big|J_{lq}-J_{lk}\big|\right)\\
		<&4\sqrt{3}\sum_{q<k}^NJ_{qk}\sum_{l=1}^N\left|J_{lq}-J_{lk}\right|\;.
	\end{split}
\end{equation}
As we did before, we can find simpler bounds by bounding the $J_{qk}$ coupling constants
\begin{equation}
	\begin{split}
		\|C_{22}\|<&4\sqrt{3}\frac{2\mu}{2N}\frac{\mu}{2N}\binom{N}{2}\sum_{l=1}^N\max_{q,k}\left|\cos(\theta_{lq})-\cos(\theta_{lk})\right|\\
		<&\sqrt{3}\mu^2\sum_{l=1}^N\max_{q,k}\left|\cos(\theta_{lq})-\cos(\theta_{lk})\right|
	\end{split}
\end{equation}
or the even simpler one
\begin{equation}
	\|C_{22}\|<\sqrt{3}\mu^2N\max_{q,k,l}\left|\cos(\theta_{lq})-\cos(\theta_{lk})\right|\;.
\end{equation}

We can now put everything together and get an error bound for a first-order Trotter decomposition. We proceed in two steps by first splitting the one and two body evolution obtaining~\cite{PhysRevX.11.011020,Amitrano:2022yyn}
\begin{equation}
	\|e^{-iHt} - e^{-itH_\nu}e^{-itH_{\nu\nu}}\|\leq \frac{t^2}{2}\|C_{12}\|\;.
\end{equation}
We then split the two body evolution into a product of evolutions over individual pairs as follows
\begin{equation}
	U_{pair}(t) = \prod_{q<k}e^{-itJ_{qk}\vec{\lambda}_q\cdot\vec{\lambda}_k}\;.
\end{equation}
The errors we incur by doing this is
\begin{equation}
	\|e^{-itH_{\nu\nu}}-U_{pair}(t)\|\leq \frac{t^2}{2}\|C_{22}\|\;.
\end{equation}
Putting all together by doing $r$ steps of size $t/r$ each we finally have
\begin{widetext}
	\begin{equation}
		\begin{split}
			\left\|e^{-iHt}-\prod_{s=1}^re^{-i\frac{t}{r}H_\nu}U_{pair}\left(\frac{t}{r}\right)\right\|&\leq r\left\|e^{-iH\frac{t}{r}}-e^{-i\frac{t}{r}H_\nu}U_{pair}\left(\frac{t}{r}\right)\right\|\\
			&\leq r\left\|e^{-iH\frac{t}{r}}-e^{-i\frac{t}{r}H_\nu}e^{-i\frac{t}{r}H_{\nu\nu}}\right\|+r\left\|e^{-i\frac{t}{r}H_\nu}e^{-i\frac{t}{r}H_{\nu\nu}}-e^{-i\frac{t}{r}H_\nu}U_{pair}\left(\frac{t}{r}\right)\right\|\\
			&\leq\frac{t^2}{2r}\left(\|C_{12}\|+\|C_{22}\|\right)\\
			&<\frac{t^2}{2r}\mu N\left(2 \max_{q,k}|\omega_k-\omega_q|+\sqrt{3}\mu\max_{q,k,l}\left|cos(\theta_{lq})-\cos(\theta_{lk})\right|\right)
		\end{split}
	\end{equation}
\end{widetext}
If we use the short hand notation
\begin{align}
	\Delta\omega_{max} \ &= \ \max_{q,k}|\omega_k-\omega_q| \\
	\Delta\theta_{max} \ &= \ \max_{q,k,l}\left|cos(\theta_{lq})-\cos(\theta_{lk})\right|\;.
\end{align}
We have that the total number of steps to guarantee an error $\epsilon$ over a total time evolution interval of size $t$ is 
\begin{equation}
	\begin{split}
		r\ &=\ \frac{t^2}{2\epsilon}\mu N\left(2 \Delta\omega_{max}+\sqrt{3} \mu \Delta\theta_{max}\right) \\
		\ &= \ \mathcal{O}\left(\frac{t^2\mu N}{\epsilon}(\Delta\omega_{max}+\mu)\right)\;.
	\end{split}
\end{equation}
By using the same strategy used in Ref.~\cite{Amitrano:2022yyn} to bound the error of the second-order formula with two flavors, we can extend this result and show a scaling linear in system size $N$ for all the higher order Trotter formulas.

\section{{Qubit quantum circuit}}
\label{app:QC}

In Sec.~\ref{subsec:qubit}, we outline our procedure to simulate three-flavor collective oscillations on qubit-based hardware. 
Here, we derive the circuits to perform these simulations in terms of elementary gates in the following order: 
First, we explore in Appendix~\ref{app:PMNSmatrix} how to transform from the flavor basis, in which we prepare our initial wave function and measure our evolved state, and the mass basis where time evolution will be performed. 
To follow up, we show how to decompose exponentials of one- and two-body terms in the Hamiltonian into elementary gates in Appendices~\ref{app:one-body} and~\ref{app:2BodyInteraction}, respectively, allowing us to perform time steps with the desired Hamiltonian model. 
These decompositions involve a number of useful algebraic identities that we prove in Appendix~\ref{app:diagonalizingRelations}.

\subsection{PMNS matrix}
\label{app:PMNSmatrix}

The unitary operator mixing neutrino flavor and mass eigenstates is the PMNS matrix:
\begin{equation}
	\begin{pmatrix} \nu_e \\ \nu_\mu \\ \nu_\tau \end{pmatrix} = U_{\rm PMNS} \times \begin{pmatrix} \nu_1 \\ \nu_2 \\ \nu_3 \end{pmatrix},
\end{equation}
\begin{equation}
	\begin{split}
		U_{\rm PMNS} =& \begin{bmatrix} 1 & 0 & 0 \\ 0 & c_{23} & s_{23} \\ 0 & -s_{23} & c_{23} \end{bmatrix} \times 
		\begin{bmatrix} c_{13} & 0 & s_{13}e^{-i\delta_\mathrm{CP}} \\ 0 & 1 & 0 \\ -s_{13}e^{i\delta_\mathrm{CP}} & 0 & c_{13} \end{bmatrix}\\
		\times& \begin{bmatrix} c_{12} & s_{12} & 0 \\ -s_{12} & c_{12} & 0 \\ 0 & 0 & 1 \end{bmatrix}
	\end{split}
\end{equation}
where $c_{ij} := \cos( \theta_{ij} ), s_{ij} :=\sin( \theta_{ij} )$. Since we are using a qubit mapping where the $\ket{00}$ state is non-physical, we can write the three rotation matrices as follows:
\begin{equation}
	\begin{pmatrix}
		1&0&0&0\\
		0&1&0&0\\
		0&0&c_{23}&s_{23}\\
		0&0&-s_{23}&c_{23}\\
	\end{pmatrix}=\vcenteredhbox{\Qcircuit @C=1em @R=1em { 
			&\ctrl{1}&\qw\\
			&\gate{R_y(-2\theta_{23})} &\qw\\
	}}
\end{equation}
\begin{equation}
	\begin{pmatrix}
		1&0&0&0\\
		0&c_{13}&0&s_{13}\\
		0&0&1&0\\
		0&-s_{13}&0&c_{13}\\
	\end{pmatrix}=\vcenteredhbox{\Qcircuit @C=1em @R=1em { 
			&\gate{R_y(-2\theta_{23})}&\qw\\
			&\ctrl{-1} &\qw\\
	}}
\end{equation}
and the Givens rotation from Ref.~\cite{Sung_2023}
\begin{equation}
	\begin{pmatrix}
		1&0&0&0\\
		0&c_{12}&s_{12}&0\\
		0&-s_{12}&c_{12}&0\\
		0&0&0&1\\
	\end{pmatrix}=
	\vcenteredhbox{
		\resizebox{0.3\textwidth}{!}{
			\Qcircuit @C=1em @R=1em {
				&\gate{R_y(\pi/2)}&\ctrl{1} &\gate{R_y(-\theta_{23})}&\ctrl{1}&\gate{R_y(-\pi/2)}&\qw\\
				&\qw      &\targ&\gate{R_y(-\theta_{23})}&\targ&\qw&\qw\\
	}}}
\end{equation}

The inclusion of a CP phase can be accomplished easily with single-qubit $Z$ rotations;
\begin{equation}
	\begin{split}
		&\begin{pmatrix}
			1&0&0&0\\
			0&c_{13}&0&s_{13}e^{-i\delta_{CP}}\\
			0&0&1&0\\
			0&-s_{13}e^{i\delta_{CP}}&0&c_{13}\\
		\end{pmatrix} \\
		&= 
		\vcenteredhbox{
			\resizebox{0.3\textwidth}{!}{
				\Qcircuit @C=1em @R=1em { 
					&\gate{R_z(-\delta_{CP})}&\gate{R_y(-2\theta_{23})}&\gate{R_z(\delta_{CP})}&\qw\\
					&\qw&\ctrl{-1} &\qw&\qw\\
		}}}
	\end{split}
\end{equation}

Alternatively, to realize the PMNS matrix with the cross-resonance gate $R_{ZX}(\theta)$, let us introduce some relations: an arbitrary $Y$ rotation can always be written as $R_y(\theta) = SHR_z(\theta)HS^\dag$, while the commutation relation between $S$ or $S^\dag$ and a CNOT are as follows
\begin{equation}
	\vcenteredhbox{\Qcircuit @C=1em @R=1em { 
			& \ctrl{1} & \qw & \qw \\
			& \targ{} & \gate{S} & \qw \\
	}}\;=
	\vcenteredhbox{\Qcircuit @C=1em @R=1em { 
			&\gate{S} &\ctrl{1}& \ctrl{1} & \qw \\
			&\gate{S} &\ctrl{0} & \targ & \qw \\
	}}\;
\end{equation}
\begin{equation}
	\vcenteredhbox{\Qcircuit @C=1em @R=1em { 
			&\qw & \ctrl{1} & \qw \\
			&\gate{S^\dag} & \targ{} & \qw \\
	}}\;
	=\vcenteredhbox{\Qcircuit @C=1em @R=1em { 
			&\ctrl{1}& \ctrl{1} & \gate{S^\dag} & \qw\\
			&\targ & \ctrl{0} & \gate{S^\dag} & \qw \\
	}}\;
\end{equation}
Since we can write a controlled-rotation using two CNOTs as
\begin{equation}
	\vcenteredhbox{\Qcircuit @C=1em @R=1em { 
			&\ctrl{1} & \qw \\
			&\gate{R_y(\theta)} & \qw \\
	}}\;=\vcenteredhbox{\Qcircuit @C=1em @R=1em { 
			&\qw & \ctrl{1} & \qw & \ctrl{1} & \qw \\
			& \gate{R_y(\theta /2)} & \targ & \gate{R_y(-\frac{\theta}{2})} & \targ &\qw \\
	}} \;
\end{equation}
we can then write it in terms of the cross-resonance gate as:
\begin{equation}
	\vcenteredhbox{\Qcircuit @C=1em @R=1em { 
			&\ctrl{1} & \qw \\
			&\gate{R_y(\theta)} & \qw \\
	}}\; = \vcenteredhbox{\Qcircuit @C=1em @R=1em { 
			&\qw & \qw & \multigate{1}{R_{ZX}(-\frac{\theta}{2})} & \qw & \qw \\
			& \gate{R_y(\frac{\theta}{2})} & \gate{S} & \ghost{R_{ZX}(-\frac{\theta}{2})} & \gate{S^\dag} & \qw \\
	}}
\end{equation}
recalling the relation between $R_{ZX}$ and $R_{ZZ}$:
\begin{equation}
	\vcenteredhbox{\Qcircuit @C=1em @R=1em { 
			& \ctrl{1} & \qw & \ctrl{1} & \qw \\
			& \targ & \gate{R_z(\theta)} & \targ & \qw \\
	}} = \vcenteredhbox{\Qcircuit @C=1em @R=1em { 
			&\qw & \multigate{1}{R_{ZX}(\theta)} & \qw & \qw \\
			&\gate{U_{1,1}} & \ghost{R_{ZX}(\theta)} & \gate{U_{1,1}} & \qw \\
	}}
\end{equation}
where $U_{1,1} = R_z(\pi/2)\sqrt{X}R_z(\pi/2)$ is the Hadamard gate $H$ up to a global phase, in particular $U_{1,1} = e^{i\pi/4}H$. As for the Givens rotation instead, we find the following circuit:
\begin{equation}
	\begin{split}
		&\vcenteredhbox{\resizebox{0.4\textwidth}{!}{
				\Qcircuit @C=1em @R=1em { 
					&\gate{R_y(\frac{\pi}{2})}&\ctrl{1} &\gate{R_y(\theta)}&\ctrl{1}&\gate{R_y(-\frac{\pi}{2})}&\qw\\
					&\qw      &\targ&\gate{R_y(\theta)}&\targ&\qw&\qw\\
		}}} = \\
		&= \vcenteredhbox{\resizebox{0.45\textwidth}{!}{
				\Qcircuit @C=1em @R=1em { 
					& \gate{R_y(\frac{\pi}{2})} & \gate{S^\dag} & \multigate{1}{R_{ZX}(\theta)} &  \gate{H} & \multigate{1}{R_{XZ}(\theta)} & \gate{H} & \gate{S} & \gate{R_Y(-\frac{\pi}{2})} & \qw \\
					&\qw & \gate{S^\dag} & \ghost{R_{ZX}(\theta)} & \gate{S} & \ghost{R_{XZ}(\theta)} & \qw & \qw & \qw & \qw \\
		}}}
	\end{split}
\end{equation}
In this way, it is possible to write the PMNS matrix either with two CNOTs plus two controlled-$R_Y$ rotations, or with four cross-resonance gates $R_{ZX}$.

\begin{figure*}
	\centering
	\vcenteredhbox{\Qcircuit @C=1em @R=1em { 
			\lstick{q_1}&\multigate{3}{D_{45}^{q_1q_2q_3q_4}} &\qw\\
			\lstick{q_2}&\ghost{D_{45}^{q_1q_2q_3q_4}}      &\qw\\
			\lstick{q_3}&\ghost{D_{45}^{q_1q_2q_3q_4}}      &\qw\\
			\lstick{q_4}&\ghost{D_{45}^{q_1q_2q_3q_4}}      &\qw\\
	}} := \quad\quad\vcenteredhbox{\Qcircuit @C=1em @R=1em { 
			\lstick{q_1}&\gate{R}&\targ  &\gate{R^\dagger}&\targ&\gate{R}&\targ  &\gate{R^\dagger}&\targ    &\gate{R}&\targ    &\gate{R^\dagger}&\targ    &\gate{R}&\targ    &\gate{R^\dagger}&\targ    &\qw\\
			\lstick{q_2}&\qw     &\qw      &\qw     &\ctrl{-1}  &\qw     &\qw    &\qw             &\ctrl{-1}&\qw     &\qw      &\qw             &\ctrl{-1}&\qw     &\qw      &\qw             &\ctrl{-1}&\qw\\
			\lstick{q_3}&\qw     &\qw      &\qw     &\qw        &\qw   &\ctrl{-2}&\qw             &\qw      &\qw     &\qw      &\qw             &\qw      &\qw     &\ctrl{-2}&\qw             &\qw      &\qw\\
			\lstick{q_4}&\qw     &\ctrl{-3}&\qw     &\qw        &\qw     &\qw    &\qw             &\qw      &\qw     &\ctrl{-3}&\qw             &\qw      &\qw     &\qw      &\qw             &\qw      &\qw\\
	}}
	\caption{Diagonal unitary required to implement  $\exp(i\alpha(Q_4\otimes Q_4+Q_5\otimes Q_5))$ in conjuction with the operator $G_{13}$ from Eq.~\eqref{eq:G_ab}. The single qubit gate $R$ is a Z rotation with angle $\alpha$}
	\label{fig:D45_1234}
\end{figure*}
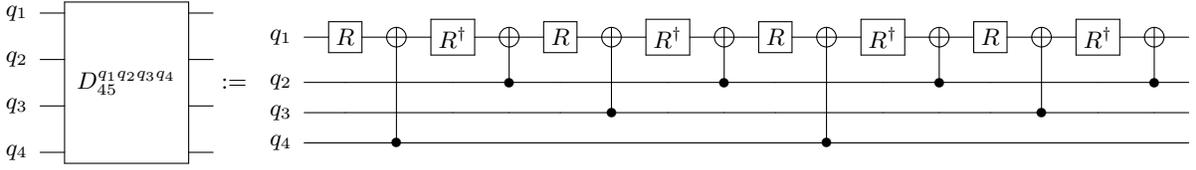

\begin{figure*}
	\centering
	\vcenteredhbox{\Qcircuit @C=1em @R=1em { 
			\lstick{q_1}&\multigate{3}{D_{12}^{q_1q_2q_3q_4}} &\qw\\
			\lstick{q_2}&\ghost{D_{12}^{q_1q_2q_3q_4}}      &\qw\\
			\lstick{q_3}&\ghost{D_{12}^{q_1q_2q_3q_4}}      &\qw\\
			\lstick{q_4}&\ghost{D_{12}^{q_1q_2q_3q_4}}      &\qw\\
	}} := \quad\quad \vcenteredhbox{\Qcircuit @C=1em @R=1em { 
			\lstick{q_1}&\gate{R}&\targ  &\gate{R}&\targ&\gate{R^\dagger}&\targ  &\gate{R}&\targ    &\gate{R^\dagger}&\targ    &\gate{R^\dagger}&\targ    &\gate{R}&\targ    &\gate{R^\dagger}&\targ    &\qw\\
			\lstick{q_2}&\qw     &\qw      &\qw     &\ctrl{-1}  &\qw     &\qw    &\qw             &\ctrl{-1}&\qw     &\qw      &\qw             &\ctrl{-1}&\qw     &\qw      &\qw             &\ctrl{-1}&\qw\\
			\lstick{q_3}&\qw     &\qw      &\qw     &\qw        &\qw   &\ctrl{-2}&\qw             &\qw      &\qw     &\qw      &\qw             &\qw      &\qw     &\ctrl{-2}&\qw             &\qw      &\qw\\
			\lstick{q_4}&\qw     &\ctrl{-3}&\qw     &\qw        &\qw     &\qw    &\qw             &\qw      &\qw     &\ctrl{-3}&\qw             &\qw      &\qw     &\qw      &\qw             &\qw      &\qw\\
	}}
	\caption{Diagonal unitary required to implement  $\exp(i\alpha(Q_1\otimes Q_1+Q_2\otimes Q_2))$ in conjuction with the operator $K$ from Eq.~\eqref{eq:K_op}. The single qubit gate $R$ is a $Z$ rotation with angle $\alpha$.}
	\label{fig:D12_1234}
\end{figure*}
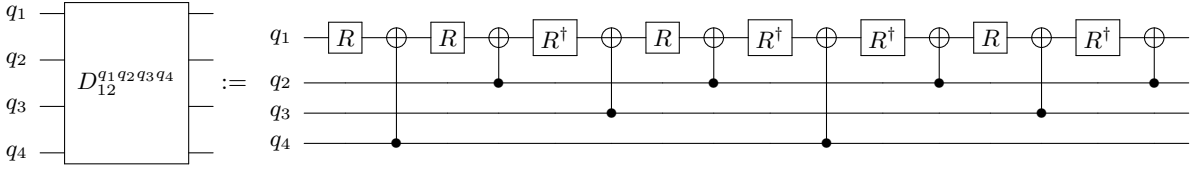

\subsection{One-body terms}
\label{app:one-body}

Next, we outline how to realize the one-body terms of Eq.~\eqref{eq:vacuum_H} with elementary gates acting on qubit hardware. 
To start, note the mass basis the one body term is, per our encoding in Sec.~\ref{sec:reps}, 
\begin{equation}
	\begin{split}
		H_{\nu}=& \sum_p {\omega}_p \vec{B}\cdot \vec{Q}_p=\sum_p \omega_p ( B_3 {Q_3}_p + B_8 {Q_8}_p ) \;,
	\end{split}
	\label{eq:app_one_body}
\end{equation}
where we can simplify the operator into
\begin{equation}
	\begin{split}
		B_3 {Q_3}_p + B_8 {Q_8}_p
		=&\left(\frac{B_3}{2} +\frac{B_8}{2\sqrt{3}}\right) (\sigma_3 \otimes \sigma_0)_p -\\
		&\left(\frac{B_3}{2} -\frac{B_8}{2\sqrt{3}}\right) (\sigma_0 \otimes \sigma_3)_p-\\
		&\frac{B_8}{\sqrt{3}} (\sigma_3 \otimes \sigma_3)_p
	\end{split}
\end{equation}

Let us denote with labels $A$ and $B$ the two qubits for a given momentum mode $p$. 
The time evolution operator for a single neutrino acting on the qubit pair is,
\begin{equation}
	\begin{split}
		U_1(t)=& R^z_{B}\left( \omega_p t\left(B_3 +\frac{B_8}{\sqrt{3}} \right)\right)\\
		&\times R^z_{A}\left(-\omega_p t\left(B_3 -\frac{B_8}{\sqrt{3}}\right) \right)\\
		&\times R^{zz}_{AB}\left(-2\omega_p \frac{B_8}{\sqrt{3}} t \right)
	\end{split}
	\label{eq:app_u1t}
\end{equation}
Using the fact that we are free to keep an arbitrary phase on the unphysical state $\ket{00}$, and everything is invariant up to a global phase, the evolution unitary we actually need to implement is the following simplification:
\begin{equation}
	\begin{split}
		\widetilde{U}_1(t)=&R^z_{B}\left( \omega_p t\left(B_3 +\frac{B_8}{\sqrt{3}}\right)\right)\\
		&\times R^z_{A}\left(-\omega_p t\left(B_3 -\frac{B_8}{\sqrt{3}}\right)\right).
	\end{split}
\end{equation}

\subsection{Two-body terms}
\label{app:2BodyInteraction}

Finally, we outline how to realize the interaction Hamiltonian of Eq.~\eqref{eq:twoBody_H} on qubits, also using the representation of Sec.~\ref{sec:reps}. 
{ We start by explicitly writing off-diagonal the pair operators $Q_{ii} + Q_{jj}$ in the Pauli basis,
	\begin{equation}
		\begin{split}
			Q_{11} + Q_{22} =& \frac{1}{4}\bigg[(X_1X_3+Y_1Y_3)(X_2X_4+Y_2Y_4)\\
			&+(X_1Y_3-Y_1X_3)(X_2Y_4-Y_2X_4)\bigg],\\
			Q_{44} + Q_{55} =&\frac{1}{4}\left(X_1X_3+Y_1Y_3\right)\left(1-Z_4-Z_2+Z_2Z_4\right),\\
			Q_{66} + Q_{77} =&\frac{1}{4}\left(X_2X_4+Y_2Y_4\right)\left(1-Z_4-Z_2+Z_2Z_4\right),\\
		\end{split}
		\label{eq:pairs}
\end{equation}}
Given this decomposition, and inspired by the circuit constructions derived in Refs.~\cite{Stetina2022simulatingeffective,PhysRevD.107.054512} using diagonalization gadgets,  we introduce the diagonalizing operators $G_{ab}$ and $K$ defined as follows:
\begin{equation}
	\label{eq:G_ab}
	\vcenteredhbox{\Qcircuit @C=1em @R=1em { 
			\lstick{q_a} &\multigate{1}{G_{ab}} &\qw\\
			\lstick{q_b} &\ghost{G_{ab}}      &\qw\\
	}}\;:=\vcenteredhbox{\Qcircuit @C=1em @R=1em { 
			&\gate{H} &\ctrl{1}&\qw\\
			&\qw      &\targ   &\qw\\
	}}\;
\end{equation}
\begin{equation}
	\label{eq:K_op}
	\vcenteredhbox{\Qcircuit @C=1em @R=1em { 
			&\multigate{3}{K} &\qw\\
			&\ghost{K}      &\qw\\
			&\ghost{K}      &\qw\\
			&\ghost{K}      &\qw\\
	}}\;:=\vcenteredhbox{\Qcircuit @C=1em @R=1em { 
			&\gate{H}&\ctrl{3}&\qw \\
			&\qw     &\targ   &\qw \\
			&\qw     &\targ   &\qw \\
			&\qw     &\targ   &\qw \\
	}}\;=\vcenteredhbox{\Qcircuit @C=1em @R=1em { 
			&\gate{H}&\ctrl{3}&\ctrl{2}&\ctrl{1}&\qw \\
			&\qw     &\qw     &\qw   &\targ &\qw \\
			&\qw     &\qw     &\targ &\qw   &\qw\\
			&\qw     &\targ   &\qw   &\qw   &\qw\\
	}}\;
\end{equation}
Indeed, these operators diagonalize the following terms 
(remember our notation $Q_i \otimes Q_i =: Q_{ii}$):
\begin{equation}
	\begin{split}
		G_{13} \left( Q_{44}+Q_{55} \right)G^\dagger_{13} &= \frac{Z_1-Z_1Z_3}{4}\left(1-Z_4-Z_2+Z_2Z_4\right) \\
		K(Q_{11} + Q_{22})K^{\dagger} &= \frac{Z_1 - Z_1Z_2}{4}(1-Z_3 + Z_4 - Z_3Z_4) \\
		G_{24} \left(Q_{66}+Q_{77}\right)G^\dagger_{24} &= \frac{Z_2-Z_2Z_4}{4}\left(1-Z_1-Z_3+Z_3Z_1\right) \\
	\end{split}
	\label{eq:diagonalizing_operators}
\end{equation}
The proof of these equalities is shown in Appendix \ref{app:diagonalizingRelations}. We can then use the fact that $Gf(\Vec{Q})G^\dag = f(G\Vec{Q}G^\dag)$ for every $f$ whose Taylor expansion is well-defined. So, from the first equality of Eq.~\eqref{eq:diagonalizing_operators} we find that the implementation of $G_{13}e^{i\alpha\left(Q_4\otimes Q_4+Q_5\otimes Q_5\right)}G^\dagger_{13}$ can be achieved with the diagonal circuit $D_{45}^{1234}$ (where the superscript denotes the order of qubits) shown in Fig.~\ref{fig:D45_1234}.

\begin{figure*}
	\centering
	\vcenteredhbox{\Qcircuit @C=1em @R=1em {
			\lstick{1} & \ctrl{2}&\gate{H}&\multigate{3}{D^{1234}_{45}} & \gate{H}& \ctrl{3}& \gate{H}& \multigate{3}{D_{12}^{1234}} & \gate{H}& \ctrl{3}& \multigate{3}{O_4R^2O_4} & \qw& \qw& \multigate{3}{D^{2143}_{45}} & \qw& \qw& \rstick{1} \qw \\
			\lstick{2} & \qw& \qw&\ghost{D^{1234}_{45}} & \qw& \targ& \qw& \ghost{D_{12}^{1234}} & \qw& \targ& \ghost{O_4R^2O_4} & \ctrl{2}& \gate{H}& \ghost{ D^{2143}_{45}} & \gate{H}& \ctrl{2}& \rstick{2} \qw \\
			\lstick{3} & \targ& \qw&\ghost{D^{1234}_{45}} & \gate{R^2}& \qw& \qw& \ghost{D_{12}^{1234}} & \qw& \targ& \ghost{O_4R^2O_4} & \qw&  \qw&\ghost{ D^{2143}_{45} } & \qw& \qw& \rstick{3} \qw \\
			\lstick{4} & \qw& \qw&\ghost{D^{1234}_{45}} & \qw& \targ& \qw& \ghost{D_{12}^{1234}} & \qw& \targ& \ghost{O_4R^2O_4} & \targ&  \qw&\ghost{ D^{2143}_{45} } & \gate{R^2}& \targ& \rstick{4} \qw \\
	}}
	\caption{Circuit implementation of the full two-body evolution operator with all-to-all connectivity.}
	\label{fig:twob_all_to_all}
\end{figure*}
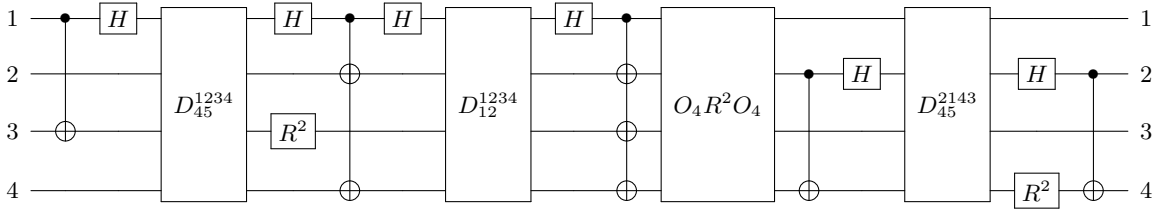
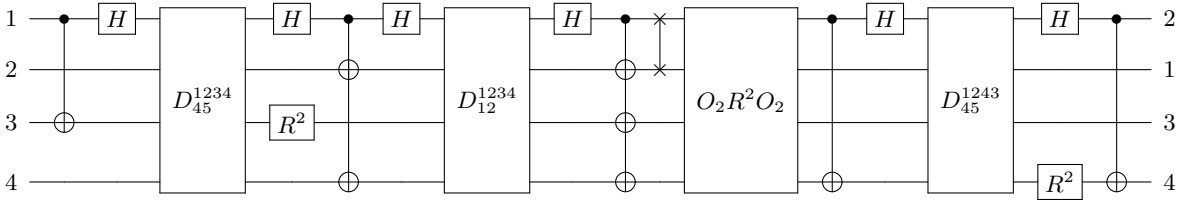
\begin{figure*}
	\centering
	\vcenteredhbox{\Qcircuit @C=1em @R=1em {
			\lstick{1} & \ctrl{2}&\gate{H}&\multigate{3}{D^{1234}_{45}} & \gate{H}& \ctrl{3}& \gate{H}& \multigate{3}{D_{12}^{1234}} & \gate{H}& \ctrl{3}& \qswap& \multigate{3}{O_2R^2O_2} & \ctrl{3}& \gate{H}& \multigate{3}{D^{1243}_{45}} & \gate{H}& \ctrl{3}& \rstick{2} \qw \\
			\lstick{2} & \qw& \qw&\ghost{D^{1234}_{45}} & \qw& \targ& \qw& \ghost{D_{12}^{1234}} & \qw& \targ& \qswap \qwx& \ghost{O_4R^2O_4} & \qw& \qw& \ghost{ D^{1243}_{45}} & \qw& \qw& \rstick{1} \qw \\
			\lstick{3} & \targ& \qw&\ghost{D^{1234}_{45}} & \gate{R^2}& \qw& \qw& \ghost{D_{12}^{1234}} & \qw& \targ& \qw& \ghost{O_4R^2O_4} & \qw&  \qw&\ghost{ D^{1243}_{45} } & \qw& \qw& \rstick{3} \qw \\
			\lstick{4} & \qw& \qw&\ghost{D^{1234}_{45}} & \qw& \targ& \qw& \ghost{D_{12}^{1234}} & \qw& \targ& \qw& \ghost{O_4R^2O_4} & \targ&  \qw&\ghost{ D^{1243}_{45} } & \gate{R^2}& \targ& \rstick{4} \qw \\
	}}
	\caption{Circuit implementation of the full two-body evolution operator with T connectivity.}
	\label{fig:twob_t_conn}
\end{figure*}

With a permutation of qubits $(1,3)\rightarrow(2,4)$ we find that $G_{24}e^{i\alpha\left(Q_6\otimes Q_6+Q_7\otimes Q_7\right)}G^\dagger_{24}$ can be implemented by the same circuit, which with the notation of before would be $D_{45}^{2143}$. Both of these terms can be done with $10$ CNOT each for all-to-all connectivity. The unitary operator corresponding to $Ke^{i\alpha\left(Q_1\otimes Q_1+Q_2\otimes Q_2\right)}K^\dag$ is slightly different, and it is implemented by the circuit shown in Fig.~\ref{fig:D12_1234}. Together with the cost of implementing the unitary $K$, and its inverse, the cost of the full evolution for this term is $14$ CNOT gates with all-to-all connectivity. { We denote rotation gate on the Z-axis as $R=\exp(-\frac{I\alpha}{2} Z)$.} 

The last operation we need to perform is the evolution under the two diagonal generators:
\begin{equation}
	Q_{33}+Q_{88} = -\frac{2}{3}\mathbb{1} +2\left(Z_1Z_3+Z_2Z_4+Z_1Z_2Z_3Z_4\right)\;
\end{equation}
This expression for $Q_3\otimes Q_3+Q_8\otimes Q_8$ was obtained by adding operators that only modify the unphysical Hilbert space. Since { $Z_1Z_3$ and $Z_2Z_4$ commute with $Q_{11}+Q_{22}$, $Q_{44}+Q_{55}$, and $Q_{66}+Q_{77}$ based on Eq.\eqref{eq:pairs}}, and $G_{13}Z_1Z_3G_{13}^\dag=Z_3$ (and the same holds for $Z_2Z_4$ with $G_{24}$) we can implement this operator by simply adding a single qubit rotation { $R^2$} when we perform the diagonal pieces $D_{45}$ shown in Fig.~\ref{fig:D45_1234}. As for the $Z_1Z_2Z_3Z_4$ term instead, we can implement it by introducing the new operator $O_i$, where the $i$ index denotes the target qubit, which gives $O_i Z_1Z_2Z_3Z_4 O_i = Z_i$. (Note $O^\dagger=O$.) This operation is simple to perform with three CNOT gates, for instance $O_1$ can be implemented as
\begin{equation}
	\vcenteredhbox{\Qcircuit @C=1em @R=1em { 
			\lstick{q_1} &\multigate{3}{O_1} &\qw\\
			\lstick{q_2} &\ghost{O_1}      &\qw\\
			\lstick{q_3} &\ghost{O_1}      &\qw\\
			\lstick{q_4} &\ghost{O_1}      &\qw\\
	}}\;=\vcenteredhbox{\Qcircuit @C=1em @R=1em { 
			&\targ    &\targ    &\targ     &\qw \\
			&\ctrl{-1}&\qw      &\qw       &\qw \\
			&\qw      &\ctrl{-2}&\qw       &\qw \\
			&\qw      &\qw      &\ctrl{-3} &\qw \\
	}}\;
\end{equation}
{ Then, the resulting operation is $O_iR^2O_i$.}
In summary, with an all-to-all connectivity, we can implement the full two-body interaction term with the circuit shown in Fig.~\ref{fig:twob_all_to_all}. 
In the shown implementation we simplified together two CNOT gates shared between $G_{13}$ and $K^\dag$ reducing the gate count to 38 CNOTs. In addition, four more CNOT gates can be removed simplifying gates shared with $O_4$ bringing the final CNOT count to $34$. We can also consider the case, like on the IBM device we employed for the simulations shown in the main text, of $4$ qubits with T connectivity where we have one qubit that interacts with the other $3$, but where the other $3$ qubits interact only with the one in the center. In this case, we see already that the diagonal unitaries in Fig.~\ref{fig:D45_1234} and Fig.~\ref{fig:D12_1234} are compatible with the T connectivity provided $q_1$ is the central qubit. We then start with qubit 1 in the middle and swap $1\leftrightarrow2$ in order to perform the evolution generated by $Q_6\otimes Q_6+Q_7\otimes Q_7$. By choosing { $O_2R^2O_2$} for the four-qubit rotation, as shown in Fig.~\ref{fig:twob_t_conn}, we can then cancel two of the CNOT gates in the middle SWAP with neighboring gates resulting in a total of 39 CNOT gates. At the end of the step the order of the first and second qubits is reversed.

\subsection{Diagonalizing-relations}
\label{app:diagonalizingRelations}

\begin{equation}
	\vcenteredhbox{\Qcircuit @C=1em @R=1em { 
			&\gate{H} &\ctrl{1}&\gate{X}&\ctrl{1}&\gate{H}&\qw\\
			&\qw      &\targ   &\gate{X}&\targ&\qw&\qw\\
	}}\;=\vcenteredhbox{\Qcircuit @C=1em @R=1em { 
			&\gate{Z} &\qw\\
			&\qw      &\qw\\
	}}
\end{equation}
\begin{equation}
	\vcenteredhbox{\Qcircuit @C=1em @R=1em { 
			&\gate{H} &\ctrl{1}&\gate{Y}&\ctrl{1}&\gate{H}&\qw\\
			&\qw      &\targ   &\gate{Y}&\targ&\qw&\qw\\
	}}\;=\vcenteredhbox{\Qcircuit @C=1em @R=1em { 
			&\gate{Z} &\qw\\
			&\gate{-Z}      &\qw\\
	}}
\end{equation}
\begin{equation}
	\vcenteredhbox{\Qcircuit @C=1em @R=1em { 
			&\gate{H} &\ctrl{1}&\gate{Z}&\ctrl{1}&\gate{H}&\qw\\
			&\qw      &\targ   &\gate{Z}&\targ&\qw&\qw\\
	}}\;=\vcenteredhbox{\Qcircuit @C=1em @R=1em { 
			&\qw &\qw\\
			&\gate{Z}      &\qw\\
	}}
\end{equation}

With those relations we can easily prove the following relations:
\begin{equation}
	G_{13}\left(Q_{44}+Q_{55}\right)G^\dagger_{13}= \frac{Z_1-Z_1Z_3}{4}\left(1-Z_4-Z_2+Z_2Z_4\right)
\end{equation}
\begin{equation}
	G_{24}\left(Q_{66}+Q_{77} \right)G^\dagger_{24}=\frac{Z_2-Z_2Z_4}{4}\left(1-Z_1-Z_3+Z_3Z_1\right)
\end{equation}

\begin{equation}
	\vcenteredhbox{\Qcircuit @C=1em @R=1em {
			&\gate{H}&\ctrl{3} &\gate{Z} &\ctrl{3} &\gate{H} &\qw \\
			&\qw     &\targ    &\gate{Z} &\targ    &\qw &\qw \\
			&\qw     &\targ    &\gate{Z} &\targ    &\qw &\qw \\
			&\qw     &\targ    &\gate{Z} &\targ    &\qw &\qw \\
	}}\;=\vcenteredhbox{\Qcircuit @C=1em @R=1em { 
			&\qw &\qw\\
			&\gate{Z} &\qw    \\
			&\gate{Z} &\qw    \\
			&\gate{Z} &\qw    \\
	}}\;
\end{equation}
\begin{equation}
	\vcenteredhbox{\Qcircuit @C=1em @R=1em {
			&\gate{H}&\ctrl{3} &\gate{X} &\ctrl{3} &\gate{H} &\qw \\
			&\qw     &\targ    &\gate{X} &\targ    &\qw &\qw \\
			&\qw     &\targ    &\gate{Y} &\targ    &\qw &\qw \\
			&\qw     &\targ    &\gate{Y} &\targ    &\qw &\qw \\
	}}\;=\vcenteredhbox{\Qcircuit @C=1em @R=1em { 
			&\gate{Z} &\qw\\
			&\qw &\qw    \\
			&\gate{-Z} &\qw    \\
			&\gate{Z} &\qw    \\
	}}\;
\end{equation}
\begin{equation}
	\vcenteredhbox{\Qcircuit @C=1em @R=1em {
			&\gate{H}&\ctrl{3} &\gate{Y} &\ctrl{3} &\gate{H} &\qw \\
			&\qw     &\targ    &\gate{Y} &\targ    &\qw &\qw \\
			&\qw     &\targ    &\gate{X} &\targ    &\qw &\qw \\
			&\qw     &\targ    &\gate{X} &\targ    &\qw &\qw \\
	}}\;=\vcenteredhbox{\Qcircuit @C=1em @R=1em { 
			&\gate{Z} &\qw\\
			&\gate{-Z} &\qw    \\
			&\qw &\qw    \\
			&\qw &\qw    \\
	}}\;
\end{equation}
\begin{equation}
	\vcenteredhbox{\Qcircuit @C=1em @R=1em {
			&\gate{H}&\ctrl{3} &\gate{Y} &\ctrl{3} &\gate{H} &\qw \\
			&\qw     &\targ    &\gate{Y} &\targ    &\qw &\qw \\
			&\qw     &\targ    &\gate{Y} &\targ    &\qw &\qw \\
			&\qw     &\targ    &\gate{Y} &\targ    &\qw &\qw \\
	}}\;=\vcenteredhbox{\Qcircuit @C=1em @R=1em { 
			&\gate{Z} &\qw\\
			&\gate{Z} &\qw    \\
			&\gate{Z} &\qw    \\
			&\gate{Z} &\qw    \\
	}}\;
\end{equation}
\begin{equation}
	\vcenteredhbox{\Qcircuit @C=1em @R=1em {
			&\gate{H}&\ctrl{3} &\gate{X} &\ctrl{3} &\gate{H} &\qw \\
			&\qw     &\targ    &\gate{X} &\targ    &\qw &\qw \\
			&\qw     &\targ    &\gate{X} &\targ    &\qw &\qw \\
			&\qw     &\targ    &\gate{X} &\targ    &\qw &\qw \\
	}}\;=\vcenteredhbox{\Qcircuit @C=1em @R=1em { 
			&\gate{Z} &\qw\\
			&\qw &\qw    \\
			&\qw &\qw    \\
			&\qw &\qw    \\
	}}\;
\end{equation}

With those operators we have that:
\begin{equation}
	K(Q_1 \otimes Q_1)K^{\dagger} = \frac{Z_1}{4} (1 - Z_2 - Z_3Z_4 + Z_2Z_3Z_4)
\end{equation}

\begin{equation}
	\vcenteredhbox{\Qcircuit @C=1em @R=1em {
			&\gate{H}&\ctrl{3} &\gate{Y} &\ctrl{3} &\gate{H} &\qw \\
			&\qw     &\targ    &\gate{X} &\targ    &\qw &\qw \\
			&\qw     &\targ    &\gate{Y} &\targ    &\qw &\qw \\
			&\qw     &\targ    &\gate{X} &\targ    &\qw &\qw \\
	}}\;=\vcenteredhbox{\Qcircuit @C=1em @R=1em { 
			&\gate{Z} &\qw\\
			&\qw &\qw    \\
			&\gate{-Z} &\qw    \\
			&\qw &\qw    \\
	}}\;
\end{equation}

\begin{equation}
	\vcenteredhbox{\Qcircuit @C=1em @R=1em {
			&\gate{H}&\ctrl{3} &\gate{Y} &\ctrl{3} &\gate{H} &\qw \\
			&\qw     &\targ    &\gate{X} &\targ    &\qw &\qw \\
			&\qw     &\targ    &\gate{X} &\targ    &\qw &\qw \\
			&\qw     &\targ    &\gate{Y} &\targ    &\qw &\qw \\
	}}\;=\vcenteredhbox{\Qcircuit @C=1em @R=1em {
			&\gate{Z} &\qw\\
			&\qw &\qw    \\
			&\qw &\qw    \\
			&\gate{-Z} &\qw    \\
	}}\;
\end{equation}

\begin{equation}
	\vcenteredhbox{\Qcircuit @C=1em @R=1em {
			&\gate{H}&\ctrl{3} &\gate{X} &\ctrl{3} &\gate{H} &\qw \\
			&\qw     &\targ    &\gate{Y} &\targ    &\qw &\qw \\
			&\qw     &\targ    &\gate{X} &\targ    &\qw &\qw \\
			&\qw     &\targ    &\gate{Y} &\targ    &\qw &\qw \\
	}}\;=\vcenteredhbox{\Qcircuit @C=1em @R=1em { 
			&\gate{Z} &\qw\\
			&\gate{Z} &\qw    \\
			&\qw &\qw    \\
			&\gate{-Z} &\qw    \\
	}}\;
\end{equation}

\begin{equation}
	\vcenteredhbox{\Qcircuit @C=1em @R=1em {
			&\gate{H}&\ctrl{3} &\gate{X} &\ctrl{3} &\gate{H} &\qw \\
			&\qw     &\targ    &\gate{Y} &\targ    &\qw &\qw \\
			&\qw     &\targ    &\gate{Y} &\targ    &\qw &\qw \\
			&\qw     &\targ    &\gate{X} &\targ    &\qw &\qw \\
	}}\;=\vcenteredhbox{\Qcircuit @C=1em @R=1em { 
			&\gate{Z} &\qw\\
			&\gate{Z} &\qw    \\
			&\gate{-Z} &\qw    \\
			&\qw &\qw    \\
	}}\;
\end{equation}
So that:
\begin{equation}
	K(Q_2 \otimes Q_2)K^{\dagger} = \frac{Z_1}{4} (-Z_3 + Z_4 + Z_2Z_3 - Z_2Z_4)
\end{equation}
and we prove with this that the operator $K$ diagonalizes the term $Q_1 \otimes Q_1 + Q_2 \otimes Q_2$:
\begin{equation}
	K(Q_{11} + Q_{22})K^{\dagger} = \frac{Z_1 - Z_1Z_2}{4} (1-Z_3 + Z_4 - Z_3Z_4)
\end{equation}

\section{Complete hardware results}
\label{app:complete_results}

\begin{figure*}
	\includegraphics[width=1.0\textwidth]{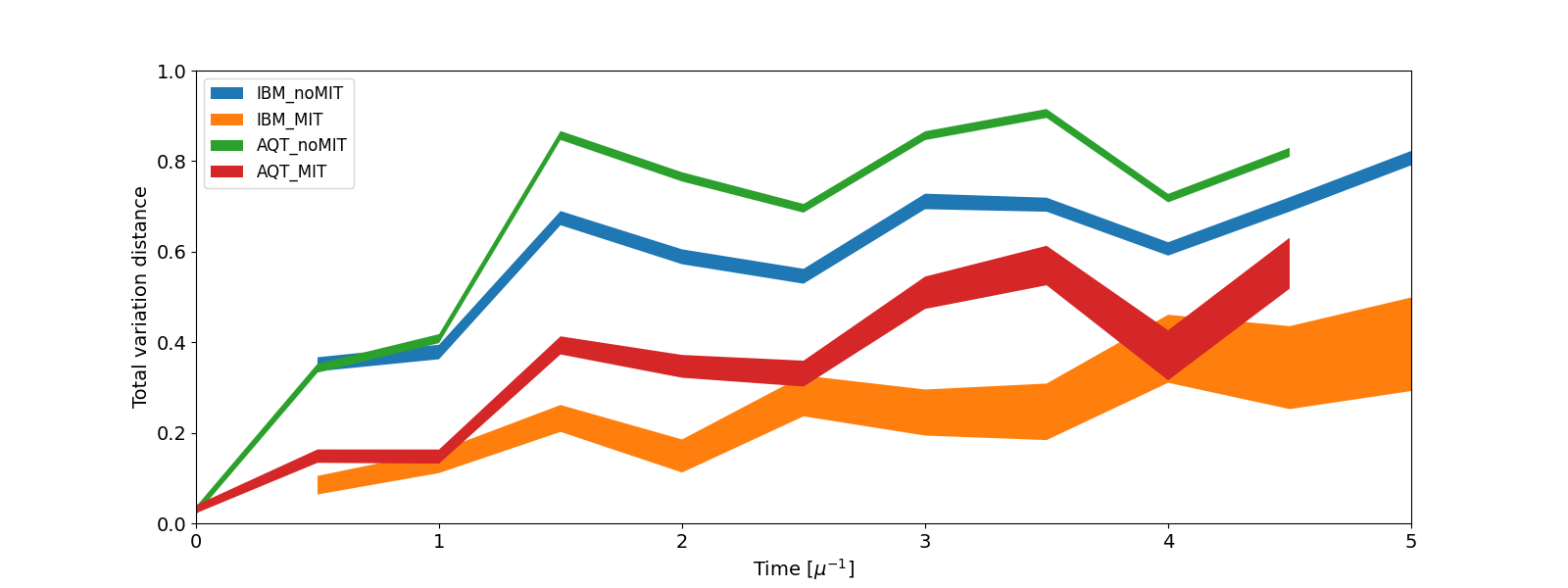}
	\caption{Total variation distances between the expected flavor probability distribution over flavor states and the distribution estimated from quantum computations. We show in blue and orange the qubit result without (IBM\_noMIT) and with (IBM\_MIT) error mitigation respectively. Similarly, we report the qutrit result without error mitigation (AQT\_noMIT) in green and with error mitigation (AQT\_MIT) in red. The bands correspond to 90\% confidence intervals.}
	\label{fig:tvd}
\end{figure*}

{
	In this section we present the results for {all states of the two neutrino systems}. We start by presenting in Fig.~\ref{fig:tvd} the total variation distance (TVD)
	\begin{equation}
		\delta(P_{ref},P_{meas})=\frac{1}{2}\sum_x\left|P_{ref}(x)-P_{meas}(x)\right|\;,
	\end{equation}
	between the expected distribution $P_{ref}$ over the nine flavor states of two neutrinos and the measured distribution $P_{meas}$ obtained from the hardware runs, with and without error mitigation. 
	The bands are $90\%$ confidence intervals obtained by resampling the experimental results assuming they follow a normal distribution with the estimated means and standard deviations. From these results, the beneficial effect of error mitigation is clear, consistently lowering the TVD in all cases. This analysis also shows that the qubit simulations, both with and without error mitigation, achieve a smaller TVD than the respective calculations on qutrit hardware even though after mitigation the spread appears reduced.
	
	In Fig.~\ref{fig:full_results_nomit}, data before error mitigation are shown. Each plot depicts the evolving probability of a given definite-flavor state, except the first panel which is a superposition of all other plots to underline the relative magnitude. 
	In each case we compare qubit and qutrit with the exact result. 
	Without any error mitigation, it is clear that hardware noise dominates our results, and data are far from the exact result after a couple of Trotter steps. 
	
	{However, one difference can be found in the probabilities on hardware for states that should have a very small theoretical probability during runtime (see the plots for the $\{\ket{\mu\mu},\ \ket{\mu\tau},\ \ket{\tau\mu},\ \ket{\tau\tau}\}$ states). The time evolution operator plays almost no role in the evolution of these states, and the major contribution is the depolarizing noise, which should converge to $1/\mathrm{Tr}[\mathbb{1}]$ where $\mathrm{Tr}[\mathbb{1}]$ is the total number of global states for (16 or 9 for qubits and qutrits, respectively). Since for qubits we have the presence of an nonphysical state, the depolarizing noise will converge to $1/16 \approx 0.06$. In the case of qutrits instead, the noise will converge to $1/9 \approx 0.1$. Indeed, unmitigated hardware results converge close to these values as function of time.} This difference is taken into account during error mitigation, as explained in more detail in Sec.~\ref{sec:experiments}. 
	
	{As already mentioned in Sec.~\ref{sec:conclusion}, this distinction} can give us some intuition about the impact of depolarizing noise with system scaling. Considering the qubit encoding for $N$ neutrinos, the fraction of physical states over the total number of states is $( 3/4 )^N$, which vanishes as the number of neutrinos increases. Since probability amplitudes will increasingly populate unphysical states due to the depolarizing noise in the qubit case, we can expect the qutrit encoding to scale better when considering only error-mitigation strategies. In practice, the problem with unphysical state might not be so impactful if one is only interested in local, small-weight observables.

	We show data after error-mitigation in Fig.~\ref{fig:full_results_mit}. As before, every plot displays the probability of being in a particular final state, while the plot in the first panel shows the relative magnitude between different states. 
Both qubit and qutrit results are close to the exact evolution, and even information in data that looked like fully depolarized have been recovered with error mitigation.  {There is no clear advantage between the two methods for this small system, but scaling up the particle number should lead to the differences outlined above.}

\begin{figure*}
	\includegraphics[width=1.0\textwidth]{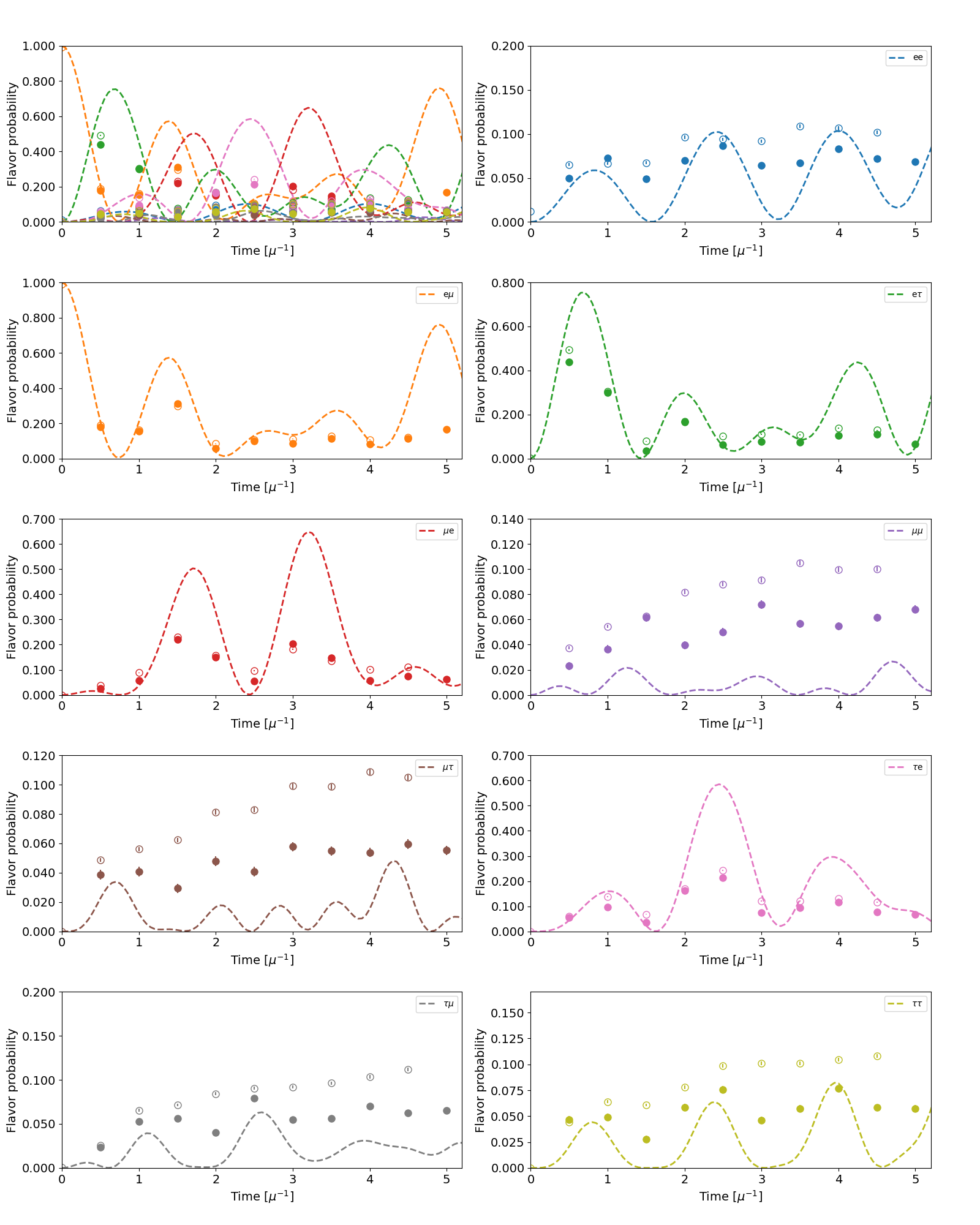}
	\caption{The vertical axis shows the probability of occupying a given final state. We present one plot for each possible two-flavor neutrino state, while the first panel shows a summary, highlighting the relative magnitudes. We also quantify the uncertainty using standard error through the vertical error bars. The dashed line is the exact calculation, while solid and empty dots represent the qubit (IBM) and qutrit (AQT) data respectively, before error mitigation.}
	\label{fig:full_results_nomit}
\end{figure*}

\begin{figure*}
	\includegraphics[width=1.0\textwidth]{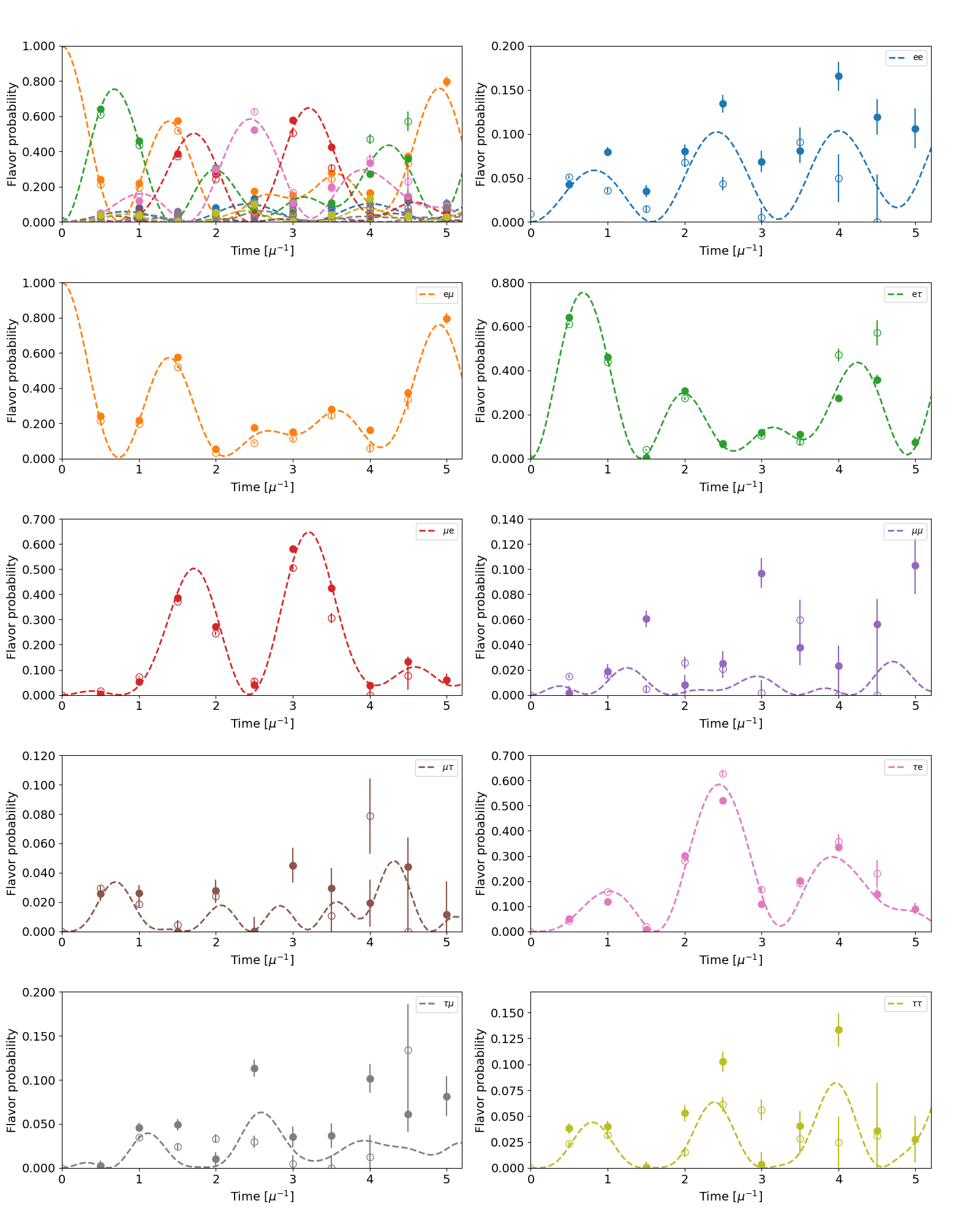}
	\caption{The vertical axis shows the probability of occupying a given final state. We present one plot for each possible two-flavor neutrino state, while the first panel shows a summary, highlighting the relative magnitudes. The dashed line is the exact calculation, while solid and empty dots represent the qubit (IBM) and qutrit (AQT) data respectively, after error mitigation. We also quantify the uncertainty using standard error through the vertical error bars. As expected error mitigation increases the variance observed.}
	\label{fig:full_results_mit}
\end{figure*}


\begin{thebibliography}{73}%
	\makeatletter
	\providecommand \@ifxundefined [1]{%
		\@ifx{#1\undefined}
	}%
	\providecommand \@ifnum [1]{%
		\ifnum #1\expandafter \@firstoftwo
		\else \expandafter \@secondoftwo
		\fi
	}%
	\providecommand \@ifx [1]{%
		\ifx #1\expandafter \@firstoftwo
		\else \expandafter \@secondoftwo
		\fi
	}%
	\providecommand \natexlab [1]{#1}%
	\providecommand \enquote  [1]{``#1''}%
	\providecommand \bibnamefont  [1]{#1}%
	\providecommand \bibfnamefont [1]{#1}%
	\providecommand \citenamefont [1]{#1}%
	\providecommand \href@noop [0]{\@secondoftwo}%
	\providecommand \href [0]{\begingroup \@sanitize@url \@href}%
	\providecommand \@href[1]{\@@startlink{#1}\@@href}%
	\providecommand \@@href[1]{\endgroup#1\@@endlink}%
	\providecommand \@sanitize@url [0]{\catcode `\\12\catcode `\$12\catcode
		`\&12\catcode `\#12\catcode `\^12\catcode `\_12\catcode `\%12\relax}%
	\providecommand \@@startlink[1]{}%
	\providecommand \@@endlink[0]{}%
	\providecommand \url  [0]{\begingroup\@sanitize@url \@url }%
	\providecommand \@url [1]{\endgroup\@href {#1}{\urlprefix }}%
	\providecommand \urlprefix  [0]{URL }%
	\providecommand \Eprint [0]{\href }%
	\providecommand \doibase [0]{https://doi.org/}%
	\providecommand \selectlanguage [0]{\@gobble}%
	\providecommand \bibinfo  [0]{\@secondoftwo}%
	\providecommand \bibfield  [0]{\@secondoftwo}%
	\providecommand \translation [1]{[#1]}%
	\providecommand \BibitemOpen [0]{}%
	\providecommand \bibitemStop [0]{}%
	\providecommand \bibitemNoStop [0]{.\EOS\space}%
	\providecommand \EOS [0]{\spacefactor3000\relax}%
	\providecommand \BibitemShut  [1]{\csname bibitem#1\endcsname}%
	\let\auto@bib@innerbib\@empty
	\bibitem [{\citenamefont {Baha~Balantekin}\ and\ \citenamefont
		{Kayser}(2018)}]{Balantekin:2018ppj}%
	\BibitemOpen
	\bibfield  {author} {\bibinfo {author} {\bibfnamefont {A.}~\bibnamefont
			{Baha~Balantekin}}\ and\ \bibinfo {author} {\bibfnamefont {B.}~\bibnamefont
			{Kayser}},\ }\bibfield  {title} {\bibinfo {title} {{On the Properties of
				Neutrinos}},\ }\href {https://doi.org/10.1146/annurev-nucl-101916-123044}
	{\bibfield  {journal} {\bibinfo  {journal} {Ann. Rev. Nucl. Part. Sci.}\
		}\textbf {\bibinfo {volume} {68}},\ \bibinfo {pages} {313} (\bibinfo {year}
		{2018})}\BibitemShut {NoStop}%
	\bibitem [{\citenamefont {Huber}\ \emph {et~al.}(2022)\citenamefont {Huber}
		\emph {et~al.}}]{Huber:2022lpm}%
	\BibitemOpen
	\bibfield  {author} {\bibinfo {author} {\bibfnamefont {P.}~\bibnamefont
			{Huber}} \emph {et~al.},\ }\bibfield  {title} {\bibinfo {title} {{Snowmass
				Neutrino Frontier Report}},\ }in\ \href@noop {} {\emph {\bibinfo {booktitle}
			{{Snowmass 2021}}}}\ (\bibinfo {year} {2022})\ \Eprint
	{https://arxiv.org/abs/2211.08641} {arXiv:2211.08641 [hep-ex]} \BibitemShut
	{NoStop}%
	\bibitem [{\citenamefont {Mirizzi}\ \emph {et~al.}(2016)\citenamefont
		{Mirizzi}, \citenamefont {Tamborra}, \citenamefont {Janka}, \citenamefont
		{Saviano}, \citenamefont {Scholberg}, \citenamefont {Bollig}, \citenamefont
		{Hudepohl},\ and\ \citenamefont {Chakraborty}}]{Mirizzi:2015eza}%
	\BibitemOpen
	\bibfield  {author} {\bibinfo {author} {\bibfnamefont {A.}~\bibnamefont
			{Mirizzi}}, \bibinfo {author} {\bibfnamefont {I.}~\bibnamefont {Tamborra}},
		\bibinfo {author} {\bibfnamefont {H.-T.}\ \bibnamefont {Janka}}, \bibinfo
		{author} {\bibfnamefont {N.}~\bibnamefont {Saviano}}, \bibinfo {author}
		{\bibfnamefont {K.}~\bibnamefont {Scholberg}}, \bibinfo {author}
		{\bibfnamefont {R.}~\bibnamefont {Bollig}}, \bibinfo {author} {\bibfnamefont
			{L.}~\bibnamefont {Hudepohl}},\ and\ \bibinfo {author} {\bibfnamefont
			{S.}~\bibnamefont {Chakraborty}},\ }\bibfield  {title} {\bibinfo {title}
		{{Supernova Neutrinos: Production, Oscillations and Detection}},\ }\href
	{https://doi.org/10.1393/ncr/i2016-10120-8} {\bibfield  {journal} {\bibinfo
			{journal} {Riv. Nuovo Cim.}\ }\textbf {\bibinfo {volume} {39}},\ \bibinfo
		{pages} {1} (\bibinfo {year} {2016})}\BibitemShut {NoStop}%
	\bibitem [{\citenamefont {Janka}(2017)}]{Janka:2017vlw}%
	\BibitemOpen
	\bibfield  {author} {\bibinfo {author} {\bibfnamefont {H.-T.}\ \bibnamefont
			{Janka}},\ }\bibinfo {title} {Neutrino emission from supernovae},\ in\ \href
	{https://doi.org/10.1007/978-3-319-21846-5_4} {\emph {\bibinfo {booktitle}
			{Handbook of Supernovae}}},\ \bibinfo {editor} {edited by\ \bibinfo {editor}
		{\bibfnamefont {A.~W.}\ \bibnamefont {Alsabti}}\ and\ \bibinfo {editor}
		{\bibfnamefont {P.}~\bibnamefont {Murdin}}}\ (\bibinfo  {publisher} {Springer
		International Publishing},\ \bibinfo {address} {Cham},\ \bibinfo {year}
	{2017})\ pp.\ \bibinfo {pages} {1575--1604},\ \Eprint
	{https://arxiv.org/abs/1702.08713} {arXiv:1702.08713 [astro-ph.HE]}
	\BibitemShut {NoStop}%
	\bibitem [{\citenamefont {Burrows}\ and\ \citenamefont
		{Vartanyan}(2021)}]{Burrows:2020qrp}%
	\BibitemOpen
	\bibfield  {author} {\bibinfo {author} {\bibfnamefont {A.}~\bibnamefont
			{Burrows}}\ and\ \bibinfo {author} {\bibfnamefont {D.}~\bibnamefont
			{Vartanyan}},\ }\bibfield  {title} {\bibinfo {title} {{Core-Collapse
				Supernova Explosion Theory}},\ }\href
	{https://doi.org/10.1038/s41586-020-03059-w} {\bibfield  {journal} {\bibinfo
			{journal} {Nature}\ }\textbf {\bibinfo {volume} {589}},\ \bibinfo {pages}
		{29} (\bibinfo {year} {2021})}\BibitemShut {NoStop}%
	\bibitem [{\citenamefont {Duan}\ \emph {et~al.}(2010)\citenamefont {Duan},
		\citenamefont {Fuller},\ and\ \citenamefont {Qian}}]{Duan:2010bg}%
	\BibitemOpen
	\bibfield  {author} {\bibinfo {author} {\bibfnamefont {H.}~\bibnamefont
			{Duan}}, \bibinfo {author} {\bibfnamefont {G.~M.}\ \bibnamefont {Fuller}},\
		and\ \bibinfo {author} {\bibfnamefont {Y.-Z.}\ \bibnamefont {Qian}},\
	}\bibfield  {title} {\bibinfo {title} {{Collective Neutrino Oscillations}},\
	}\href {https://doi.org/10.1146/annurev.nucl.012809.104524} {\bibfield
		{journal} {\bibinfo  {journal} {Ann. Rev. Nucl. Part. Sci.}\ }\textbf
		{\bibinfo {volume} {60}},\ \bibinfo {pages} {569} (\bibinfo {year}
		{2010})}\BibitemShut {NoStop}%
	\bibitem [{\citenamefont {Chakraborty}\ \emph {et~al.}(2016)\citenamefont
		{Chakraborty}, \citenamefont {Hansen}, \citenamefont {Izaguirre},\ and\
		\citenamefont {Raffelt}}]{Chakraborty:2016yeg}%
	\BibitemOpen
	\bibfield  {author} {\bibinfo {author} {\bibfnamefont {S.}~\bibnamefont
			{Chakraborty}}, \bibinfo {author} {\bibfnamefont {R.}~\bibnamefont {Hansen}},
		\bibinfo {author} {\bibfnamefont {I.}~\bibnamefont {Izaguirre}},\ and\
		\bibinfo {author} {\bibfnamefont {G.}~\bibnamefont {Raffelt}},\ }\bibfield
	{title} {\bibinfo {title} {{Collective neutrino flavor conversion: Recent
				developments}},\ }\href {https://doi.org/10.1016/j.nuclphysb.2016.02.012}
	{\bibfield  {journal} {\bibinfo  {journal} {Nucl. Phys. B}\ }\textbf
		{\bibinfo {volume} {908}},\ \bibinfo {pages} {366} (\bibinfo {year}
		{2016})}\BibitemShut {NoStop}%
	\bibitem [{\citenamefont {Balantekin}(2018)}]{Balantekin:2018mpq}%
	\BibitemOpen
	\bibfield  {author} {\bibinfo {author} {\bibfnamefont {A.~B.}\ \bibnamefont
			{Balantekin}},\ }\bibfield  {title} {\bibinfo {title} {{Symmetries and
				Algebraic Methods in Neutrino Physics}},\ }\href
	{https://doi.org/10.1088/1361-6471/aae3d8} {\bibfield  {journal} {\bibinfo
			{journal} {J. Phys. G}\ }\textbf {\bibinfo {volume} {45}},\ \bibinfo {pages}
		{113001} (\bibinfo {year} {2018})}\BibitemShut {NoStop}%
	\bibitem [{\citenamefont {Tamborra}\ and\ \citenamefont
		{Shalgar}(2021)}]{Tamborra:2020cul}%
	\BibitemOpen
	\bibfield  {author} {\bibinfo {author} {\bibfnamefont {I.}~\bibnamefont
			{Tamborra}}\ and\ \bibinfo {author} {\bibfnamefont {S.}~\bibnamefont
			{Shalgar}},\ }\bibfield  {title} {\bibinfo {title} {{New Developments in
				Flavor Evolution of a Dense Neutrino Gas}},\ }\href
	{https://doi.org/10.1146/annurev-nucl-102920-050505} {\bibfield  {journal}
		{\bibinfo  {journal} {Ann. Rev. Nucl. Part. Sci.}\ }\textbf {\bibinfo
			{volume} {71}},\ \bibinfo {pages} {165} (\bibinfo {year} {2021})}\BibitemShut
	{NoStop}%
	\bibitem [{\citenamefont {Richers}\ and\ \citenamefont
		{Sen}(2022)}]{Richers:2022zug}%
	\BibitemOpen
	\bibfield  {author} {\bibinfo {author} {\bibfnamefont {S.}~\bibnamefont
			{Richers}}\ and\ \bibinfo {author} {\bibfnamefont {M.}~\bibnamefont {Sen}},\
	}\bibinfo {title} {{Fast Flavor Transformations}},\ in\ \href
	{https://doi.org/10.1007/978-981-15-8818-1_125-1} {\emph {\bibinfo
			{booktitle} {{Handbook of Nuclear Physics}}}},\ \bibinfo {editor} {edited by\
		\bibinfo {editor} {\bibfnamefont {I.}~\bibnamefont {Tanihata}}, \bibinfo
		{editor} {\bibfnamefont {H.}~\bibnamefont {Toki}},\ and\ \bibinfo {editor}
		{\bibfnamefont {T.}~\bibnamefont {Kajino}}}\ (\bibinfo {year} {2022})\ pp.\
	\bibinfo {pages} {1--17}\BibitemShut {NoStop}%
	\bibitem [{\citenamefont {Balantekin}\ \emph {et~al.}(2023)\citenamefont
		{Balantekin}, \citenamefont {Cervia}, \citenamefont {Patwardhan},
		\citenamefont {Rrapaj},\ and\ \citenamefont {Siwach}}]{Balantekin:2023qvm}%
	\BibitemOpen
	\bibfield  {author} {\bibinfo {author} {\bibfnamefont {A.~B.}\ \bibnamefont
			{Balantekin}}, \bibinfo {author} {\bibfnamefont {M.~J.}\ \bibnamefont
			{Cervia}}, \bibinfo {author} {\bibfnamefont {A.~V.}\ \bibnamefont
			{Patwardhan}}, \bibinfo {author} {\bibfnamefont {E.}~\bibnamefont {Rrapaj}},\
		and\ \bibinfo {author} {\bibfnamefont {P.}~\bibnamefont {Siwach}},\
	}\bibfield  {title} {\bibinfo {title} {{Quantum information and quantum
				simulation of neutrino physics}},\ }\href
	{https://doi.org/10.1140/epja/s10050-023-01092-7} {\bibfield  {journal}
		{\bibinfo  {journal} {Eur. Phys. J. A}\ }\textbf {\bibinfo {volume} {59}},\
		\bibinfo {pages} {186} (\bibinfo {year} {2023})}\BibitemShut {NoStop}%
	\bibitem [{\citenamefont {Volpe}(2024)}]{Volpe:2023met}%
	\BibitemOpen
	\bibfield  {author} {\bibinfo {author} {\bibfnamefont {M.~C.}\ \bibnamefont
			{Volpe}},\ }\bibfield  {title} {\bibinfo {title} {{Neutrinos from dense
				environments: Flavor mechanisms, theoretical approaches, observations, and
				new directions}},\ }\href {https://doi.org/10.1103/RevModPhys.96.025004}
	{\bibfield  {journal} {\bibinfo  {journal} {Rev. Mod. Phys.}\ }\textbf
		{\bibinfo {volume} {96}},\ \bibinfo {pages} {025004} (\bibinfo {year}
		{2024})}\BibitemShut {NoStop}%
	\bibitem [{\citenamefont {Pantaleone}(1992)}]{PANTALEONE1992128}%
	\BibitemOpen
	\bibfield  {author} {\bibinfo {author} {\bibfnamefont {J.}~\bibnamefont
			{Pantaleone}},\ }\bibfield  {title} {\bibinfo {title} {Neutrino oscillations
			at high densities},\ }\href
	{https://doi.org/https://doi.org/10.1016/0370-2693(92)91887-F} {\bibfield
		{journal} {\bibinfo  {journal} {Physics Letters B}\ }\textbf {\bibinfo
			{volume} {287}},\ \bibinfo {pages} {128} (\bibinfo {year}
		{1992})}\BibitemShut {NoStop}%
	\bibitem [{\citenamefont {Balantekin}\ and\ \citenamefont
		{Pehlivan}(2007)}]{Balantekin:2006tg}%
	\BibitemOpen
	\bibfield  {author} {\bibinfo {author} {\bibfnamefont {A.~B.}\ \bibnamefont
			{Balantekin}}\ and\ \bibinfo {author} {\bibfnamefont {Y.}~\bibnamefont
			{Pehlivan}},\ }\bibfield  {title} {\bibinfo {title} {{Neutrino-Neutrino
				Interactions and Flavor Mixing in Dense Matter}},\ }\href
	{https://doi.org/10.1088/0954-3899/34/1/004} {\bibfield  {journal} {\bibinfo
			{journal} {J. Phys. G}\ }\textbf {\bibinfo {volume} {34}},\ \bibinfo {pages}
		{47} (\bibinfo {year} {2007})}\BibitemShut {NoStop}%
	\bibitem [{\citenamefont {Bell}\ \emph {et~al.}(2003)\citenamefont {Bell},
		\citenamefont {Rawlinson},\ and\ \citenamefont {Sawyer}}]{Bell:2003mg}%
	\BibitemOpen
	\bibfield  {author} {\bibinfo {author} {\bibfnamefont {N.~F.}\ \bibnamefont
			{Bell}}, \bibinfo {author} {\bibfnamefont {A.~A.}\ \bibnamefont
			{Rawlinson}},\ and\ \bibinfo {author} {\bibfnamefont {R.~F.}\ \bibnamefont
			{Sawyer}},\ }\bibfield  {title} {\bibinfo {title} {{Speedup through
				entanglement: Many body effects in neutrino processes}},\ }\href
	{https://doi.org/10.1016/j.physletb.2003.08.035} {\bibfield  {journal}
		{\bibinfo  {journal} {Phys. Lett. B}\ }\textbf {\bibinfo {volume} {573}},\
		\bibinfo {pages} {86} (\bibinfo {year} {2003})}\BibitemShut {NoStop}%
	\bibitem [{\citenamefont {Friedland}\ and\ \citenamefont
		{Lunardini}(2003{\natexlab{a}})}]{Friedland:2003dv}%
	\BibitemOpen
	\bibfield  {author} {\bibinfo {author} {\bibfnamefont {A.}~\bibnamefont
			{Friedland}}\ and\ \bibinfo {author} {\bibfnamefont {C.}~\bibnamefont
			{Lunardini}},\ }\bibfield  {title} {\bibinfo {title} {{Neutrino flavor
				conversion in a neutrino background: Single particle versus multiparticle
				description}},\ }\href {https://doi.org/10.1103/PhysRevD.68.013007}
	{\bibfield  {journal} {\bibinfo  {journal} {Phys. Rev. D}\ }\textbf {\bibinfo
			{volume} {68}},\ \bibinfo {pages} {013007} (\bibinfo {year}
		{2003}{\natexlab{a}})}\BibitemShut {NoStop}%
	\bibitem [{\citenamefont {Friedland}\ and\ \citenamefont
		{Lunardini}(2003{\natexlab{b}})}]{Friedland:2003eh}%
	\BibitemOpen
	\bibfield  {author} {\bibinfo {author} {\bibfnamefont {A.}~\bibnamefont
			{Friedland}}\ and\ \bibinfo {author} {\bibfnamefont {C.}~\bibnamefont
			{Lunardini}},\ }\bibfield  {title} {\bibinfo {title} {{Do many particle
				neutrino interactions cause a novel coherent effect?}},\ }\href
	{https://doi.org/10.1088/1126-6708/2003/10/043} {\bibfield  {journal}
		{\bibinfo  {journal} {JHEP}\ }\textbf {\bibinfo {volume} {10}},\ \bibinfo
		{pages} {043}}\BibitemShut {NoStop}%
	\bibitem [{\citenamefont {Friedland}\ \emph {et~al.}(2006)\citenamefont
		{Friedland}, \citenamefont {McKellar},\ and\ \citenamefont
		{Okuniewicz}}]{Friedland:2006ke}%
	\BibitemOpen
	\bibfield  {author} {\bibinfo {author} {\bibfnamefont {A.}~\bibnamefont
			{Friedland}}, \bibinfo {author} {\bibfnamefont {B.~H.~J.}\ \bibnamefont
			{McKellar}},\ and\ \bibinfo {author} {\bibfnamefont {I.}~\bibnamefont
			{Okuniewicz}},\ }\bibfield  {title} {\bibinfo {title} {{Construction and
				analysis of a simplified many-body neutrino model}},\ }\href
	{https://doi.org/10.1103/PhysRevD.73.093002} {\bibfield  {journal} {\bibinfo
			{journal} {Phys. Rev. D}\ }\textbf {\bibinfo {volume} {73}},\ \bibinfo
		{pages} {093002} (\bibinfo {year} {2006})}\BibitemShut {NoStop}%
	\bibitem [{\citenamefont {McKellar}\ \emph {et~al.}(2009)\citenamefont
		{McKellar}, \citenamefont {Okuniewicz},\ and\ \citenamefont
		{Quach}}]{McKellar:2009py}%
	\BibitemOpen
	\bibfield  {author} {\bibinfo {author} {\bibfnamefont {B.~H.~J.}\
			\bibnamefont {McKellar}}, \bibinfo {author} {\bibfnamefont {I.}~\bibnamefont
			{Okuniewicz}},\ and\ \bibinfo {author} {\bibfnamefont {J.}~\bibnamefont
			{Quach}},\ }\bibfield  {title} {\bibinfo {title} {{Non-Boltzmann behaviour in
				models of interacting neutrinos}},\ }\href
	{https://doi.org/10.1103/PhysRevD.80.013011} {\bibfield  {journal} {\bibinfo
			{journal} {Phys. Rev. D}\ }\textbf {\bibinfo {volume} {80}},\ \bibinfo
		{pages} {013011} (\bibinfo {year} {2009})}\BibitemShut {NoStop}%
	\bibitem [{\citenamefont {Pehlivan}\ \emph {et~al.}(2011)\citenamefont
		{Pehlivan}, \citenamefont {Balantekin}, \citenamefont {Kajino},\ and\
		\citenamefont {Yoshida}}]{Pehlivan:2011hp}%
	\BibitemOpen
	\bibfield  {author} {\bibinfo {author} {\bibfnamefont {Y.}~\bibnamefont
			{Pehlivan}}, \bibinfo {author} {\bibfnamefont {A.~B.}\ \bibnamefont
			{Balantekin}}, \bibinfo {author} {\bibfnamefont {T.}~\bibnamefont {Kajino}},\
		and\ \bibinfo {author} {\bibfnamefont {T.}~\bibnamefont {Yoshida}},\
	}\bibfield  {title} {\bibinfo {title} {{Invariants of Collective Neutrino
				Oscillations}},\ }\href {https://doi.org/10.1103/PhysRevD.84.065008}
	{\bibfield  {journal} {\bibinfo  {journal} {Phys. Rev. D}\ }\textbf {\bibinfo
			{volume} {84}},\ \bibinfo {pages} {065008} (\bibinfo {year}
		{2011})}\BibitemShut {NoStop}%
	\bibitem [{\citenamefont {Birol}\ \emph {et~al.}(2018)\citenamefont {Birol},
		\citenamefont {Pehlivan}, \citenamefont {Balantekin},\ and\ \citenamefont
		{Kajino}}]{Birol:2018qhx}%
	\BibitemOpen
	\bibfield  {author} {\bibinfo {author} {\bibfnamefont {S.}~\bibnamefont
			{Birol}}, \bibinfo {author} {\bibfnamefont {Y.}~\bibnamefont {Pehlivan}},
		\bibinfo {author} {\bibfnamefont {A.~B.}\ \bibnamefont {Balantekin}},\ and\
		\bibinfo {author} {\bibfnamefont {T.}~\bibnamefont {Kajino}},\ }\bibfield
	{title} {\bibinfo {title} {{Neutrino Spectral Split in the Exact Many Body
				Formalism}},\ }\href {https://doi.org/10.1103/PhysRevD.98.083002} {\bibfield
		{journal} {\bibinfo  {journal} {Phys. Rev. D}\ }\textbf {\bibinfo {volume}
			{98}},\ \bibinfo {pages} {083002} (\bibinfo {year} {2018})}\BibitemShut
	{NoStop}%
	\bibitem [{\citenamefont {Patwardhan}\ \emph {et~al.}(2019)\citenamefont
		{Patwardhan}, \citenamefont {Cervia},\ and\ \citenamefont
		{Baha~Balantekin}}]{Patwardhan:2019zta}%
	\BibitemOpen
	\bibfield  {author} {\bibinfo {author} {\bibfnamefont {A.~V.}\ \bibnamefont
			{Patwardhan}}, \bibinfo {author} {\bibfnamefont {M.~J.}\ \bibnamefont
			{Cervia}},\ and\ \bibinfo {author} {\bibfnamefont {A.}~\bibnamefont
			{Baha~Balantekin}},\ }\bibfield  {title} {\bibinfo {title} {{Eigenvalues and
				eigenstates of the many-body collective neutrino oscillation problem}},\
	}\href {https://doi.org/10.1103/PhysRevD.99.123013} {\bibfield  {journal}
		{\bibinfo  {journal} {Phys. Rev. D}\ }\textbf {\bibinfo {volume} {99}},\
		\bibinfo {pages} {123013} (\bibinfo {year} {2019})}\BibitemShut {NoStop}%
	\bibitem [{\citenamefont {Cervia}\ \emph {et~al.}(2019)\citenamefont {Cervia},
		\citenamefont {Patwardhan}, \citenamefont {Balantekin}, \citenamefont
		{Coppersmith},\ and\ \citenamefont {Johnson}}]{Cervia:2019res}%
	\BibitemOpen
	\bibfield  {author} {\bibinfo {author} {\bibfnamefont {M.~J.}\ \bibnamefont
			{Cervia}}, \bibinfo {author} {\bibfnamefont {A.~V.}\ \bibnamefont
			{Patwardhan}}, \bibinfo {author} {\bibfnamefont {A.~B.}\ \bibnamefont
			{Balantekin}}, \bibinfo {author} {\bibfnamefont {t.~S.~N.}\ \bibnamefont
			{Coppersmith}},\ and\ \bibinfo {author} {\bibfnamefont {C.~W.}\ \bibnamefont
			{Johnson}},\ }\bibfield  {title} {\bibinfo {title} {{Entanglement and
				collective flavor oscillations in a dense neutrino gas}},\ }\href
	{https://doi.org/10.1103/PhysRevD.100.083001} {\bibfield  {journal} {\bibinfo
			{journal} {Phys. Rev. D}\ }\textbf {\bibinfo {volume} {100}},\ \bibinfo
		{pages} {083001} (\bibinfo {year} {2019})}\BibitemShut {NoStop}%
	\bibitem [{\citenamefont {Rrapaj}(2020)}]{Rrapaj:2019pxz}%
	\BibitemOpen
	\bibfield  {author} {\bibinfo {author} {\bibfnamefont {E.}~\bibnamefont
			{Rrapaj}},\ }\bibfield  {title} {\bibinfo {title} {{Exact solution of
				multiangle quantum many-body collective neutrino-flavor oscillations}},\
	}\href {https://doi.org/10.1103/PhysRevC.101.065805} {\bibfield  {journal}
		{\bibinfo  {journal} {Phys. Rev. C}\ }\textbf {\bibinfo {volume} {101}},\
		\bibinfo {pages} {065805} (\bibinfo {year} {2020})}\BibitemShut {NoStop}%
	\bibitem [{\citenamefont {Roggero}(2021{\natexlab{a}})}]{Roggero:2021asb}%
	\BibitemOpen
	\bibfield  {author} {\bibinfo {author} {\bibfnamefont {A.}~\bibnamefont
			{Roggero}},\ }\bibfield  {title} {\bibinfo {title} {{Entanglement and
				many-body effects in collective neutrino oscillations}},\ }\href
	{https://doi.org/10.1103/PhysRevD.104.103016} {\bibfield  {journal} {\bibinfo
			{journal} {Phys. Rev. D}\ }\textbf {\bibinfo {volume} {104}},\ \bibinfo
		{pages} {103016} (\bibinfo {year} {2021}{\natexlab{a}})}\BibitemShut
	{NoStop}%
	\bibitem [{\citenamefont {Roggero}(2021{\natexlab{b}})}]{Roggero:2021fyo}%
	\BibitemOpen
	\bibfield  {author} {\bibinfo {author} {\bibfnamefont {A.}~\bibnamefont
			{Roggero}},\ }\bibfield  {title} {\bibinfo {title} {{Dynamical phase
				transitions in models of collective neutrino oscillations}},\ }\href
	{https://doi.org/10.1103/PhysRevD.104.123023} {\bibfield  {journal} {\bibinfo
			{journal} {Phys. Rev. D}\ }\textbf {\bibinfo {volume} {104}},\ \bibinfo
		{pages} {123023} (\bibinfo {year} {2021}{\natexlab{b}})}\BibitemShut
	{NoStop}%
	\bibitem [{\citenamefont {Xiong}(2022)}]{Xiong:2021evk}%
	\BibitemOpen
	\bibfield  {author} {\bibinfo {author} {\bibfnamefont {Z.}~\bibnamefont
			{Xiong}},\ }\bibfield  {title} {\bibinfo {title} {{Many-body effects of
				collective neutrino oscillations}},\ }\href
	{https://doi.org/10.1103/PhysRevD.105.103002} {\bibfield  {journal} {\bibinfo
			{journal} {Phys. Rev. D}\ }\textbf {\bibinfo {volume} {105}},\ \bibinfo
		{pages} {103002} (\bibinfo {year} {2022})}\BibitemShut {NoStop}%
	\bibitem [{\citenamefont {Martin}\ \emph {et~al.}(2022)\citenamefont {Martin},
		\citenamefont {Roggero}, \citenamefont {Duan}, \citenamefont {Carlson},\ and\
		\citenamefont {Cirigliano}}]{Martin:2021bri}%
	\BibitemOpen
	\bibfield  {author} {\bibinfo {author} {\bibfnamefont {J.~D.}\ \bibnamefont
			{Martin}}, \bibinfo {author} {\bibfnamefont {A.}~\bibnamefont {Roggero}},
		\bibinfo {author} {\bibfnamefont {H.}~\bibnamefont {Duan}}, \bibinfo {author}
		{\bibfnamefont {J.}~\bibnamefont {Carlson}},\ and\ \bibinfo {author}
		{\bibfnamefont {V.}~\bibnamefont {Cirigliano}},\ }\bibfield  {title}
	{\bibinfo {title} {{Classical and quantum evolution in a simple coherent
				neutrino problem}},\ }\href {https://doi.org/10.1103/PhysRevD.105.083020}
	{\bibfield  {journal} {\bibinfo  {journal} {Phys. Rev. D}\ }\textbf {\bibinfo
			{volume} {105}},\ \bibinfo {pages} {083020} (\bibinfo {year}
		{2022})}\BibitemShut {NoStop}%
	\bibitem [{\citenamefont {Patwardhan}\ \emph {et~al.}(2021)\citenamefont
		{Patwardhan}, \citenamefont {Cervia},\ and\ \citenamefont
		{Balantekin}}]{Patwardhan:2021rej}%
	\BibitemOpen
	\bibfield  {author} {\bibinfo {author} {\bibfnamefont {A.~V.}\ \bibnamefont
			{Patwardhan}}, \bibinfo {author} {\bibfnamefont {M.~J.}\ \bibnamefont
			{Cervia}},\ and\ \bibinfo {author} {\bibfnamefont {A.~B.}\ \bibnamefont
			{Balantekin}},\ }\bibfield  {title} {\bibinfo {title} {{Spectral splits and
				entanglement entropy in collective neutrino oscillations}},\ }\href
	{https://doi.org/10.1103/PhysRevD.104.123035} {\bibfield  {journal} {\bibinfo
			{journal} {Phys. Rev. D}\ }\textbf {\bibinfo {volume} {104}},\ \bibinfo
		{pages} {123035} (\bibinfo {year} {2021})}\BibitemShut {NoStop}%
	\bibitem [{\citenamefont {Roggero}\ \emph {et~al.}(2022)\citenamefont
		{Roggero}, \citenamefont {Rrapaj},\ and\ \citenamefont
		{Xiong}}]{Roggero:2022hpy}%
	\BibitemOpen
	\bibfield  {author} {\bibinfo {author} {\bibfnamefont {A.}~\bibnamefont
			{Roggero}}, \bibinfo {author} {\bibfnamefont {E.}~\bibnamefont {Rrapaj}},\
		and\ \bibinfo {author} {\bibfnamefont {Z.}~\bibnamefont {Xiong}},\ }\bibfield
	{title} {\bibinfo {title} {{Entanglement and correlations in fast collective
				neutrino flavor oscillations}},\ }\href
	{https://doi.org/10.1103/PhysRevD.106.043022} {\bibfield  {journal} {\bibinfo
			{journal} {Phys. Rev. D}\ }\textbf {\bibinfo {volume} {106}},\ \bibinfo
		{pages} {043022} (\bibinfo {year} {2022})}\BibitemShut {NoStop}%
	\bibitem [{\citenamefont {Cervia}\ \emph {et~al.}(2022)\citenamefont {Cervia},
		\citenamefont {Siwach}, \citenamefont {Patwardhan}, \citenamefont
		{Balantekin}, \citenamefont {Coppersmith},\ and\ \citenamefont
		{Johnson}}]{Cervia:2022pro}%
	\BibitemOpen
	\bibfield  {author} {\bibinfo {author} {\bibfnamefont {M.~J.}\ \bibnamefont
			{Cervia}}, \bibinfo {author} {\bibfnamefont {P.}~\bibnamefont {Siwach}},
		\bibinfo {author} {\bibfnamefont {A.~V.}\ \bibnamefont {Patwardhan}},
		\bibinfo {author} {\bibfnamefont {A.~B.}\ \bibnamefont {Balantekin}},
		\bibinfo {author} {\bibfnamefont {S.~N.}\ \bibnamefont {Coppersmith}},\ and\
		\bibinfo {author} {\bibfnamefont {C.~W.}\ \bibnamefont {Johnson}},\
	}\bibfield  {title} {\bibinfo {title} {{Collective neutrino oscillations with
				tensor networks using a time-dependent variational principle}},\ }\href
	{https://doi.org/10.1103/PhysRevD.105.123025} {\bibfield  {journal} {\bibinfo
			{journal} {Phys. Rev. D}\ }\textbf {\bibinfo {volume} {105}},\ \bibinfo
		{pages} {123025} (\bibinfo {year} {2022})}\BibitemShut {NoStop}%
	\bibitem [{\citenamefont {Illa}\ and\ \citenamefont
		{Savage}(2023{\natexlab{a}})}]{Illa:2022zgu}%
	\BibitemOpen
	\bibfield  {author} {\bibinfo {author} {\bibfnamefont {M.}~\bibnamefont
			{Illa}}\ and\ \bibinfo {author} {\bibfnamefont {M.~J.}\ \bibnamefont
			{Savage}},\ }\bibfield  {title} {\bibinfo {title} {{Multi-Neutrino
				Entanglement and Correlations in Dense Neutrino Systems}},\ }\href
	{https://doi.org/10.1103/PhysRevLett.130.221003} {\bibfield  {journal}
		{\bibinfo  {journal} {Phys. Rev. Lett.}\ }\textbf {\bibinfo {volume} {130}},\
		\bibinfo {pages} {221003} (\bibinfo {year} {2023}{\natexlab{a}})}\BibitemShut
	{NoStop}%
	\bibitem [{\citenamefont {Lacroix}\ \emph {et~al.}(2022)\citenamefont
		{Lacroix}, \citenamefont {Balantekin}, \citenamefont {Cervia}, \citenamefont
		{Patwardhan},\ and\ \citenamefont {Siwach}}]{Lacroix:2022krq}%
	\BibitemOpen
	\bibfield  {author} {\bibinfo {author} {\bibfnamefont {D.}~\bibnamefont
			{Lacroix}}, \bibinfo {author} {\bibfnamefont {A.~B.}\ \bibnamefont
			{Balantekin}}, \bibinfo {author} {\bibfnamefont {M.~J.}\ \bibnamefont
			{Cervia}}, \bibinfo {author} {\bibfnamefont {A.~V.}\ \bibnamefont
			{Patwardhan}},\ and\ \bibinfo {author} {\bibfnamefont {P.}~\bibnamefont
			{Siwach}},\ }\bibfield  {title} {\bibinfo {title} {{Role of non-Gaussian
				quantum fluctuations in neutrino entanglement}},\ }\href
	{https://doi.org/10.1103/PhysRevD.106.123006} {\bibfield  {journal} {\bibinfo
			{journal} {Phys. Rev. D}\ }\textbf {\bibinfo {volume} {106}},\ \bibinfo
		{pages} {123006} (\bibinfo {year} {2022})}\BibitemShut {NoStop}%
	\bibitem [{\citenamefont {Bhaskar}\ \emph {et~al.}(2024)\citenamefont
		{Bhaskar}, \citenamefont {Roggero},\ and\ \citenamefont
		{Savage}}]{PhysRevC.110.045801}%
	\BibitemOpen
	\bibfield  {author} {\bibinfo {author} {\bibfnamefont {R.}~\bibnamefont
			{Bhaskar}}, \bibinfo {author} {\bibfnamefont {A.}~\bibnamefont {Roggero}},\
		and\ \bibinfo {author} {\bibfnamefont {M.~J.}\ \bibnamefont {Savage}},\
	}\bibfield  {title} {\bibinfo {title} {Timescales in many-body
			fast-neutrino-flavor conversion},\ }\href
	{https://doi.org/10.1103/PhysRevC.110.045801} {\bibfield  {journal} {\bibinfo
			{journal} {Phys. Rev. C}\ }\textbf {\bibinfo {volume} {110}},\ \bibinfo
		{pages} {045801} (\bibinfo {year} {2024})}\BibitemShut {NoStop}%
	\bibitem [{\citenamefont {Martin}\ \emph
		{et~al.}(2023{\natexlab{a}})\citenamefont {Martin}, \citenamefont {Neill},
		\citenamefont {Roggero}, \citenamefont {Duan},\ and\ \citenamefont
		{Carlson}}]{Martin:2023gbo}%
	\BibitemOpen
	\bibfield  {author} {\bibinfo {author} {\bibfnamefont {J.~D.}\ \bibnamefont
			{Martin}}, \bibinfo {author} {\bibfnamefont {D.}~\bibnamefont {Neill}},
		\bibinfo {author} {\bibfnamefont {A.}~\bibnamefont {Roggero}}, \bibinfo
		{author} {\bibfnamefont {H.}~\bibnamefont {Duan}},\ and\ \bibinfo {author}
		{\bibfnamefont {J.}~\bibnamefont {Carlson}},\ }\bibfield  {title} {\bibinfo
		{title} {{Equilibration of quantum many-body fast neutrino flavor
				oscillations}},\ }\href {https://doi.org/10.1103/PhysRevD.108.123010}
	{\bibfield  {journal} {\bibinfo  {journal} {Phys. Rev. D}\ }\textbf {\bibinfo
			{volume} {108}},\ \bibinfo {pages} {123010} (\bibinfo {year}
		{2023}{\natexlab{a}})}\BibitemShut {NoStop}%
	\bibitem [{\citenamefont {Martin}\ \emph
		{et~al.}(2023{\natexlab{b}})\citenamefont {Martin}, \citenamefont {Roggero},
		\citenamefont {Duan},\ and\ \citenamefont {Carlson}}]{Martin:2023ljq}%
	\BibitemOpen
	\bibfield  {author} {\bibinfo {author} {\bibfnamefont {J.~D.}\ \bibnamefont
			{Martin}}, \bibinfo {author} {\bibfnamefont {A.}~\bibnamefont {Roggero}},
		\bibinfo {author} {\bibfnamefont {H.}~\bibnamefont {Duan}},\ and\ \bibinfo
		{author} {\bibfnamefont {J.}~\bibnamefont {Carlson}},\ }\bibfield  {title}
	{\bibinfo {title} {{Many-body neutrino flavor entanglement in a simple
				dynamic model}},\ }\Eprint {https://arxiv.org/abs/2301.07049}
	{arXiv:2301.07049 [hep-ph]}  (\bibinfo {year}
	{2023}{\natexlab{b}})\BibitemShut {NoStop}%
	\bibitem [{\citenamefont {Lacroix}\ \emph {et~al.}(2024)\citenamefont
		{Lacroix}, \citenamefont {Bauge}, \citenamefont {Yilmaz}, \citenamefont
		{Mangin-Brinet}, \citenamefont {Roggero},\ and\ \citenamefont
		{Balantekin}}]{PhysRevD.110.103027}%
	\BibitemOpen
	\bibfield  {author} {\bibinfo {author} {\bibfnamefont {D.}~\bibnamefont
			{Lacroix}}, \bibinfo {author} {\bibfnamefont {A.}~\bibnamefont {Bauge}},
		\bibinfo {author} {\bibfnamefont {B.}~\bibnamefont {Yilmaz}}, \bibinfo
		{author} {\bibfnamefont {M.}~\bibnamefont {Mangin-Brinet}}, \bibinfo {author}
		{\bibfnamefont {A.}~\bibnamefont {Roggero}},\ and\ \bibinfo {author}
		{\bibfnamefont {A.~B.}\ \bibnamefont {Balantekin}},\ }\bibfield  {title}
	{\bibinfo {title} {Phase-space methods for neutrino oscillations: Extension
			to multibeams},\ }\href {https://doi.org/10.1103/PhysRevD.110.103027}
	{\bibfield  {journal} {\bibinfo  {journal} {Phys. Rev. D}\ }\textbf {\bibinfo
			{volume} {110}},\ \bibinfo {pages} {103027} (\bibinfo {year}
		{2024})}\BibitemShut {NoStop}%
	\bibitem [{\citenamefont {Shalgar}\ and\ \citenamefont
		{Tamborra}(2023)}]{Shalgar:2023ooi}%
	\BibitemOpen
	\bibfield  {author} {\bibinfo {author} {\bibfnamefont {S.}~\bibnamefont
			{Shalgar}}\ and\ \bibinfo {author} {\bibfnamefont {I.}~\bibnamefont
			{Tamborra}},\ }\bibfield  {title} {\bibinfo {title} {{Do we have enough
				evidence to invalidate the mean-field approximation adopted to model
				collective neutrino oscillations?}},\ }\href
	{https://doi.org/10.1103/PhysRevD.107.123004} {\bibfield  {journal} {\bibinfo
			{journal} {Phys. Rev. D}\ }\textbf {\bibinfo {volume} {107}},\ \bibinfo
		{pages} {123004} (\bibinfo {year} {2023})}\BibitemShut {NoStop}%
	\bibitem [{\citenamefont {Kost}\ \emph {et~al.}(2024)\citenamefont {Kost},
		\citenamefont {Johns},\ and\ \citenamefont {Duan}}]{Kost:2024esc}%
	\BibitemOpen
	\bibfield  {author} {\bibinfo {author} {\bibfnamefont {A.}~\bibnamefont
			{Kost}}, \bibinfo {author} {\bibfnamefont {L.}~\bibnamefont {Johns}},\ and\
		\bibinfo {author} {\bibfnamefont {H.}~\bibnamefont {Duan}},\ }\bibfield
	{title} {\bibinfo {title} {{Once-in-a-lifetime encounter models for neutrino
				media: From coherent oscillations to flavor equilibration}},\ }\href
	{https://doi.org/10.1103/PhysRevD.109.103037} {\bibfield  {journal} {\bibinfo
			{journal} {Phys. Rev. D}\ }\textbf {\bibinfo {volume} {109}},\ \bibinfo
		{pages} {103037} (\bibinfo {year} {2024})}\BibitemShut {NoStop}%
	\bibitem [{\citenamefont {Johns}(2023)}]{Johns:2023ewj}%
	\BibitemOpen
	\bibfield  {author} {\bibinfo {author} {\bibfnamefont {L.}~\bibnamefont
			{Johns}},\ }\bibfield  {title} {\bibinfo {title} {{Neutrino many-body
				correlations}},\ }\Eprint {https://arxiv.org/abs/2305.04916}
	{arXiv:2305.04916 [hep-ph]}  (\bibinfo {year} {2023})\BibitemShut {NoStop}%
	\bibitem [{\citenamefont {Cirigliano}\ \emph {et~al.}(2024)\citenamefont
		{Cirigliano}, \citenamefont {Sen},\ and\ \citenamefont
		{Yamauchi}}]{PhysRevD.110.123028}%
	\BibitemOpen
	\bibfield  {author} {\bibinfo {author} {\bibfnamefont {V.}~\bibnamefont
			{Cirigliano}}, \bibinfo {author} {\bibfnamefont {S.}~\bibnamefont {Sen}},\
		and\ \bibinfo {author} {\bibfnamefont {Y.}~\bibnamefont {Yamauchi}},\
	}\bibfield  {title} {\bibinfo {title} {Neutrino many-body flavor evolution:
			The full hamiltonian},\ }\href {https://doi.org/10.1103/PhysRevD.110.123028}
	{\bibfield  {journal} {\bibinfo  {journal} {Phys. Rev. D}\ }\textbf {\bibinfo
			{volume} {110}},\ \bibinfo {pages} {123028} (\bibinfo {year}
		{2024})}\BibitemShut {NoStop}%
	\bibitem [{\citenamefont {Goimil-Garc\'\i{}a}\ \emph
		{et~al.}(2024)\citenamefont {Goimil-Garc\'\i{}a}, \citenamefont {Shalgar},\
		and\ \citenamefont {Tamborra}}]{Goimil-Garcia:2024wgw}%
	\BibitemOpen
	\bibfield  {author} {\bibinfo {author} {\bibfnamefont {M.}~\bibnamefont
			{Goimil-Garc\'\i{}a}}, \bibinfo {author} {\bibfnamefont {S.}~\bibnamefont
			{Shalgar}},\ and\ \bibinfo {author} {\bibfnamefont {I.}~\bibnamefont
			{Tamborra}},\ }\bibfield  {title} {\bibinfo {title} {{Pauli blocking: probing
				beyond-mean-field effects in neutrino flavor evolution}},\ }\Eprint
	{https://arxiv.org/abs/2412.12268} {arXiv:2412.12268 [astro-ph.HE]}
	(\bibinfo {year} {2024})\BibitemShut {NoStop}%
	\bibitem [{\citenamefont {Pehlivan}\ \emph {et~al.}(2014)\citenamefont
		{Pehlivan}, \citenamefont {Balantekin},\ and\ \citenamefont
		{Kajino}}]{Pehlivan:2014zua}%
	\BibitemOpen
	\bibfield  {author} {\bibinfo {author} {\bibfnamefont {Y.}~\bibnamefont
			{Pehlivan}}, \bibinfo {author} {\bibfnamefont {A.~B.}\ \bibnamefont
			{Balantekin}},\ and\ \bibinfo {author} {\bibfnamefont {T.}~\bibnamefont
			{Kajino}},\ }\bibfield  {title} {\bibinfo {title} {{Neutrino Magnetic Moment,
				CP Violation and Flavor Oscillations in Matter}},\ }\href
	{https://doi.org/10.1103/PhysRevD.90.065011} {\bibfield  {journal} {\bibinfo
			{journal} {Phys. Rev. D}\ }\textbf {\bibinfo {volume} {90}},\ \bibinfo
		{pages} {065011} (\bibinfo {year} {2014})}\BibitemShut {NoStop}%
	\bibitem [{\citenamefont {Siwach}\ \emph
		{et~al.}(2023{\natexlab{a}})\citenamefont {Siwach}, \citenamefont {Suliga},\
		and\ \citenamefont {Balantekin}}]{Siwach:2022xhx}%
	\BibitemOpen
	\bibfield  {author} {\bibinfo {author} {\bibfnamefont {P.}~\bibnamefont
			{Siwach}}, \bibinfo {author} {\bibfnamefont {A.~M.}\ \bibnamefont {Suliga}},\
		and\ \bibinfo {author} {\bibfnamefont {A.~B.}\ \bibnamefont {Balantekin}},\
	}\bibfield  {title} {\bibinfo {title} {{Entanglement in three-flavor
				collective neutrino oscillations}},\ }\href
	{https://doi.org/10.1103/PhysRevD.107.023019} {\bibfield  {journal} {\bibinfo
			{journal} {Phys. Rev. D}\ }\textbf {\bibinfo {volume} {107}},\ \bibinfo
		{pages} {023019} (\bibinfo {year} {2023}{\natexlab{a}})}\BibitemShut
	{NoStop}%
	\bibitem [{\citenamefont {Chernyshev}\ \emph {et~al.}(2024)\citenamefont
		{Chernyshev}, \citenamefont {Robin},\ and\ \citenamefont
		{Savage}}]{Chernyshev:2024}%
	\BibitemOpen
	\bibfield  {author} {\bibinfo {author} {\bibfnamefont {I.}~\bibnamefont
			{Chernyshev}}, \bibinfo {author} {\bibfnamefont {C.~E.~P.}\ \bibnamefont
			{Robin}},\ and\ \bibinfo {author} {\bibfnamefont {M.~J.}\ \bibnamefont
			{Savage}},\ }\href {https://arxiv.org/abs/2411.04203} {\bibinfo {title}
		{Quantum magic and computational complexity in the neutrino sector}}
	(\bibinfo {year} {2024}),\ \Eprint {https://arxiv.org/abs/2411.04203}
	{arXiv:2411.04203 [quant-ph]} \BibitemShut {NoStop}%
	\bibitem [{\citenamefont {Klco}\ \emph {et~al.}(2022)\citenamefont {Klco},
		\citenamefont {Roggero},\ and\ \citenamefont {Savage}}]{Klco_2022}%
	\BibitemOpen
	\bibfield  {author} {\bibinfo {author} {\bibfnamefont {N.}~\bibnamefont
			{Klco}}, \bibinfo {author} {\bibfnamefont {A.}~\bibnamefont {Roggero}},\ and\
		\bibinfo {author} {\bibfnamefont {M.~J.}\ \bibnamefont {Savage}},\ }\bibfield
	{title} {\bibinfo {title} {Standard model physics and the digital quantum
			revolution: thoughts about the interface},\ }\href
	{https://doi.org/10.1088/1361-6633/ac58a4} {\bibfield  {journal} {\bibinfo
			{journal} {Reports on Progress in Physics}\ }\textbf {\bibinfo {volume}
			{85}},\ \bibinfo {pages} {064301} (\bibinfo {year} {2022})}\BibitemShut
	{NoStop}%
	\bibitem [{\citenamefont {Bauer}\ \emph {et~al.}(2023)\citenamefont {Bauer}
		\emph {et~al.}}]{Bauer:2022hpo}%
	\BibitemOpen
	\bibfield  {author} {\bibinfo {author} {\bibfnamefont {C.~W.}\ \bibnamefont
			{Bauer}} \emph {et~al.},\ }\bibfield  {title} {\bibinfo {title} {{Quantum
				Simulation for High-Energy Physics}},\ }\href
	{https://doi.org/10.1103/PRXQuantum.4.027001} {\bibfield  {journal} {\bibinfo
			{journal} {PRX Quantum}\ }\textbf {\bibinfo {volume} {4}},\ \bibinfo {pages}
		{027001} (\bibinfo {year} {2023})}\BibitemShut {NoStop}%
	\bibitem [{\citenamefont {Di~Meglio}\ \emph {et~al.}(2024)\citenamefont
		{Di~Meglio} \emph {et~al.}}]{PRXQuantum.5.037001}%
	\BibitemOpen
	\bibfield  {author} {\bibinfo {author} {\bibfnamefont {A.}~\bibnamefont
			{Di~Meglio}} \emph {et~al.},\ }\bibfield  {title} {\bibinfo {title} {Quantum
			computing for high-energy physics: State of the art and challenges},\ }\href
	{https://doi.org/10.1103/PRXQuantum.5.037001} {\bibfield  {journal} {\bibinfo
			{journal} {PRX Quantum}\ }\textbf {\bibinfo {volume} {5}},\ \bibinfo {pages}
		{037001} (\bibinfo {year} {2024})}\BibitemShut {NoStop}%
	\bibitem [{\citenamefont {Hall}\ \emph {et~al.}(2021)\citenamefont {Hall},
		\citenamefont {Roggero}, \citenamefont {Baroni},\ and\ \citenamefont
		{Carlson}}]{Hall:2021rbv}%
	\BibitemOpen
	\bibfield  {author} {\bibinfo {author} {\bibfnamefont {B.}~\bibnamefont
			{Hall}}, \bibinfo {author} {\bibfnamefont {A.}~\bibnamefont {Roggero}},
		\bibinfo {author} {\bibfnamefont {A.}~\bibnamefont {Baroni}},\ and\ \bibinfo
		{author} {\bibfnamefont {J.}~\bibnamefont {Carlson}},\ }\bibfield  {title}
	{\bibinfo {title} {{Simulation of collective neutrino oscillations on a
				quantum computer}},\ }\href {https://doi.org/10.1103/PhysRevD.104.063009}
	{\bibfield  {journal} {\bibinfo  {journal} {Phys. Rev. D}\ }\textbf {\bibinfo
			{volume} {104}},\ \bibinfo {pages} {063009} (\bibinfo {year}
		{2021})}\BibitemShut {NoStop}%
	\bibitem [{\citenamefont {Yeter-Aydeniz}\ \emph {et~al.}(2022)\citenamefont
		{Yeter-Aydeniz}, \citenamefont {Bangar}, \citenamefont {Siopsis},\ and\
		\citenamefont {Pooser}}]{Yeter-Aydeniz:2021olz}%
	\BibitemOpen
	\bibfield  {author} {\bibinfo {author} {\bibfnamefont {K.}~\bibnamefont
			{Yeter-Aydeniz}}, \bibinfo {author} {\bibfnamefont {S.}~\bibnamefont
			{Bangar}}, \bibinfo {author} {\bibfnamefont {G.}~\bibnamefont {Siopsis}},\
		and\ \bibinfo {author} {\bibfnamefont {R.~C.}\ \bibnamefont {Pooser}},\
	}\bibfield  {title} {\bibinfo {title} {{Collective neutrino oscillations on a
				quantum computer}},\ }\href {https://doi.org/10.1007/s11128-021-03348-x}
	{\bibfield  {journal} {\bibinfo  {journal} {Quant. Inf. Proc.}\ }\textbf
		{\bibinfo {volume} {21}},\ \bibinfo {pages} {84} (\bibinfo {year}
		{2022})}\BibitemShut {NoStop}%
	\bibitem [{\citenamefont {Illa}\ and\ \citenamefont
		{Savage}(2022)}]{PhysRevA.106.052605}%
	\BibitemOpen
	\bibfield  {author} {\bibinfo {author} {\bibfnamefont {M.}~\bibnamefont
			{Illa}}\ and\ \bibinfo {author} {\bibfnamefont {M.~J.}\ \bibnamefont
			{Savage}},\ }\bibfield  {title} {\bibinfo {title} {Basic elements for
			simulations of standard-model physics with quantum annealers: Multigrid and
			clock states},\ }\href {https://doi.org/10.1103/PhysRevA.106.052605}
	{\bibfield  {journal} {\bibinfo  {journal} {Phys. Rev. A}\ }\textbf {\bibinfo
			{volume} {106}},\ \bibinfo {pages} {052605} (\bibinfo {year}
		{2022})}\BibitemShut {NoStop}%
	\bibitem [{\citenamefont {Amitrano}\ \emph {et~al.}(2023)\citenamefont
		{Amitrano}, \citenamefont {Roggero}, \citenamefont {Luchi}, \citenamefont
		{Turro}, \citenamefont {Vespucci},\ and\ \citenamefont
		{Pederiva}}]{Amitrano:2022yyn}%
	\BibitemOpen
	\bibfield  {author} {\bibinfo {author} {\bibfnamefont {V.}~\bibnamefont
			{Amitrano}}, \bibinfo {author} {\bibfnamefont {A.}~\bibnamefont {Roggero}},
		\bibinfo {author} {\bibfnamefont {P.}~\bibnamefont {Luchi}}, \bibinfo
		{author} {\bibfnamefont {F.}~\bibnamefont {Turro}}, \bibinfo {author}
		{\bibfnamefont {L.}~\bibnamefont {Vespucci}},\ and\ \bibinfo {author}
		{\bibfnamefont {F.}~\bibnamefont {Pederiva}},\ }\bibfield  {title} {\bibinfo
		{title} {{Trapped-ion quantum simulation of collective neutrino
				oscillations}},\ }\href {https://doi.org/10.1103/PhysRevD.107.023007}
	{\bibfield  {journal} {\bibinfo  {journal} {Phys. Rev. D}\ }\textbf {\bibinfo
			{volume} {107}},\ \bibinfo {pages} {023007} (\bibinfo {year}
		{2023})}\BibitemShut {NoStop}%
	\bibitem [{\citenamefont {Illa}\ and\ \citenamefont
		{Savage}(2023{\natexlab{b}})}]{PhysRevLett.130.221003}%
	\BibitemOpen
	\bibfield  {author} {\bibinfo {author} {\bibfnamefont {M.}~\bibnamefont
			{Illa}}\ and\ \bibinfo {author} {\bibfnamefont {M.~J.}\ \bibnamefont
			{Savage}},\ }\bibfield  {title} {\bibinfo {title} {Multi-neutrino
			entanglement and correlations in dense neutrino systems},\ }\href
	{https://doi.org/10.1103/PhysRevLett.130.221003} {\bibfield  {journal}
		{\bibinfo  {journal} {Phys. Rev. Lett.}\ }\textbf {\bibinfo {volume} {130}},\
		\bibinfo {pages} {221003} (\bibinfo {year} {2023}{\natexlab{b}})}\BibitemShut
	{NoStop}%
	\bibitem [{\citenamefont {Siwach}\ \emph
		{et~al.}(2023{\natexlab{b}})\citenamefont {Siwach}, \citenamefont
		{Harrison},\ and\ \citenamefont {Balantekin}}]{Siwach:2023wzy}%
	\BibitemOpen
	\bibfield  {author} {\bibinfo {author} {\bibfnamefont {P.}~\bibnamefont
			{Siwach}}, \bibinfo {author} {\bibfnamefont {K.}~\bibnamefont {Harrison}},\
		and\ \bibinfo {author} {\bibfnamefont {A.~B.}\ \bibnamefont {Balantekin}},\
	}\bibfield  {title} {\bibinfo {title} {{Collective neutrino oscillations on a
				quantum computer with hybrid quantum-classical algorithm}},\ }\href
	{https://doi.org/10.1103/PhysRevD.108.083039} {\bibfield  {journal} {\bibinfo
			{journal} {Phys. Rev. D}\ }\textbf {\bibinfo {volume} {108}},\ \bibinfo
		{pages} {083039} (\bibinfo {year} {2023}{\natexlab{b}})}\BibitemShut
	{NoStop}%
	\bibitem [{\citenamefont {Turro}\ \emph {et~al.}(2025)\citenamefont {Turro},
		\citenamefont {Chernyshev}, \citenamefont {Bhaskar},\ and\ \citenamefont
		{Illa}}]{Turro:2025}%
	\BibitemOpen
	\bibfield  {author} {\bibinfo {author} {\bibfnamefont {F.}~\bibnamefont
			{Turro}}, \bibinfo {author} {\bibfnamefont {I.~A.}\ \bibnamefont
			{Chernyshev}}, \bibinfo {author} {\bibfnamefont {R.}~\bibnamefont
			{Bhaskar}},\ and\ \bibinfo {author} {\bibfnamefont {M.}~\bibnamefont
			{Illa}},\ }\href {https://arxiv.org/abs/2407.13914} {\bibinfo {title} {Qutrit
			and qubit circuits for three-flavor collective neutrino oscillations}}
	(\bibinfo {year} {2025}),\ \Eprint {https://arxiv.org/abs/2407.13914}
	{arXiv:2407.13914 [quant-ph]} \BibitemShut {NoStop}%
	\bibitem [{\citenamefont {Chernyshev}(2025)}]{PhysRevD.111.043017}%
	\BibitemOpen
	\bibfield  {author} {\bibinfo {author} {\bibfnamefont {I.~A.}\ \bibnamefont
			{Chernyshev}},\ }\bibfield  {title} {\bibinfo {title} {Three-flavor
			collective neutrino oscillation simulations on a qubit quantum annealer},\
	}\href {https://doi.org/10.1103/PhysRevD.111.043017} {\bibfield  {journal}
		{\bibinfo  {journal} {Phys. Rev. D}\ }\textbf {\bibinfo {volume} {111}},\
		\bibinfo {pages} {043017} (\bibinfo {year} {2025})}\BibitemShut {NoStop}%
	\bibitem [{\citenamefont {Childs}\ \emph {et~al.}(2021)\citenamefont {Childs},
		\citenamefont {Su}, \citenamefont {Tran}, \citenamefont {Wiebe},\ and\
		\citenamefont {Zhu}}]{PhysRevX.11.011020}%
	\BibitemOpen
	\bibfield  {author} {\bibinfo {author} {\bibfnamefont {A.~M.}\ \bibnamefont
			{Childs}}, \bibinfo {author} {\bibfnamefont {Y.}~\bibnamefont {Su}}, \bibinfo
		{author} {\bibfnamefont {M.~C.}\ \bibnamefont {Tran}}, \bibinfo {author}
		{\bibfnamefont {N.}~\bibnamefont {Wiebe}},\ and\ \bibinfo {author}
		{\bibfnamefont {S.}~\bibnamefont {Zhu}},\ }\bibfield  {title} {\bibinfo
		{title} {Theory of trotter error with commutator scaling},\ }\href
	{https://doi.org/10.1103/PhysRevX.11.011020} {\bibfield  {journal} {\bibinfo
			{journal} {Phys. Rev. X}\ }\textbf {\bibinfo {volume} {11}},\ \bibinfo
		{pages} {011020} (\bibinfo {year} {2021})}\BibitemShut {NoStop}%
	\bibitem [{\citenamefont {Molewski}\ and\ \citenamefont
		{Jones}(2022)}]{PhysRevD.105.056024}%
	\BibitemOpen
	\bibfield  {author} {\bibinfo {author} {\bibfnamefont {M.~J.}\ \bibnamefont
			{Molewski}}\ and\ \bibinfo {author} {\bibfnamefont {B.~J.~P.}\ \bibnamefont
			{Jones}},\ }\bibfield  {title} {\bibinfo {title} {Scalable qubit
			representations of neutrino mixing matrices},\ }\href
	{https://doi.org/10.1103/PhysRevD.105.056024} {\bibfield  {journal} {\bibinfo
			{journal} {Phys. Rev. D}\ }\textbf {\bibinfo {volume} {105}},\ \bibinfo
		{pages} {056024} (\bibinfo {year} {2022})}\BibitemShut {NoStop}%
	\bibitem [{\citenamefont {Wang}\ \emph {et~al.}(2020)\citenamefont {Wang},
		\citenamefont {Hu}, \citenamefont {Sanders},\ and\ \citenamefont
		{Kais}}]{Wang_2020}%
	\BibitemOpen
	\bibfield  {author} {\bibinfo {author} {\bibfnamefont {Y.}~\bibnamefont
			{Wang}}, \bibinfo {author} {\bibfnamefont {Z.}~\bibnamefont {Hu}}, \bibinfo
		{author} {\bibfnamefont {B.~C.}\ \bibnamefont {Sanders}},\ and\ \bibinfo
		{author} {\bibfnamefont {S.}~\bibnamefont {Kais}},\ }\bibfield  {title}
	{\bibinfo {title} {Qudits and high-dimensional quantum computing},\
	}\bibfield  {journal} {\bibinfo  {journal} {Frontiers in Physics}\ }\textbf
	{\bibinfo {volume} {8}},\ \href {https://doi.org/10.3389/fphy.2020.589504}
	{10.3389/fphy.2020.589504} (\bibinfo {year} {2020})\BibitemShut {NoStop}%
	\bibitem [{\citenamefont {Morvan}\ \emph {et~al.}(2021)\citenamefont {Morvan},
		\citenamefont {Ramasesh}, \citenamefont {Blok}, \citenamefont {Kreikebaum},
		\citenamefont {O'Brien}, \citenamefont {Chen}, \citenamefont {Mitchell},
		\citenamefont {Naik}, \citenamefont {Santiago},\ and\ \citenamefont
		{Siddiqi}}]{PhysRevLett.126.210504}%
	\BibitemOpen
	\bibfield  {author} {\bibinfo {author} {\bibfnamefont {A.}~\bibnamefont
			{Morvan}}, \bibinfo {author} {\bibfnamefont {V.~V.}\ \bibnamefont
			{Ramasesh}}, \bibinfo {author} {\bibfnamefont {M.~S.}\ \bibnamefont {Blok}},
		\bibinfo {author} {\bibfnamefont {J.~M.}\ \bibnamefont {Kreikebaum}},
		\bibinfo {author} {\bibfnamefont {K.}~\bibnamefont {O'Brien}}, \bibinfo
		{author} {\bibfnamefont {L.}~\bibnamefont {Chen}}, \bibinfo {author}
		{\bibfnamefont {B.~K.}\ \bibnamefont {Mitchell}}, \bibinfo {author}
		{\bibfnamefont {R.~K.}\ \bibnamefont {Naik}}, \bibinfo {author}
		{\bibfnamefont {D.~I.}\ \bibnamefont {Santiago}},\ and\ \bibinfo {author}
		{\bibfnamefont {I.}~\bibnamefont {Siddiqi}},\ }\bibfield  {title} {\bibinfo
		{title} {Qutrit randomized benchmarking},\ }\href
	{https://doi.org/10.1103/PhysRevLett.126.210504} {\bibfield  {journal}
		{\bibinfo  {journal} {Phys. Rev. Lett.}\ }\textbf {\bibinfo {volume} {126}},\
		\bibinfo {pages} {210504} (\bibinfo {year} {2021})}\BibitemShut {NoStop}%
	\bibitem [{\citenamefont {de~Guise}\ \emph {et~al.}(2018)\citenamefont
		{de~Guise}, \citenamefont {Di~Matteo},\ and\ \citenamefont
		{S\'anchez-Soto}}]{PhysRevA.97.022328}%
	\BibitemOpen
	\bibfield  {author} {\bibinfo {author} {\bibfnamefont {H.}~\bibnamefont
			{de~Guise}}, \bibinfo {author} {\bibfnamefont {O.}~\bibnamefont
			{Di~Matteo}},\ and\ \bibinfo {author} {\bibfnamefont {L.~L.}\ \bibnamefont
			{S\'anchez-Soto}},\ }\bibfield  {title} {\bibinfo {title} {Simple
			factorization of unitary transformations},\ }\href
	{https://doi.org/10.1103/PhysRevA.97.022328} {\bibfield  {journal} {\bibinfo
			{journal} {Phys. Rev. A}\ }\textbf {\bibinfo {volume} {97}},\ \bibinfo
		{pages} {022328} (\bibinfo {year} {2018})}\BibitemShut {NoStop}%
	\bibitem [{\citenamefont {Goss}\ \emph {et~al.}(2022)\citenamefont {Goss},
		\citenamefont {Morvan}, \citenamefont {Marinelli}, \citenamefont {Mitchell},
		\citenamefont {Nguyen}, \citenamefont {Naik}, \citenamefont {Chen},
		\citenamefont {J{\"u}nger}, \citenamefont {Kreikebaum}, \citenamefont
		{Santiago}, \citenamefont {Wallman},\ and\ \citenamefont
		{Siddiqi}}]{goss-high}%
	\BibitemOpen
	\bibfield  {author} {\bibinfo {author} {\bibfnamefont {N.}~\bibnamefont
			{Goss}}, \bibinfo {author} {\bibfnamefont {A.}~\bibnamefont {Morvan}},
		\bibinfo {author} {\bibfnamefont {B.}~\bibnamefont {Marinelli}}, \bibinfo
		{author} {\bibfnamefont {B.~K.}\ \bibnamefont {Mitchell}}, \bibinfo {author}
		{\bibfnamefont {L.~B.}\ \bibnamefont {Nguyen}}, \bibinfo {author}
		{\bibfnamefont {R.~K.}\ \bibnamefont {Naik}}, \bibinfo {author}
		{\bibfnamefont {L.}~\bibnamefont {Chen}}, \bibinfo {author} {\bibfnamefont
			{C.}~\bibnamefont {J{\"u}nger}}, \bibinfo {author} {\bibfnamefont {J.~M.}\
			\bibnamefont {Kreikebaum}}, \bibinfo {author} {\bibfnamefont {D.~I.}\
			\bibnamefont {Santiago}}, \bibinfo {author} {\bibfnamefont {J.~J.}\
			\bibnamefont {Wallman}},\ and\ \bibinfo {author} {\bibfnamefont
			{I.}~\bibnamefont {Siddiqi}},\ }\bibfield  {title} {\bibinfo {title}
		{High-fidelity qutrit entangling gates for superconducting circuits},\ }\href
	{https://doi.org/10.1038/s41467-022-34851-z} {\bibfield  {journal} {\bibinfo
			{journal} {Nature Communications}\ }\textbf {\bibinfo {volume} {13}},\
		\bibinfo {pages} {7481} (\bibinfo {year} {2022})}\BibitemShut {NoStop}%
	\bibitem [{\citenamefont {Nguyen}\ \emph {et~al.}(2024)\citenamefont {Nguyen},
		\citenamefont {Goss}, \citenamefont {Siva}, \citenamefont {Kim},
		\citenamefont {Younis}, \citenamefont {Qing}, \citenamefont {Hashim},
		\citenamefont {Santiago},\ and\ \citenamefont {Siddiqi}}]{empowering}%
	\BibitemOpen
	\bibfield  {author} {\bibinfo {author} {\bibfnamefont {L.~B.}\ \bibnamefont
			{Nguyen}}, \bibinfo {author} {\bibfnamefont {N.}~\bibnamefont {Goss}},
		\bibinfo {author} {\bibfnamefont {K.}~\bibnamefont {Siva}}, \bibinfo {author}
		{\bibfnamefont {Y.}~\bibnamefont {Kim}}, \bibinfo {author} {\bibfnamefont
			{E.}~\bibnamefont {Younis}}, \bibinfo {author} {\bibfnamefont
			{B.}~\bibnamefont {Qing}}, \bibinfo {author} {\bibfnamefont {A.}~\bibnamefont
			{Hashim}}, \bibinfo {author} {\bibfnamefont {D.~I.}\ \bibnamefont
			{Santiago}},\ and\ \bibinfo {author} {\bibfnamefont {I.}~\bibnamefont
			{Siddiqi}},\ }\bibfield  {title} {\bibinfo {title} {Empowering a qudit-based
			quantum processor by traversing the dual bosonic ladder},\ }\href
	{https://doi.org/10.1038/s41467-024-51434-2} {\bibfield  {journal} {\bibinfo
			{journal} {Nature Communications}\ }\textbf {\bibinfo {volume} {15}},\
		\bibinfo {pages} {7117} (\bibinfo {year} {2024})}\BibitemShut {NoStop}%
	\bibitem [{\citenamefont {Blok}\ \emph {et~al.}(2021)\citenamefont {Blok},
		\citenamefont {Ramasesh}, \citenamefont {Schuster}, \citenamefont {O'Brien},
		\citenamefont {Kreikebaum}, \citenamefont {Dahlen}, \citenamefont {Morvan},
		\citenamefont {Yoshida}, \citenamefont {Yao},\ and\ \citenamefont
		{Siddiqi}}]{Blok2021}%
	\BibitemOpen
	\bibfield  {author} {\bibinfo {author} {\bibfnamefont {M.~S.}\ \bibnamefont
			{Blok}}, \bibinfo {author} {\bibfnamefont {V.~V.}\ \bibnamefont {Ramasesh}},
		\bibinfo {author} {\bibfnamefont {T.}~\bibnamefont {Schuster}}, \bibinfo
		{author} {\bibfnamefont {K.}~\bibnamefont {O'Brien}}, \bibinfo {author}
		{\bibfnamefont {J.~M.}\ \bibnamefont {Kreikebaum}}, \bibinfo {author}
		{\bibfnamefont {D.}~\bibnamefont {Dahlen}}, \bibinfo {author} {\bibfnamefont
			{A.}~\bibnamefont {Morvan}}, \bibinfo {author} {\bibfnamefont
			{B.}~\bibnamefont {Yoshida}}, \bibinfo {author} {\bibfnamefont {N.~Y.}\
			\bibnamefont {Yao}},\ and\ \bibinfo {author} {\bibfnamefont {I.}~\bibnamefont
			{Siddiqi}},\ }\bibfield  {title} {\bibinfo {title} {Quantum information
			scrambling on a superconducting qutrit processor},\ }\href
	{https://doi.org/10.1103/PhysRevX.11.021010} {\bibfield  {journal} {\bibinfo
			{journal} {Phys. Rev. X}\ }\textbf {\bibinfo {volume} {11}},\ \bibinfo
		{pages} {021010} (\bibinfo {year} {2021})}\BibitemShut {NoStop}%
	\bibitem [{\citenamefont {Urbanek}\ \emph {et~al.}(2021)\citenamefont
		{Urbanek}, \citenamefont {Nachman}, \citenamefont {Pascuzzi}, \citenamefont
		{He}, \citenamefont {Bauer},\ and\ \citenamefont
		{de~Jong}}]{PhysRevLett.127.270502}%
	\BibitemOpen
	\bibfield  {author} {\bibinfo {author} {\bibfnamefont {M.}~\bibnamefont
			{Urbanek}}, \bibinfo {author} {\bibfnamefont {B.}~\bibnamefont {Nachman}},
		\bibinfo {author} {\bibfnamefont {V.~R.}\ \bibnamefont {Pascuzzi}}, \bibinfo
		{author} {\bibfnamefont {A.}~\bibnamefont {He}}, \bibinfo {author}
		{\bibfnamefont {C.~W.}\ \bibnamefont {Bauer}},\ and\ \bibinfo {author}
		{\bibfnamefont {W.~A.}\ \bibnamefont {de~Jong}},\ }\bibfield  {title}
	{\bibinfo {title} {Mitigating depolarizing noise on quantum computers with
			noise-estimation circuits},\ }\href
	{https://doi.org/10.1103/PhysRevLett.127.270502} {\bibfield  {journal}
		{\bibinfo  {journal} {Phys. Rev. Lett.}\ }\textbf {\bibinfo {volume} {127}},\
		\bibinfo {pages} {270502} (\bibinfo {year} {2021})}\BibitemShut {NoStop}%
	\bibitem [{\citenamefont {A~Rahman}\ \emph {et~al.}(2022)\citenamefont
		{A~Rahman}, \citenamefont {Lewis}, \citenamefont {Mendicelli},\ and\
		\citenamefont {Powell}}]{PhysRevD.106.074502}%
	\BibitemOpen
	\bibfield  {author} {\bibinfo {author} {\bibfnamefont {S.}~\bibnamefont
			{A~Rahman}}, \bibinfo {author} {\bibfnamefont {R.}~\bibnamefont {Lewis}},
		\bibinfo {author} {\bibfnamefont {E.}~\bibnamefont {Mendicelli}},\ and\
		\bibinfo {author} {\bibfnamefont {S.}~\bibnamefont {Powell}},\ }\bibfield
	{title} {\bibinfo {title} {Self-mitigating trotter circuits for su(2) lattice
			gauge theory on a quantum computer},\ }\href
	{https://doi.org/10.1103/PhysRevD.106.074502} {\bibfield  {journal} {\bibinfo
			{journal} {Phys. Rev. D}\ }\textbf {\bibinfo {volume} {106}},\ \bibinfo
		{pages} {074502} (\bibinfo {year} {2022})}\BibitemShut {NoStop}%
	\bibitem [{\citenamefont {Farrell}\ \emph {et~al.}(2024)\citenamefont
		{Farrell}, \citenamefont {Illa}, \citenamefont {Ciavarella},\ and\
		\citenamefont {Savage}}]{PRXQuantum.5.020315}%
	\BibitemOpen
	\bibfield  {author} {\bibinfo {author} {\bibfnamefont {R.~C.}\ \bibnamefont
			{Farrell}}, \bibinfo {author} {\bibfnamefont {M.}~\bibnamefont {Illa}},
		\bibinfo {author} {\bibfnamefont {A.~N.}\ \bibnamefont {Ciavarella}},\ and\
		\bibinfo {author} {\bibfnamefont {M.~J.}\ \bibnamefont {Savage}},\ }\bibfield
	{title} {\bibinfo {title} {Scalable circuits for preparing ground states on
			digital quantum computers: The schwinger model vacuum on 100 qubits},\ }\href
	{https://doi.org/10.1103/PRXQuantum.5.020315} {\bibfield  {journal} {\bibinfo
			{journal} {PRX Quantum}\ }\textbf {\bibinfo {volume} {5}},\ \bibinfo {pages}
		{020315} (\bibinfo {year} {2024})}\BibitemShut {NoStop}%
	\bibitem [{\citenamefont {Kiss}\ \emph {et~al.}(2025)\citenamefont {Kiss},
		\citenamefont {Grossi},\ and\ \citenamefont {Roggero}}]{PhysRevD.111.034504}%
	\BibitemOpen
	\bibfield  {author} {\bibinfo {author} {\bibfnamefont {O.}~\bibnamefont
			{Kiss}}, \bibinfo {author} {\bibfnamefont {M.}~\bibnamefont {Grossi}},\ and\
		\bibinfo {author} {\bibfnamefont {A.}~\bibnamefont {Roggero}},\ }\bibfield
	{title} {\bibinfo {title} {Quantum error mitigation for fourier moment
			computation},\ }\href {https://doi.org/10.1103/PhysRevD.111.034504}
	{\bibfield  {journal} {\bibinfo  {journal} {Phys. Rev. D}\ }\textbf {\bibinfo
			{volume} {111}},\ \bibinfo {pages} {034504} (\bibinfo {year}
		{2025})}\BibitemShut {NoStop}%
	\bibitem [{\citenamefont {Hashim}\ \emph {et~al.}(2021)\citenamefont {Hashim},
		\citenamefont {Naik}, \citenamefont {Morvan}, \citenamefont {Ville},
		\citenamefont {Mitchell}, \citenamefont {Kreikebaum}, \citenamefont {Davis},
		\citenamefont {Smith}, \citenamefont {Iancu}, \citenamefont {O'Brien},
		\citenamefont {Hincks}, \citenamefont {Wallman}, \citenamefont {Emerson},\
		and\ \citenamefont {Siddiqi}}]{PhysRevX.11.041039}%
	\BibitemOpen
	\bibfield  {author} {\bibinfo {author} {\bibfnamefont {A.}~\bibnamefont
			{Hashim}}, \bibinfo {author} {\bibfnamefont {R.~K.}\ \bibnamefont {Naik}},
		\bibinfo {author} {\bibfnamefont {A.}~\bibnamefont {Morvan}}, \bibinfo
		{author} {\bibfnamefont {J.-L.}\ \bibnamefont {Ville}}, \bibinfo {author}
		{\bibfnamefont {B.}~\bibnamefont {Mitchell}}, \bibinfo {author}
		{\bibfnamefont {J.~M.}\ \bibnamefont {Kreikebaum}}, \bibinfo {author}
		{\bibfnamefont {M.}~\bibnamefont {Davis}}, \bibinfo {author} {\bibfnamefont
			{E.}~\bibnamefont {Smith}}, \bibinfo {author} {\bibfnamefont
			{C.}~\bibnamefont {Iancu}}, \bibinfo {author} {\bibfnamefont {K.~P.}\
			\bibnamefont {O'Brien}}, \bibinfo {author} {\bibfnamefont {I.}~\bibnamefont
			{Hincks}}, \bibinfo {author} {\bibfnamefont {J.~J.}\ \bibnamefont {Wallman}},
		\bibinfo {author} {\bibfnamefont {J.}~\bibnamefont {Emerson}},\ and\ \bibinfo
		{author} {\bibfnamefont {I.}~\bibnamefont {Siddiqi}},\ }\bibfield  {title}
	{\bibinfo {title} {Randomized compiling for scalable quantum computing on a
			noisy superconducting quantum processor},\ }\href
	{https://doi.org/10.1103/PhysRevX.11.041039} {\bibfield  {journal} {\bibinfo
			{journal} {Phys. Rev. X}\ }\textbf {\bibinfo {volume} {11}},\ \bibinfo
		{pages} {041039} (\bibinfo {year} {2021})}\BibitemShut {NoStop}%
	\bibitem [{\citenamefont {Goss}\ \emph {et~al.}(2024)\citenamefont {Goss},
		\citenamefont {Ferracin}, \citenamefont {Hashim}, \citenamefont
		{Carignan-Dugas}, \citenamefont {Kreikebaum}, \citenamefont {Naik},
		\citenamefont {Santiago},\ and\ \citenamefont {Siddiqi}}]{extending}%
	\BibitemOpen
	\bibfield  {author} {\bibinfo {author} {\bibfnamefont {N.}~\bibnamefont
			{Goss}}, \bibinfo {author} {\bibfnamefont {S.}~\bibnamefont {Ferracin}},
		\bibinfo {author} {\bibfnamefont {A.}~\bibnamefont {Hashim}}, \bibinfo
		{author} {\bibfnamefont {A.}~\bibnamefont {Carignan-Dugas}}, \bibinfo
		{author} {\bibfnamefont {J.~M.}\ \bibnamefont {Kreikebaum}}, \bibinfo
		{author} {\bibfnamefont {R.~K.}\ \bibnamefont {Naik}}, \bibinfo {author}
		{\bibfnamefont {D.~I.}\ \bibnamefont {Santiago}},\ and\ \bibinfo {author}
		{\bibfnamefont {I.}~\bibnamefont {Siddiqi}},\ }\bibfield  {title} {\bibinfo
		{title} {Extending the computational reach of a superconducting qutrit
			processor},\ }\href {https://doi.org/10.1038/s41534-024-00892-z} {\bibfield
		{journal} {\bibinfo  {journal} {npj Quantum Information}\ }\textbf {\bibinfo
			{volume} {10}},\ \bibinfo {pages} {101} (\bibinfo {year} {2024})}\BibitemShut
	{NoStop}%
	\bibitem [{\citenamefont {Sung}\ \emph {et~al.}(2023)\citenamefont {Sung},
		\citenamefont {Rančić}, \citenamefont {Lanes},\ and\ \citenamefont
		{Bronn}}]{Sung_2023}%
	\BibitemOpen
	\bibfield  {author} {\bibinfo {author} {\bibfnamefont {K.~J.}\ \bibnamefont
			{Sung}}, \bibinfo {author} {\bibfnamefont {M.~J.}\ \bibnamefont {Rančić}},
		\bibinfo {author} {\bibfnamefont {O.~T.}\ \bibnamefont {Lanes}},\ and\
		\bibinfo {author} {\bibfnamefont {N.~T.}\ \bibnamefont {Bronn}},\ }\bibfield
	{title} {\bibinfo {title} {Simulating majorana zero modes on a noisy quantum
			processor},\ }\href {https://doi.org/10.1088/2058-9565/acb796} {\bibfield
		{journal} {\bibinfo  {journal} {Quantum Science and Technology}\ }\textbf
		{\bibinfo {volume} {8}},\ \bibinfo {pages} {025010} (\bibinfo {year}
		{2023})}\BibitemShut {NoStop}%
	\bibitem [{\citenamefont {Stetina}\ \emph {et~al.}(2022)\citenamefont
		{Stetina}, \citenamefont {Ciavarella}, \citenamefont {Li},\ and\
		\citenamefont {Wiebe}}]{Stetina2022simulatingeffective}%
	\BibitemOpen
	\bibfield  {author} {\bibinfo {author} {\bibfnamefont {T.~F.}\ \bibnamefont
			{Stetina}}, \bibinfo {author} {\bibfnamefont {A.}~\bibnamefont {Ciavarella}},
		\bibinfo {author} {\bibfnamefont {X.}~\bibnamefont {Li}},\ and\ \bibinfo
		{author} {\bibfnamefont {N.}~\bibnamefont {Wiebe}},\ }\bibfield  {title}
	{\bibinfo {title} {Simulating {E}ffective {QED} on {Q}uantum {C}omputers},\
	}\href {https://doi.org/10.22331/q-2022-01-18-622} {\bibfield  {journal}
		{\bibinfo  {journal} {{Quantum}}\ }\textbf {\bibinfo {volume} {6}},\ \bibinfo
		{pages} {622} (\bibinfo {year} {2022})}\BibitemShut {NoStop}%
	\bibitem [{\citenamefont {Farrell}\ \emph {et~al.}(2023)\citenamefont
		{Farrell}, \citenamefont {Chernyshev}, \citenamefont {Powell}, \citenamefont
		{Zemlevskiy}, \citenamefont {Illa},\ and\ \citenamefont
		{Savage}}]{PhysRevD.107.054512}%
	\BibitemOpen
	\bibfield  {author} {\bibinfo {author} {\bibfnamefont {R.~C.}\ \bibnamefont
			{Farrell}}, \bibinfo {author} {\bibfnamefont {I.~A.}\ \bibnamefont
			{Chernyshev}}, \bibinfo {author} {\bibfnamefont {S.~J.~M.}\ \bibnamefont
			{Powell}}, \bibinfo {author} {\bibfnamefont {N.~A.}\ \bibnamefont
			{Zemlevskiy}}, \bibinfo {author} {\bibfnamefont {M.}~\bibnamefont {Illa}},\
		and\ \bibinfo {author} {\bibfnamefont {M.~J.}\ \bibnamefont {Savage}},\
	}\bibfield  {title} {\bibinfo {title} {Preparations for quantum simulations
			of quantum chromodynamics in $1+1$ dimensions. i. axial gauge},\ }\href
	{https://doi.org/10.1103/PhysRevD.107.054512} {\bibfield  {journal} {\bibinfo
			{journal} {Phys. Rev. D}\ }\textbf {\bibinfo {volume} {107}},\ \bibinfo
		{pages} {054512} (\bibinfo {year} {2023})}\BibitemShut {NoStop}%
\end{thebibliography}
\end{document}